\documentclass[letterpaper,11pt]{article}

\usepackage{amsmath,amsthm,amssymb,amsfonts}
\usepackage{hyperref,lineno}
\usepackage{etex}
\usepackage{natbib}
\usepackage{mathpazo}
\usepackage{multimedia}
\usepackage{graphicx, color}
\usepackage{mathtools}
\usepackage{esint}
\usepackage{url}
\usepackage{graphicx}
\usepackage{relsize}
\usepackage{amsfonts}
\usepackage{bm}
\usepackage{fancyheadings}
\usepackage{float}
\usepackage{color}
\usepackage{mathrsfs}
\usepackage{setspace}
\usepackage[mathscr]{euscript}
\usepackage{caption}
\usepackage{subcaption}
\usepackage{pdflscape}
\usepackage{booktabs}
\usepackage[utf8]{inputenc}
\usepackage[T1]{fontenc}
\usepackage{geometry}
\usepackage[dvipsnames]{xcolor}
\usepackage{tikz}
\usepackage{caption}
\usepackage{subcaption}
\usepackage[bottom]{footmisc}
\usepackage{comment}
\usepackage{siunitx} 

\hypersetup{
    colorlinks,
    linkcolor={red!50!black},
    citecolor={blue!50!black},
    urlcolor={blue!80!black}
}
\defcitealias{GMV}{GMV}
\defcitealias{DFH}{DFH}
\usetikzlibrary{decorations.pathreplacing}
\usetikzlibrary{shapes}
\usetikzlibrary{arrows.meta,arrows}
\usetikzlibrary{fit,positioning}
\usetikzlibrary {intersections}
\tikzset{>=latex}

\newtheorem{assumption}{Assumption}
\newtheorem{proposition}{Proposition}[section]
\newtheorem{lemma}{Lemma}[section]
\newtheorem{corollary}{Corollary}[section]
\newtheorem{theorem}{Theorem}

\theoremstyle{definition}
\newtheorem{remark}{Remark}[section]
\newtheorem{example}{Example}[section]

\usepackage{nicefrac}

\title{Characterizing Optimality in Dynamic Settings: A Monotonicity-based Approach
\thanks{We thank participants at various seminars as well as Federico Echenique. Usual disclaimer applies. }}

\author{Zhuokai Huang\thanks{Harvard University, zhuokaihuang@fas.harvard.edu} \and
  Demian Pouzo\thanks{University of California, Berkeley, dpouzo@berkeley.edu} \and
  Andrés Rodríguez-Clare\thanks{University of California, Berkeley, CEPR, and NBER, arc@berkeley.edu}
}

\date{August 2025}

\clearpage

\begin{document}  

\maketitle

\vspace{2em}

\begin{abstract}

\noindent We develop a novel analytical method for studying optimal paths in dynamic optimization problems under general monotonicity conditions. The method centers on a \emph{locator function}—a simple object constructed directly from the model’s primitives—whose roots identify interior steady states and whose slope determines their local stability. Under strict concavity of the payoff function, the locator function also characterizes basins of attraction, yielding a complete description of qualitative dynamics. Without concavity, it can still deliver sharp results: if the function is single crossing from above, its root identifies a globally stable steady state; if the locator function is inverted-U-shaped with two interior roots (a typical case), only the higher root can be a locally stable interior steady state. The locator function further enables comparative statics of steady states with respect to parameters through direct analysis of its derivatives. These results are obtained without solving the full dynamic program. We illustrate the approach using a generalized neoclassical growth model, a rational (un)fitness model, and a learning-by-doing economy.
\end{abstract}

\thispagestyle{empty}

\clearpage

\setcounter{page}{1}

  \section{Introduction}
\label{sec:intro}

Dynamic optimization problems are central to many areas of economic analysis, from growth and consumption to investment and environmental policy. In many such settings, the goal is not to compute the full optimal path—that is, the sequence of state and control variables over time that solves the dynamic optimization problem—but rather to understand its qualitative features—such as the location and stability of steady states, and the direction in which the system evolves from a given initial condition. Standard methods based on dynamic programming characterize optimal behavior via the Bellman equation, but solving this equation typically requires numerical methods and yields limited insight into these qualitative dynamics, especially in the presence of multiple steady states. In this paper, we develop an analytical approach for characterizing optimal paths in a class of dynamic optimization problems satisfying a set of monotonicity conditions. Our method relies on direct analysis of the model’s primitives and provides sharp results on the set of steady states, their local stability, and the global behavior of the system. 

The key innovation in our analysis is the introduction of the \emph{locator function}, defined as $$ \mathcal{L}(s)  = \pi_2(s, s) + \delta \pi_1(s, s),$$ which is a simple scalar function of the state $s$ constructed from the primitives of the model: the per-period payoff, $\pi$, and the discount factor $\delta$.\footnote{$\pi_{l}$ denotes the partial derivative with respect to the $l$-th arguments.}  As shown in our main theorem, the roots of this function identify the interior steady states, and their local stability is determined by the sign of the derivative at those points: a negative derivative indicates local stability, while a positive derivative implies instability. 
		
Building on this result, we show that the locator function provides easy-to-verify sufficient conditions for the existence and the location of a globally stable steady state. In particular, if the locator function is single crossing from above---i.e., it has a unique root and is positive to the left and negative to the right---then any optimal path converges globally to that root from any interior initial condition. This allows applied researchers to verify global stability using only properties of the payoff function and its partial derivatives, without solving the whole dynamic programming problem. 
	
We also consider what is arguably the typical non-convex case: a locator function that is inverted U-shaped with two interior roots. In this setting, we show that the only possible locally stable steady state in the interior of the state space is the highest root. Moreover, we show that this is indeed the case if the payoff function is strictly concave in its second argument and the myopic policy function—which is trivial to characterize analytically—has an interior steady state. 
	


If the payoff function is strictly concave --- the standard textbook assumption (e.g., \citet{stokey1989recursive}) --- then, under a mild condition on the payoff function, the locator function not only identifies the interior steady states but also fully characterizes their basins of attraction. That is, we obtain a complete description of the qualitative dynamics: for any initial condition, we can determine the direction of motion and the limiting behavior of the optimal path. This result should be compared with the standard local approach, which relies on linearizing the system around a steady state and assessing stability through the eigenvalues of the Jacobian. That method typically establishes only the existence of basins of attraction, without characterizing them.
	
Finally, we demonstrate that the locator function provides a tractable tool for conducting comparative statics with respect to payoff parameters in stable interior steady states. Focusing on stable steady states is natural, since these are the only long-run outcomes attained by optimal paths, whereas unstable steady states cannot be reached except from trivial initial conditions (the steady state itself). 

Our approach relies on two key monotonicity conditions imposed on the payoff function. First, we assume that $\pi$ is increasing in the current state, so that higher values of the state are beneficial. Second, we assume that the current and future states are complementary in the sense that the marginal payoff from a higher future state increases with the level of the current state—that is, $\pi_2(s, s')$ is increasing in $s$. Importantly, in contrast to the textbook analysis, our approach does not rely on assuming strict concavity of the payoff function.

We illustrate our approach with one example and two applications, each allowing for multiple steady states. This permits cases where, for instance, a low initial value of the state variable makes it optimal to converge to a steady state associated with a lower payoff. The example is the neoclassical growth model with a production function that may feature non-convexities—as in \citet{skiba1978optimal}—and is further extended to allow for kinks. For the case of a smooth convex-concave production function, the optimal path for capital was fully characterized by \citet{dechert2012complete}, but their analysis relies on mathematical tools that may not easily extend to other settings. We show how our more general and simpler approach reproduces the key results more directly.

Our first application draws on the rational addiction framework of \citet{becker1988theory}, adapting it to health and exercise. Agents decide whether and how much to exercise, accounting for its impact on fitness and, in turn, on future utility. A key feature is the \emph{endorphins effect}: a direct utility gain from exercising that increases with current fitness. When sufficiently strong, this effect violates strict concavity, showing that the assumption is not merely technical but rules out economically relevant behaviors. In particular, it makes $\pi_2(s, s')$ increasing in $s$, creating the possibility of multiple steady states—from a low-fitness “couch potato” state to a high-fitness active state. This is consistent with evidence in \citet{charness2009incentives} that short-term interventions can trigger lasting behavioral change.

Our second application connects to a large literature studying how learning-by-doing externalities can give rise to multiple steady states and inefficient specialization in market economies absent industrial policies (see \citet{krugman1987}, \citet{lucas1988}, and \citet{young1991}). Instead of considering market outcomes, we take a step back and analyze the problem from the perspective of a benevolent social planner, as in \citet{bardhan1971} and \citet{melitz2005}. Using the locator function, we derive a simple condition for the existence of a globally stable interior steady state: the locator function must be single crossing from above. When this condition fails, either the lowest possible state is globally stable, or there may be multiple steady states, with only the highest interior root of the locator function potentially locally stable. 

The remainder of the paper proceeds as follows. Section \ref{sec:environment} presents the general setup and assumptions. Section \ref{sec:analytical} introduces the locator function, presents our main result, and illustrates its usefulness by applying it to the convex-concave neoclassical growth model. Section \ref{sec:refinements} provides refinements of this result under further assumptions. Section \ref{sec:applications} presents our two applications, namely a model of rational (un)fitness and a model of learning by doing. Section \ref{sec:remarks} discusses the relationship between our approach and existing methods in the literature. Section \ref{sec:conclusion} briefly concludes.

  \section{Environment}
\label{sec:environment}

In this section we define the planner's problem. The planner chooses a sequence of the aggregate state, $(s_{t})_{t=1}^{\infty}$, given an initial value, $s_{0}$, to maximize $\sum_{t=0}^{\infty} \delta^{t} \pi(s_{t},s_{t+1}) $ subject to $s_{t+1} \in \Upsilon(s_{t})$, where $\delta \in [0,1)$ is the discount factor, $\pi : \mathbb{S}^{2} \rightarrow \mathbb{R}_{+}$ is the per-period payoff, and $ \Upsilon : \mathbb{S} \rightrightarrows \mathbb{R}_{+}$ is the constraint correspondence. The state space $\mathbb{S}$ is assumed to be a bounded, convex subset of $\mathbb{R}$.\footnote{In Appendix \ref{app:state.unbounded} we extend the framework to allow for unbounded state spaces. Under standard "growth conditions" on the per-period payoff the results in the paper go through.} 

This problem can be cast recursively with the Bellman equation, 
\begin{align*}
	V(s,\delta) = \max_{s' \in \Upsilon(s)} \pi(s,s') + \delta V(s',\delta),
\end{align*}
where $V : \mathbb{S} \times [0,1) \rightarrow \mathbb{R}_{+}$ is the value function. The optimal policy correspondence is denoted by $\Gamma(\cdot, \delta) : \mathbb{S} \rightrightarrows \mathbb{R}_{+}$ and is given by
\begin{align*}
	s \mapsto \Gamma(s, \delta) : = \arg\max_{s' \in \Upsilon(s)} \pi(s,s') + \delta V(s',\delta).
\end{align*}

We now introduce some technical definitions and list the assumptions used throughout the paper.
For any set $S$, $S^{o}$ denotes its interior. We say a function $x \mapsto f(x,y)$ is uniformly right differentiable at $(x,y)$ if, for any sequence $(y_{n})_{n}$ converging to $y$ and any positive sequence $(\Delta_{n})_{n}$ converging to zero,
	\begin{align*}
	& \lim_{n \rightarrow \infty} \left| \frac{f(x+\Delta_{n},y_{n}) - f(x,y_{n}) }{\Delta_{n}} - \frac{f(x+\Delta_{n},y) - f(x,y) }{\Delta_{n}} \right|  = 0~and \\
	&  \partial^{+}_{1}f(x,y) : = \lim_{n \rightarrow \infty} \frac{f(x+\Delta_{n},y) - f(x,y) }{\Delta_{n}} ~exists~and~is~continuous~(x,y).
\end{align*}
It is uniformly left differentiable at $(x,y)$ if the same holds but with $(\Delta_n)_{n}$ being negative and is denoted as $\partial^{-}_{1}f(x,y)$. We say $x \mapsto f(x,y)$ is 
uniformly smooth almost everywhere (a.e.) if it is (uniformly) right and left differentiable at every point and these derivatives coincide \emph{except possibly in a finite set}. We say a function $x \mapsto f(x,y)$ is smooth if it is right and left differentiable at every point and these derivatives coincide \emph{everywhere}.

For the per-period payoff we make the following assumption:

\begin{assumption}
	\label{ass:pi.properties} $\pi$ is continuous and (i) $s \mapsto \pi(s,s')$ is increasing, uniformly smooth a.e. with $\partial^{+}_{1} \pi \geq \partial^{-}_{1} \pi$; (ii) $s' \mapsto \pi(s,s')$ is smooth with derivative denoted as $\pi_{2}$; (iii) $s \mapsto \pi_{2}(s,s')$ and  $s' \mapsto \partial^{+}_{1} \pi (s,s') $,  $s' \mapsto \partial^{-}_{1} \pi (s,s') $   are increasing. 
\end{assumption}

Part (i) states that the per-period payoff is increasing in the current state, and allows for a finite number of "kinks" in the first derivative. However, the right and left partial derivatives of the per-period payoff function w.r.t. the first argument are restricted to ensure monotonicity and differentiability of the value function, which in turn will ensure that the optimal solution satisfies the FOC. The "uniform" aspect is technical and used to prove differentiability of the value functions in the absence of concavity. It essentially strengthens the assumption of differentiability in one variable to hold \emph{uniformly} with respect to the other variables. Part (ii) is standard. Part (iii) ensures that there is complementarity between the current and next period's states, in the sense that the marginal return of next period's state is increasing in the current state. If $\pi$ is smooth, then this condition translates to $\pi_{12} > 0$. 

Parts (i) and (iii) are both critical assumptions of our framework, and can be seen as "replacements" of the standard strict concavity of $\pi$ --- while not innocuous, they still encompass many applications of interest. We refer the reader to Section \ref{sec:remarks} for a more thorough discussion, including ways in which 
 Assumption \ref{ass:pi.properties}(iii) 
may be relaxed.

We impose the following restrictions on the constraint correspondence:

\begin{assumption}\label{ass:C.properties}
	(i) $s \mapsto \Upsilon(s)$ is continuous, convex-/compact-valued, with non-empty interior over $\mathbb{S}^{o}$; (ii) $s \mapsto \Upsilon(s)$ is non-decreasing in the inclusion sense; (iii) $s \mapsto \Upsilon(s)$ is non-decreasing in the strong set order sense.\footnote{That is, take any $Y \in \Upsilon(s)$ and any $Y' \in \Upsilon(s')$ with $s'>s$, then $\min\{Y,Y'\} \in \Upsilon(s)$ and $\max\{Y,Y'\} \in \Upsilon(s')$. }
\end{assumption}

Part (i) is standard. Part (ii) is used solely for establishing monotonicity of the value function. While standard (cf \cite{stokey1989recursive} Assumption 4.6) it could be too strong for some applications (e.g., the NCG model and application \ref{exa:fit} below). In view of this, Appendix \ref{app:VF.mono} provides two alternative approaches for  establishing monotonicity of the value function that dispense with this assumption. One uses essentially the same insights as the standard approach but with a weaker assumption. The other approach relies on a completely different approach that hinges on a generalized version of the mean value theorem for a.e. smooth functions --- to our knowledge, this approach is novel and might be of independent interest. Part (iii) is used to invoke the celebrated Milgrom-Shannon Theorem (\cite{milgrom1994monotone}), which is in turn used to establish monotonicity of the optimal policy correspondence.

We now illustrate our assumptions in a generalization of the Neo-Classical Growth (NCG) model.

\setcounter{example}{0}
\renewcommand{\theexample}{generalized NCG model}

\begin{example}
	\label{exa:NCG}
	Following the seminal paper by \cite{skiba1978optimal}, several papers have extended the classical Ramsey-Cass-Koopmans one-sector growth model to allow for non-concave production function, $f$. A typical result in these papers is that there are multiple steady state capital levels. 
	
	A strand of this literature considers a case where the production function is smooth but non-concave (e.g., \cite{majumdar1982intertemporal,kamihigashi2007nonsmooth,dechert2012complete} and references therein). Formally, there exists a level of capital $s_{I}$ such that $s \mapsto f'(s)$ is increasing at any $s <s_{I}$ and decreasing at any $s > s_{I}$.  A different strand of the literature considers a more extreme failure of concavity wherein the production function exhibits kinks (see \cite{kamihigashi2007nonsmooth}). Formally, there exists a $s_{I}$ such that $f$ is increasing, and piece-wise concave with $s_{I}$ being the "kink", i.e., $\partial^{-} f (s_I) \leq \partial^{+} f  (s_I)$.
	
	All these cases are encompassed by our framework, with the constraint correspondence given by $s \mapsto \Upsilon(s) : = [0,f(s)]$ and the payoff function given by $(s,s') \mapsto \pi(s,s') : = u(f(s) - s')$, where $u$ is smooth, increasing and concave. The production function is increasing and smooth almost everywhere. There exists a kink, $s_{I}$, such that $f''$ can either change signs (from positive to negative), or $\partial^{-} f (s_I)< \partial^{+} f  (s_I)$ --- the standard case where $f$ is everywhere smooth and concave, is obviously also allowed. In principle the state space could be all of $\mathbb{R}_{+}$, but given the features of the model it is essentially without loss of generality to assume $\mathbb{S} = [0,\bar{s}]$ for some $\bar{s} \geq s_{max}$, where $s_{max}$ is the level of capital for which $f(s_{max}) = s_{max}$, $f'(s_{max}) \leq 1$, and $f(s) < s$ for all $s>s_{max}$.
	
	We now verify our assumptions. Observe that $\partial^{+}_{1}\pi(s,s') = u'(f(s) - s')\partial^{+}_{1} f(s)$ and $\partial^{-}_{1}\pi(s,s') = u'(f(s) - s')\partial^{-}_{1} f(s)$, which satisfy $\partial^{-}_{1}\pi \leq \partial^{+}_{1}\pi$ as $u'>0$ and $\partial^{-} f \leq \partial^{+} f$. Moreover, it is easy to see that differentiability of $\pi$ holds in the uniform sense. Finally, $s \mapsto \pi_{2}(s,s') = - u'(f(s) - s')$ is increasing because $f$ is increasing and $-u'$ is too ($u''<0$ by concavity). So, Assumption \ref{ass:pi.properties} holds. 
	
	We conclude this example with technical remark about this literature. \cite{dechert2012complete} only consider a feasible correspondence of the form $s \mapsto \Upsilon(s) : = [0,f(s)]$, which implies either full depreciation or allows for negative investment.\footnote{Investment can attain negative values if we adjust the production function so that output includes undepreciated capital -- i.e., if $\tilde{f}$ is the production function and $d$ is depreciation then we proceed as if the production function is $ f(s) = \tilde{f}(s) + (1-d) s$ and we proceed as if there were no depreciation. In this case, if consumption exceeds $\tilde{f}(s)$ then $s' =  f(s) - c = \tilde{f}(s) +(1-d)s - c < (1-d)s$. Combined with $s' = i + (1-d)s$, we have $i < 0$.} 
	In either case, this assumption rules out cases of interest. Thus, it is fruitful to consider the case $s \mapsto \Upsilon(s) : = [(1-d) s,f(s) + (1-d) s]$ which reflects an explicit depreciation rate and 
imposes that investment be non-negative.	The technical issue with this formulation is that Assumption \ref{ass:C.properties}(ii) is not met. The only usage of this assumption, however, is to establish monotonicity of the value function, so in order to apply our result it is sufficient to establish monotonicity by other methods. As it turns out, the condition in Lemma \ref{lem:VF.incr.v2} in Appendix \ref{app:VF.mono} is met (see the Remark below the lemma) and the value function can be shown to be increasing.  Therefore, our theory could also be applied to the case $s \mapsto \Upsilon(s) : = [(1-d) s,f(s) + (1-d) s]$.
\end{example}

  \section{Analytical Results}
\label{sec:analytical}


This section presents the main analytical results of the paper. Subsection \ref{sec:results.VF+PF} establishes basic properties of the value function and the optimal correspondence, including a differentiability result for the value function that extends existing results in the literature and may be of independent interest. Section \ref{sec:planner} then presents the main results on the analytical characterization of the steady state and the dynamics.

\subsection{Properties of the Value Function and Optimal Correspondence}
\label{sec:results.VF+PF}


Our analysis of the planner’s dynamic problem relies on establishing monotonicity and differentiability properties of the value function and the optimal policy correspondence. We establish these results in this subsection. However, due to the potential non-concavity and non-smoothness of $\pi$, standard “textbook” arguments do not apply, and alternative technical approaches are required. We refer the reader to Appendices \ref{app:VF.properties} and \ref{app:PF.properties} for details.

\begin{lemma}\label{lem:VF.prop.text}
	The value function $V$ is continuous, bounded, and increasing as a function of $s$.
\end{lemma}

\begin{proof}
	See Appendix \ref{app:results.VF+PF}.
\end{proof}

\begin{lemma}\label{lem:PF.prop.text}
	The optimal policy correspondence $\Gamma$ is non-empty, compact-valued, UHC, function-like, and non-decreasing in the sense that $\max \Gamma(s, \delta) \leq \min \Gamma(s', \delta)$ for any $s \leq s'$ and $\max \Gamma(s, \delta) \leq \min \Gamma(s, \delta')$ for any $\delta \leq \delta'$.\footnote{A correspondence is function-like if its graph has an empty interior.} 
\end{lemma}

\begin{proof}
	See Appendix \ref{app:results.VF+PF}.
\end{proof}

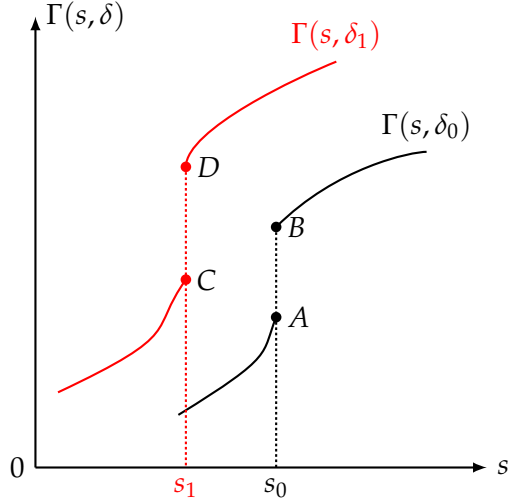
\begin{figure}[h]
\centering
	\begin{tikzpicture}[thick]
  		\draw[->] (0,0) node[anchor=east] {$0$} -- (6,0) node[anchor=west] {$s$};
 		\draw[->] (0,0) -- (0,6) node[anchor=west] {$\Gamma(s, \delta)$};

		\draw [name path = G1] (1.9, 0.7) .. controls (3.2, 1.5) and (3, 1.5) .. (3.2, 2);
		\fill[black] (3.2, 2) circle (2pt);
		\node[anchor=west] at (3.2, 2) {$A$};
		
		\draw [name path = G2] (3.2, 3.2) .. controls (4, 4) and (5 ,4.2) .. (5.2, 4.2) node[anchor=south] {$\Gamma(s,\delta_0)$};
		\fill[black] (3.2, 3.2) circle (2pt);
		\node[anchor=west] at (3.2, 3.2) {$B$};
		\draw[dash=on 1pt off 1pt phase 0pt] (3.2,3.2) -- (3.2,0) node[anchor=north]{$s_0$};
		
		\draw[red, name path = G3] (0.3, 1) .. controls (2, 1.8) and (1.5, 1.8) .. (2, 2.5);
		\fill[red] (2, 2.5) circle (2pt);
		\node[anchor=west] at (2, 2.5) {$C$};
		
		\draw[red, name path = G4] (2, 4) .. controls (2, 4.5) and (3.5 ,5.2) .. (4, 5.4) node[red, anchor=south] {$\Gamma(s,\delta_1)$};
		\fill[red] (2, 4) circle (2pt);
		\node[anchor=west] at (2, 4) {$D$};
		
		\draw[red, dash=on 1pt off 1pt phase 0pt] (2,4) -- (2,0) node[anchor=north]{$s_1$};
		
		
	\end{tikzpicture}
\caption{Topological properties of $\Gamma(s,\delta)$, with $\delta_1>\delta_0$}\label{fig:Gamma}
\end{figure}
Figure \ref{fig:Gamma} illustrates a typical case in which the policy correspondence is not a function, along with its monotonicity properties in $s$ and $\delta$. The failure of the policy to be a function arises solely due to the presence of Skiba points—at $s_0$ for $\delta_0$ and at $s_1$ for $\delta_1$. Away from these points, the policy correspondence is increasing and single-valued. At the Skiba points, however, there are two optimal choices for the next-period state, resulting in a correspondence rather than a function. As the discount factor increases, the Skiba points shift to the left, and the values of the policy correspondence move upward relative to those under the lower discount rate.

Even though there are no restrictions on concavity and $s \mapsto \pi(s,s')$ is not continuously differentiable, the value function still inherits the smoothness properties of $\pi$. As the next proposition shows, the left and right derivatives exist and can be characterized in terms of those of $\pi$.

\begin{proposition}\label{pro:VF.diff}
	$s \mapsto V(s,\delta)$ has left and right derivatives, which are
	\begin{align*}
		\partial^{+}_{1} V(s,\delta)  = \max_{y \in \Gamma(s, \delta)}  \partial^{+}_{1} \pi(s,y)~and~\partial^{-}_{1}  V(s,\delta)  =  \min_{y \in \Gamma(s, \delta) }  \partial^{-}_{1} \pi(s,y)
	\end{align*}
	for any $s \in \mathbb{S}$ such that $\Gamma(s, \delta) \subseteq \Upsilon^{o}(s)$.\footnote{For the boundary of $\mathbb{S}$ only one of the two derivatives are defined.} 
\end{proposition}

\begin{proof}
	See Appendix \ref{app:results.VF+PF}.
\end{proof}

This result and the fact that $\partial^{+}_{1} \pi \geq \partial^{-}_{1} \pi$ imply that 
$V$ is \emph{everywhere} smooth (i.e., continuously differentiable) over $Range \Gamma$.\footnote{For any correspondence, $F$, $Range F : = \{ y \colon \exists s \in \mathbb{S},~s.t.~F(s) \ni y  \}$.} 

\begin{corollary}\label{cor:VF.diff.eqn}
	For any $s \in  Range (\Gamma \cap \Upsilon^{o})$, $\Gamma(s, \delta)$ is a singleton and
	\begin{align*}
		\partial^{+}_{1} V(s,\delta) = \partial^{-}_{1} V(s,\delta) 	=  \partial^{+}_{1} \pi(s,\Gamma(s, \delta)) =  \partial^{-}_{1} \pi(s,\Gamma(s, \delta)) = :   \pi_{1} (s,s')~ \forall s' \in \Gamma(s, \delta)
	\end{align*}
\end{corollary}

\begin{proof}
	See  Appendix \ref{app:results.VF+PF}.
\end{proof}

This corollary generalizes Theorem 2 by \cite{cotter2006non} (see also \cite{clausen2016general}) to the case where the per-period payoff is only a.e. differentiable. Indeed, a perhaps surprising feature of this result is that even though $\pi$ is only \emph{a.e.} smooth, the value function is \emph{everywhere} smooth "along the optimal path" (i.e., for any $s \in Range \Gamma$). That is, the value function features stronger smoothness than the primitive pay-off. This follows from two key facts: first, the optimality of the value function, and second, the monotonicity assumption \ref{ass:pi.properties}(i).

	\paragraph{Application to the generalized NCG model.} An application of Corollary \ref{cor:VF.diff.eqn} shows that the
	differences in the optimal paths between the generalized NCG model presented in Section 2 and its "textbook" version
	are due to non-concavity of the production function, not to the kinks per-se --- i.e., kinks do not yield new features above and beyond those offered by the non-concavities.  To show this, note that Corollary \ref{cor:VF.diff.eqn} states that optimal paths satisfy:\footnote{Optimal paths are formally defined below, but essentially are $(s_{t})_{t}$ such that $s_{t+1} \in \Gamma(s_{t}, \delta)$.}
	\begin{align*}
	 u'(f(s_{t}) - s_{t+1}) - \delta f'(s_{t+1}) u'(f(s_{t+1}) - s_{t+2}) = 0.
	\end{align*}
This expression is the same regardless of whether $\pi$ has kinks (as in \cite{kamihigashi2007nonsmooth}) or is smooth but non-concave (as in \cite{nishimura2004discrete}).

\subsection{Analytical Results for Optimal Paths}
\label{sec:planner}



This section introduces our new approach for analyzing the steady states and dynamics of the planner’s problem. The core of the method is the locator function --- a simple object constructed directly from the model’s primitives whose roots identify interior steady states and whose slope determines their local stability. By focusing on this function, we reduce the study of dynamic behavior to the analysis of a tractable, easily computable scalar object.

%

\paragraph{Optimal paths and their steady states.}  We first formally define some concepts. An \emph{optimal path with initial condition $s_{0} \in \mathbb{S}$} is a mapping $\phi(.,s_{0}) : \mathbb{N}_{0} \rightarrow \mathbb{S}$ such that
  \begin{align*}
  	\phi(t,s_{0}) \in \Gamma(\phi(t-1,s_{0}), \delta),~\forall t \geq 1,
  \end{align*} 
  and $\phi(0,s_{0}) = s_{0}$. Let $\Phi(s_0)$ be the class of all optimal paths with initial condition $s_{0}$.

 A first step in our analysis is to characterize the limit points of optimal paths. An obvious candidate for characterizing such behavior is the set of fixed points of $\Gamma(\cdot, \delta)$,
  \begin{align*}
  	\mathcal{R}[ \bar{\Gamma}] : = \{  s \in \mathbb{S} \colon  \bar{\Gamma}(s, \delta) = 0 \},~\text{where}~s \mapsto \bar{\Gamma}(s, \delta) = : \Gamma(s, \delta) - s,
  \end{align*}
  but in principle there could be other, more complex, candidates such as cycles. The next proposition confirms this is not the case and that fixed points fully describe the asymptotic behavior of optimal paths.
  
  \begin{proposition}\label{pro:equilibria.chactarization}
  	For any $s_0 \in \mathbb{S}$ and any $\phi(.,s_0) \in \Phi(s_0)$, $\lim_{t \rightarrow \infty} \phi(t,s_{0}) \in  \mathcal{R}[\bar{\Gamma}]$.  
  \end{proposition}
  
  \begin{proof}
  	See Appendix \ref{app:app_planner}. 
  \end{proof}
  
This proposition shows that for any initial condition the limit of the path is well-defined and is a fixed point. However, this limit may not be unique. That is, for a given initial condition, there might be more than one optimal path, and thus more than one limit. This occurs if (and only if) the initial condition, $s_{0}$, is a so-called Skiba point, wherein $\Gamma(s_{0}, \delta) = \{ \gamma_{l}(s_{0}) , \gamma_{h}(s_{0}) \}$ such that $\gamma_{l}(s_{0}) < s_{0} < \gamma_{h}(s_{0})$, as illustrated in Figure \ref{fig:Gamma}. The generalized NCG model with a convex-concave production function can exhibit a Skiba point for certain parameter values, as shown in Figure \ref{fig:ncg}(c). Our rational (un)fitness model can likewise exhibit a Skiba point, as shown in Figure \ref{fig:fit}(b).  
  

In light of Proposition \ref{pro:equilibria.chactarization} we henceforth refer to the fixed points of $\Gamma(\cdot,\delta)$ as a \emph{steady state}.  Throughout the analysis we focus on steady states that are \emph{generic}. Intuitively, non-generic steady states are those that would vanish under small perturbations of $\Gamma$ --- in particular of the discount factor. Geometrically, this genericity restriction rules out steady states for which $\Gamma$ is tangent to but does not cross the \SI{45}{\degree} line, or for which $\Gamma$, when perturbed, develops a discontinuity.\footnote{Formally, a steady state $s$ is generic at $\delta$ if there exists an open neighborhood around $\delta$ and a mapping $q$ over it such that $q(\delta) =s $ and for any $\delta'$ in the neighborhood, $q(\delta')$ is a steady state at $\delta'$. In other words, genericity implies the existence of a local branch through $s$ at $\delta$ without bifurcations. This assumption is tightly related to the standard assumption of regularity in general equilibrium theory and Game theory \cite{debreu1970economies,dechert2012complete} Ch. 17, and hyperbolicity in dynamical systems. To see where this genericity restriction is used, we refer the reader to the proof of Lemma \ref{lem:FP.IFT} below.} 

While every steady state is a fixed point of $\Gamma(\cdot, \delta)$, this set may be too broad to characterize asymptotic behavior: not all fixed points can be reachable, in the sense that they arise as limits of optimal paths that do not start at the fixed point itself. Thus, to sharpen our analysis of steady states, we therefore introduce notions of stability and instability.
   
  \paragraph{Stability of steady states.} We say that a steady state $e \in \mathcal{R}[\bar{\Gamma}]$ is \emph{stable} if there exists a non-empty open interval containing it such that, for any $s_{0}$ in the interval and any optimal path $\phi(\cdot, s_{0}) \in \Phi(s_{0})$, $\lim_{t \rightarrow \infty} \phi(t,s_{0}) = e$.\footnote{In principle, focusing on intervals might restrict the notion of basin of attraction since the latter could be, say, a union of intervals. In our framework, thanks to the monotonicity properties of $\Gamma$, this turns out not to be the case: the basin of attraction will indeed be an interval.} The largest such interval is denoted as $\mathcal{B}(e,\delta)$ and will be referred to as the \emph{basin of attraction} of $e$. We say $e$ is \emph{unstable} if no such open interval exists --- or, with a slight abuse of notation, $\mathcal{B}(e,\delta) = \{e\}$. 

\begin{figure}[h!]
   	\centering
   		\begin{tikzpicture}[scale=0.7, >=stealth]
   		
   		\draw[->] (0,0) -- (6,0) node[right] {\large $s$};
   		\draw[->] (0,0) -- (0,6) node[left] {\large $s'$};
   		
   		\draw[dashed] (0,0) -- (6,6);
   		
   		\draw[thick, smooth] plot coordinates {
   			(0,0)  
   			(1,0.5)
   			(2,2)  
   			(3,4)  
   			(4,4.75)  
   			(5,5)  
   			(6,5.25)  
   		};
   		
   		\draw[dotted] (2,0) -- (2,2); 
   		\draw[dotted] (5,0) -- (5,5);

   		\draw (2,0) node[anchor=north,yshift=-4pt]{$u$};
   		\draw (5,0) node[anchor=north,yshift=-4pt]{$s$};
   		
   		\draw (6,5.25) node[anchor=north,yshift=-4pt]{$\Gamma(\cdot, \delta)$};

   		\draw [
   		decorate,
   		decoration={brace, mirror, raise=16pt}
   		] (2,0) -- (6,0)
   		node [
   		pos=0.5,
   		anchor=north,
   		yshift=-18pt
   		] {$\mathcal{B}(s,\delta)$};
   		
   	\end{tikzpicture}
   \caption{Basin of attraction of stable steady state $s$. The basin of attraction of the unstable one, $u$, only contains $u$ itself. \label{fig:BoA.Gamma}}
   \end{figure}
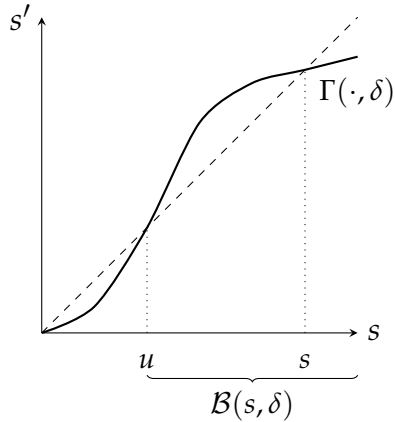

   
   A direct approach to study stability of steady states would be to study the mapping $s \mapsto \bar{\Gamma}(s, \delta) : = \Gamma(s, \delta) - s$ and analyze its zeros—where it crosses zero from above, the steady state is stable; where it crosses from below, it is unstable. This is illustrated in Figure \ref{fig:BoA.Gamma} and formally proven in Proposition \ref{pro:Basins.chactarization} in the Onlinea Appendix \ref{app:Gamma.BoA}. At this level of generality, however, this result by itself is of limited utility as  $\Gamma$ is not a primitive but rather an outcome of a dynamic programming problem. 
   
\paragraph{The locator function.} We propose an alternative approach that relies on a simple function to locate the steady states and provide information about their stability or instability. The proposed function, $\mathcal{L}(\cdot,\delta) : \mathbb{S}^{o} \rightarrow \mathbb{R}$ for any $\delta \in [0,1)$,  is dubbed the \emph{locator function} and is given by
\begin{align}
	(s,\delta) \mapsto \mathcal{L}(s,\delta) : = 	\pi_{2}(s,s) + \delta \pi_{1}(s,s),~a.e.
\end{align}
The almost everywhere (a.e.) is needed because $\pi_{1}$ may only be defined a.e.. However, $\mathcal{L}$ will only be used in neighborhoods of fixed points, for which --- by Corollary \ref{cor:VF.diff.eqn} --- $\pi_{1}$ is well-defined. The usefulness of the locator function lies on locating steady states and classifying whether they are stable or unstable. 

To understand why the locator function may be useful in locating steady states, note that any interior optimal path, $\phi(\cdot,s_{0}) \in \Phi(s_{0})$, must satisfy the first-order condition (i.e, Euler equation), 
\begin{align} \label{eqn:FOC.flow-1}
\pi_{2}(\phi(t-1,s_{0}),\phi(t,s_{0})) + \delta \pi_{1}(\phi(t,s_{0}),\phi(t+1,s_{0})) = 0 \quad \forall t \geq 1.
\end{align} 
Taking limits, it is then clear that interior steady states must be roots of the locator function.\footnote{The validity of expression \ref{eqn:FOC.flow-1} follows because, by definition, $\phi_{\delta}(t,s_{0}) \in \Gamma(\phi_{\delta}(t-1,s_{0}), \delta)$ and is interior. So, all the assumptions of Proposition \ref{pro:VF.diff} and Corollary \ref{cor:VF.diff.eqn} are met.} The perhaps surprising result is that the locator function also contains information about the stability properties of the steady state, as we shown below in our main theorem. 

To proceed, we impose the following technical condition:

\begin{assumption}\label{ass:Pi.roots}
	All roots of $s \mapsto \mathcal{L}(s,\delta)$ are well-separated and regular.\footnote{A root, $r$, is said to be regular if $\mathcal{L}_{1}(r,\delta) \ne 0$; it is well-separated if there exists a $c>0$ such that $|r-r'|\geq c$ for any roots $r,r'$.}
\end{assumption}

Regularity is a necessary condition for applying the Implicit Function Theorem below which is key for the proof of the theorem. It is a \emph{generic} property in the sense that if a root fails to be regular for some value of $\delta$, then small perturbations of $\delta$ will eliminate that root. The assumption of well-separated roots serves mainly to rule out pathological cases in which the locator function exhibits infinitely frequent oscillations. Lemma \ref{lem:separate.roots} in Appendix \ref{app:separate.roots} provides sufficient conditions, based on the derivatives of the locator function, for this to hold.

\paragraph{Main result.}  The following theorem shows that the locator function can be used not only to identify interior steady states, but also to assess their local stability by examining the sign of its derivative at interior roots: a negative derivative implies stability, while a positive one implies instability. Henceforth, a steady state is \emph{interior} if it belongs to $ \mathcal{R}^{o}[\bar{\Gamma}] : = \{ s \in  \mathcal{R}[\bar{\Gamma}] \colon s \in \Upsilon^{o}(s)  \}$.\footnote{$\mathcal{R}^{o}[\mathcal{L}]$ is defined analogously.}



\begin{theorem}\label{thm:FP.characterization}
	For any interior steady state $s$, the following are true: 
	\begin{enumerate}
		\item $s \in \mathcal{R}^{o}[\mathcal{L}]$.
		\item If $\mathcal{L}_{1}(s,\delta) < 0$, then $s$ is stable.
		\item If $\mathcal{L}_{1}(s,\delta) > 0$, then $s$ is unstable.
	\end{enumerate}
\end{theorem}

By providing a link between the (local) monotonicity of the locator function and the stability or instability of steady states, Theorem \ref{thm:FP.characterization} offers a simple way of identifying the limit points of the planner's dynamics: for stable fixed points the locator function cuts zero "from above." Importantly, unlike $\Gamma$, which is the solution of a dynamic programming problem, the locator function depends on primitives and is relatively easy to compute and study.

	

The theorem does not establish an equivalence between steady states and roots of the locator function: while every interior steady state is a root of the locator function, the converse does not hold. This is not surprising, as it reflects the fact that first-order conditions are not sufficient in the absence of concavity. This raises a natural question: under what conditions on $\pi$ does the locator function avoid generating "false positives"? We explore this question further in Section~\ref{sec:refinements}. 


What may be more surprising, however, is that the result—that interior steady states are roots of the locator function—continues to hold even when the function $s \mapsto \pi(s,s')$ is not differentiable everywhere. The reason is that the first-order condition \eqref{eqn:FOC.flow-1} is identical to the one that would arise if $s \mapsto \pi(s,s')$ were smooth. Under our assumptions—particularly the monotonicity conditions in Assumption~\ref{ass:pi.properties}—these potential "kinks" in $\pi$ do not affect the characterization of optimal paths.

\paragraph{Application to the \ref{exa:NCG}.} 
We now return to the generalized NCG model to illustrate the usefulness of Theorem \ref{thm:FP.characterization} and the locator function. The dynamics of this economy were previously analyzed in \cite{majumdar1982intertemporal} and \cite{dechert2012complete}.\footnote{These previous papers did not allow for a kink in the production function but their results extend directly to such settings via Corollary \ref{cor:VF.diff.eqn}} Our analysis offers an alternative and complementary perspective, showing in passing how the locator function helps explain the different cases presented in \cite{dechert2012complete}.

For expositional purposes, we assume that $f$ is smooth and that $f''$ changes signs at only one point, $s_{I}$; if more points like $s_{I}$ exist we simply repeat the analysis in each interval. 

The locator function is given by 
\begin{align*}
	\mathcal{L}(s,\delta) = u'(f(s) - s)(\delta f'(s) - 1),
\end{align*}
and $\mathbb{S} : = [0,s_{max}]$, where $s_{max}$ is such that $\mathcal{L}(s_{max},\delta) < 0$. Since $u'>0$, the only interior roots are those of function $s \mapsto \delta f'(s) - 1$. Since $s \mapsto f'(s)$ is increasing in $[0,s_{I})$ and decreasing in $(s_{I},s_{max}]$, there are at most two interior roots of $s \mapsto \mathcal{L}(s,\delta)$, which we denote as $s_{\ast} < s^{\ast}$.


There are three cases: (1) $\delta f'(0) -1 \geq 0$, which is dubbed as mild discounting by \cite{dechert2012complete}; (2)  $\delta f'(0) -1 < 0 < \delta f'(s_{I}) - 1$; and (3) $\delta f'(s_{I}) < 1$.

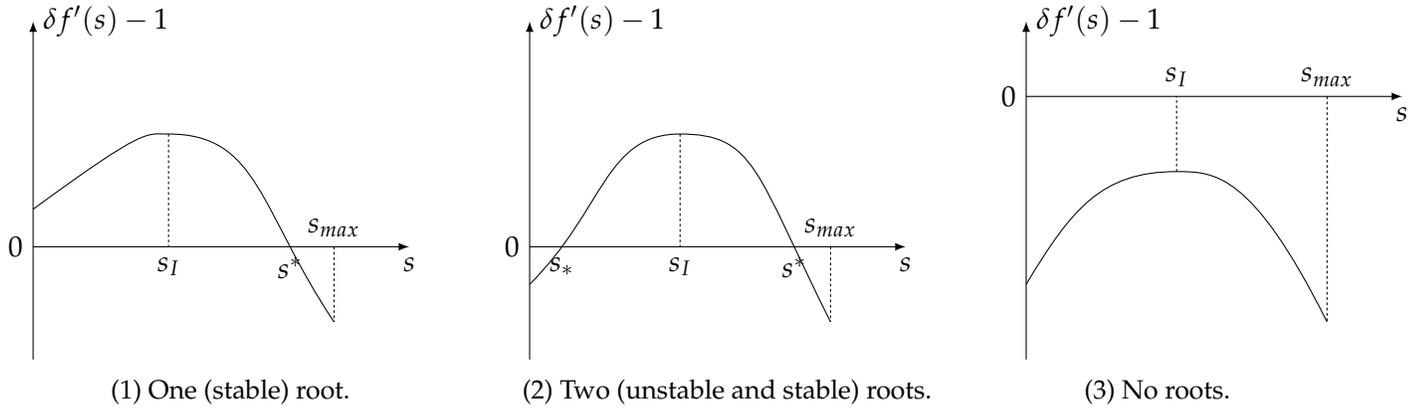
\begin{figure}
	\begin{subfigure}[b]{0.35\textwidth}  
		\begin{tikzpicture}
			
			\draw[name path=H, ->] (0,1.5) node[anchor=east] {$0$} -- (5,1.5) node[anchor=north] {$s$};
			\draw[->] (0,0) -- (0,4.5) node[anchor=west] {$\delta f'(s) -1$} ;
			
			\draw (0,2) .. controls (1.5, 3.1) and (1.5, 3) .. (1.8, 3);
			\draw [name path=curve 1] (1.8, 3)  .. controls (3,3) and (3,2)  .. (4,0.5);
			
			\draw [dash=on 1pt off 1pt phase 0pt ] (1.8, 3) -- (1.8, 1.5) node[anchor=north] {$s_I$};
			\draw [name path=V, dash=on 1pt off 1pt phase 0pt] (4,1.5) -- (4, 0.5);
			\node [anchor=south] at (4, 1.5) {$s_{max}$};
			\path [name intersections={of=curve 1 and H}];
			\node[anchor=north] at (intersection-1) {$s^*$};
		\end{tikzpicture} 
		\caption{One (stable) root.}
	\end{subfigure} 
	\hfill
	\begin{subfigure}[b]{0.35\textwidth}  
		\begin{tikzpicture}
			\draw[name path=H, ->] (0,1.5) node[anchor=east] {$0$} -- (5,1.5) node[anchor=north] {$s$};
			\draw[->] (0,0) -- (0,4.5) node[anchor=west] {$\delta f'(s)-1$} ;
			
			\draw [name path=curve 1] (0, 1)  .. controls (1,2) and (1,3)  .. (2,3);
			\draw[name path=curve 2] (2,3) .. controls (3, 3) and (3, 2.5) .. (4, 0.5);
			
			\draw [dash=on 1pt off 1pt phase 0pt ] (2, 3) -- (2, 1.5) node[anchor=north] {$s_I$};
			\draw [name path=V, dash=on 1pt off 1pt phase 0pt] (4, 1.5) -- (4, 0.5);
			\node [anchor=south] at (4, 1.5) {$s_{max}$};
			\path [name intersections={of=curve 1 and H, name=first}];
			\node[anchor=north] at (first-1) {$s_*$};
			\path [name intersections={of=curve 2 and H, name=second}];
			\node[anchor=north] at (second-1) {$s^*$};
			
		\end{tikzpicture}
		\caption{Two (unstable and stable) roots.}
	\end{subfigure}
	\hfill 
	\begin{subfigure}[b]{0.25\textwidth}  
		\begin{tikzpicture}
			\draw[name path=H,->] (0,3.5) node[anchor=east] {$0$} -- (5,3.5) node[anchor=north] {$s$};
			\draw[->] (0,0) -- (0,4.5) node[anchor=west] {$\delta f'(s)-1$};
			
			\draw [name path=curve 1] (0, 1)  .. controls (0.6,2) and (1,2.5)  .. (2,2.5);
			\draw[name path=curve 2] (2,2.5) .. controls (2.5, 2.5) and (3, 2.5) .. (4, 0.5);
			
			\draw [dash=on 1pt off 1pt phase 0pt ] (2, 2.5) -- (2, 3.5) node[anchor=south] {$s_I$};
			\draw [name path=V, dash=on 1pt off 1pt phase 0pt] (4, 3.5) -- (4, 0.5);
			\node [anchor=south] at (4, 3.5) {$s_{max}$};
			
		\end{tikzpicture}
		\caption{No roots.}
	\end{subfigure}
	\caption{$\frac{\mathcal{L}(s,\delta) }{u'(f(s)-s)}$ in \ref{exa:NCG}} \label{fig:ncg_L}
\end{figure}

 It is easy to see by Figure \ref{fig:ncg_L}(3) that in the last case there are no roots of the locator function. Consequently, by Theorem \ref{thm:FP.characterization}, there are no interior steady states and the only steady state is $s=0$, which is globally stable.\footnote{Since $\delta f'(s_{max}) < 1$, this implies that $\Gamma(s_{max},\delta) < s_{max}$. Since there are no roots, then the only steady state is $s=0$ and is globally stable.} 
In the other two cases, the locator function has either a single interior root, $s^{\ast}$, as in Figure \ref{fig:ncg_L}(1), or two interior roots, $s_{\ast}$ and $s^{\ast}$, as in Figure \ref{fig:ncg_L}(2). By Theorem \ref{thm:FP.characterization}, $s^{\ast}$ is the only candidate for a locally stable interior steady state, but in case (2) we could also have a locally unstable steady state at $s_{\ast}$.

Below we provide a more refined analysis of these cases, but we hope this example illustrates a key advantage of our approach: examining the shape of the locator function involves significantly simpler calculations while delivering results comparable to those obtained through standard methods. In addition, we can extend the non-concave Neoclassical Growth Model to accommodate any production function, not just those with kinks or s-shaped profiles. Indeed, Theorem \ref{thm:FP.characterization} implies that for any function $f$, the stable steady states must satisfy two conditions: (a) $f'$ equals $1/\delta$, and (b) $f$ is concave at that point. This generalizes the classical results for the Neoclassical Growth Model to a much broader class of settings.

\paragraph{Proof of Theorem \ref{thm:FP.characterization}}

Throughout this section it is useful to make explicit the dependence of roots and fixed points on the discount factor. Hence, we use $ \mathcal{R}^{o}[\bar{\Gamma}(\cdot,\delta)]$ and $\mathcal{R}^{o}[\mathcal{L}(\cdot,\delta)]$. Also, we define $\mathcal{E}=\{(s,\delta) \in \mathbb{S}\times [0,1) : \bar{\Gamma}(s, \delta)=0 \}$.

The first part of Theorem \ref{thm:FP.characterization} follows from the following lemma.

\begin{lemma}\label{lem:eqn.roots}
	For any $\delta \in [0,1)$, $ \mathcal{R}^{o}[\bar{\Gamma}(\cdot,\delta)] \subseteq \mathcal{R}^{o}[\mathcal{L}(\cdot,\delta)]$.
\end{lemma}

\begin{proof}
	See Appendix \ref{app:app_proofs.main}. 
\end{proof}


The proof of the second part of the theorem rests on a two-step argument. The first step links the stability of a steady state to the comparative statics of fixed points of the optimal correspondence with respect to the discount factor. The second step relates these comparative statics to the slope of the locator function evaluated at the root corresponding to the steady state.

\begin{figure} 
		\centering
		\begin{tikzpicture}
				\draw[->] (0,0) node[anchor=east]{$0$} -- (5,0) node[anchor=west]{$s$};
				\draw[->] (0,0) -- (0,5) node[anchor=east] {$s'$};
				\coordinate (A) at (4.9,4.9);
				\draw [name path = R] (0,0) -- (A) node[anchor=east]{$45\si{\degree}$};
				\coordinate (B) at (2,2);
				\coordinate (C) at (3.8, 3.8);
				\draw [name path = policy 1] (0,0) .. controls (1, 0.1) and (1.5, 0.2) .. (B)
				.. controls (2.2, 3) and (2.5, 3.4) .. (C)
				.. controls (4, 3.9) and (4, 3.9) .. (5, 4.2) node[anchor=west]{$\Gamma(\cdot, \delta)$};
				\draw [dash=on 1pt off 1pt phase 0pt] (B) -- (B|-0,0) node[anchor=north]{$u(\delta)$};
				\draw [dash=on 1pt off 1pt phase 0pt] (C) -- (C|-0,0) node[anchor=north]{$s(\delta)$};
				\coordinate (D) at (1.3,1.3);
				\coordinate (E) at (4.5, 4.5);
				\draw [name path = policy 2,dash=on 3pt off 1pt phase 0pt] (0,0) .. controls (0.5, 0.2) and (1.2, 0.6) .. (D)
				.. controls (2, 4) and (3,4) .. (E)
				.. controls (4.8, 4.6) and (4.8, 4.6) .. (5, 4.65) node[anchor=west]{$\Gamma(\cdot, \delta')$};
				\draw [dash=on 1pt off 1pt phase 0pt] (D) -- (D|-0,0) node[anchor=north]{$u(\delta')$};
				\draw [dash=on 1pt off 1pt phase 0pt] (E) -- (E|-0,0) node[anchor=north]{$s(\delta')$};
			\end{tikzpicture}
	\caption{Comparative statics of stable and unstable fixed points \label{fig:CS}}
		%
	
	\end{figure}
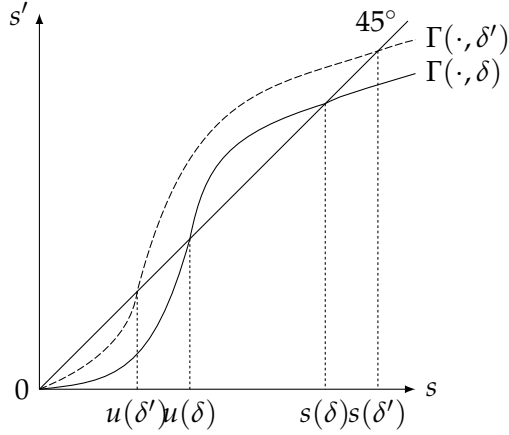

In the first step, stability is inferred from the response of steady states to small changes in the discount factor. By Lemma \ref{lem:PF.prop.text}, the mapping $\delta \mapsto \Gamma(s,\delta)$ is non-decreasing. Intuitively, as the discount factor rises the planner attaches greater weight to future payoffs, and since $s \mapsto V(s,\delta)$ is increasing, the planner is drawn toward higher values of the next-period state.\footnote{Strictly speaking, monotonicity of $s \mapsto V(s,\delta)$ is not sufficient for $s \mapsto \Gamma(s,\delta)$ to be increasing. We also require either that $\pi$ is strictly concave or that $s' \mapsto \partial^{+}_{1}\pi(s,s')$ and $s' \mapsto \partial^{-}_{1}\pi(s,s')$ are increasing; see the proof of Lemma \ref{lem:VF.prop.text}.} As illustrated in Figure \ref{fig:CS}, when the policy correspondence weakly shifts up with $\delta$, stable fixed points cannot shift left as $\delta$ increases, while unstable ones cannot shift right.

The following lemma formalizes this observation. 

\begin{lemma}\label{lem:FP.IFT}
	For any $(e,\delta) \in Graph \mathcal{E}^{o}$, consider an open neighborhood around $\delta$, $U_{q}(\delta)$, and a continuous mapping $q$ from $U_{q}(\delta)$ to $\mathbb{R}$ such that  $q(\delta) = e$ and 
 $q(\delta') \in \mathcal{R}^{o}[\bar{\Gamma}(\cdot, \delta')]$ for any $\delta' \in U_{q}(\delta)$. Then $q$ is non-decreasing at $\delta$ if $e$ is stable and non-increasing if $e$ is unstable.
\end{lemma}

\begin{proof}
	See Appendix \ref{app:app_proofs.main}. 
\end{proof}

In the second step, we connect comparative statics with respect to the discount factor to the slope of the locator function at the corresponding root. By Corollary \ref{cor:VF.diff.eqn}, $\mathcal{L}$ is continuously differentiable around each $s \in \mathcal{R}^{o}[\bar{\Gamma}(\cdot,\delta)]$, and $\mathcal{L}_{1}(s,\delta) \neq 0$ by Assumption \ref{ass:Pi.roots}. Hence, by the Implicit Function Theorem, for any $(e,\delta) \in Graph \mathcal{E}^{o}$ there exists an open neighborhood $U_{p}(\delta)$ and a unique smooth mapping $p: U_{p}(\delta) \to \mathbb{R}$ such that $p(\delta) = e$ and $\mathcal{L}(p(\delta'),\delta') = 0$ for all $\delta' \in U_{p}(\delta)$, with derivative
\begin{align}\label{eqn:IFT-1}
\frac{dp(\delta)}{d\delta} = - \frac{\pi_{1}(p(\delta),\delta)}{\mathcal{L}_{1}(p(\delta),\delta)}.
\end{align}
Since $\pi_{1} \geq 0$ by Assumption \ref{ass:pi.properties}, $p$ is non-decreasing iff $\mathcal{L}_{1}(p(\delta),\delta)<0$ and non-increasing iff $\mathcal{L}_{1}(p(\delta),\delta)>0$. Moreover, by Lemma \ref{lem:eqn.roots}, $\mathcal{R}^{o}[\bar{\Gamma}(\cdot,\delta)] \subseteq \mathcal{R}^{o}[\mathcal{L}(\cdot,\delta)]$, so the mapping in  Lemma \ref{lem:FP.IFT} must coincide with the mapping $p$, at least over $U_{q}(\delta) \cap U_{p}(\delta)$. Thus, it follows that for any stable point $s$, $p$ must be non-decreasing at $\delta$, so $\mathcal{L}_{1}(s,\delta)<0$, and for any unstable point $u$, $p$ must be non-increasing, so $\mathcal{L}_{1}(u,\delta)>0$. Since stability and instability are mutually exclusive, we conclude that $\mathcal{L}_1(s,\delta)<0$ implies stability and $\mathcal{L}_1(s,\delta)>0$ implies instability, as claimed in the second part of Theorem \ref{thm:FP.characterization}.

We conclude this section with a remark. 
The first step of the proof — the link between stability and comparative statics — is related to Samuelson’s Correspondence Principle \citep{samuelson1983foundations}. Samuelson’s principle uses local dynamic stability (e.g., under tâtonnement dynamics) to sign comparative statics, such as concluding that equilibrium price rises with an outward demand shift because the demand curve must be flatter than the supply curve at a locally stable equilibrium. Our approach runs in the opposite direction: we use comparative statics with respect to the discount factor to infer local stability of the model’s underlying dynamics.

  \section{Refinements and Implications of Theorem \ref{thm:FP.characterization}}
\label{sec:refinements}

While Theorem \ref{thm:FP.characterization} provides a general method for identifying and classifying the stability of interior steady states using the locator function, this section highlights important cases in which additional analysis yields sharper insights. We begin by showing that when the locator function crosses zero only once and from above, the unique interior root corresponds to a globally stable steady state. We then examine the case in which the locator function has an inverted U shape with two interior roots, demonstrating that meaningful conclusions about stability can still be drawn. Next, we consider the special case of a strictly concave per-period payoff function and show that, under an additional condition, the locator function fully characterizes all interior steady states and their basins of attraction. Finally, we show how the locator function can be used to do comparative statics for stable steady states. 
Together, these results clarify the scope and limitations of the main theorem and offer guidance for its application across a range of economic environments.

Throughout this section, to facilitate the exposition of the results, we maintain the assumption of smoothness of the locator function and that $s \in \Upsilon^{o}(s)$ for all $s \in \mathbb{S}^{o}$.

\subsection{Locator function is single crossing from above} 


In some applications, it is useful to impose conditions that guarantee the existence of a globally stable steady state. This subsection presents one such condition on the locator function—namely, that it be single crossing from above. By this, we mean that there exists a point $c \in \mathbb{S}^{o}$ such that $\mathcal{L}(s,\delta) > 0$ for all $s < c$, $\mathcal{L}(s,\delta) < 0$ for all $s > c$, and $\mathcal{L}(c,\delta) = 0$.

\begin{proposition}\label{pro:root.unique}
	Suppose the locator function satisfies single crossing from above. Then there exist a unique interior steady state, given by the root of the locator function, and it is globally stable over $\mathbb{S}^{o}$.\footnote{The qualifier "over $\mathbb{S}^{o}$" means that its basin of attraction is the whole of $\mathbb{S}^{o}$. That is, the result does not rule out \emph{unstable} steady states at the boundary of $\mathbb{S}$.} 
\end{proposition}

\begin{proof}
	See Appendix \ref{app:root.unique}. 
\end{proof}


\paragraph{Application to the generalized NCG model.} Consider case (1) above, in which $\delta f'(0) > 1$. 
This inequality implies that $\mathcal{L}(0,\delta) > 0$, and since $s \mapsto f'(s)$ is increasing on $[0, s_I)$ and decreasing on $(s_I, s_{\max}]$, it follows that $s \mapsto \mathcal{L}(s,\delta)$ is single crossing decreasing from above. Thus, by Proposition \ref{pro:root.unique}, the root $s^{\ast}$ 
is the unique interior steady state and is globally stable over $\mathbb{S}^{o}$.

\subsection{Locator function has two interior roots}

In some situations the locator function will have multiple roots and thus Proposition \ref{pro:root.unique} cannot be used. The next proposition shows that the locator function can still identify stable interior steady states in this case.

\begin{proposition}\label{pro:root.two}
	Suppose the locator has two roots, $s_{\ast}  < s^{\ast}$ and $\mathcal{L}(\underbar{s},\delta), \mathcal{L}(\bar{s},\delta)  < 0$.\footnote{Here and throughout, $\underline{s}$ and $\overline{s}$ are the lower and upper bounds of the state space. The case where $\mathcal{L}(\underbar{s},\delta), \mathcal{L}(\bar{s},\delta)  > 0$ is completely analogous and thus omitted. } Then either
	\begin{enumerate}
		\item  $\underbar{s}$ is the unique globally stable steady state, or
		\item  $s^{\ast}$ is the only locally stable interior steady state.
	\end{enumerate}

Moreover, if  $s' \mapsto \pi(s,s')$ is strictly concave  and there exists $s \in \mathbb{S}^{o}$ such that  $\pi_{2}(s,s) = 0$, then case 2 is the only possibility.
\end{proposition}

\begin{proof}
	See Appendix \ref{app:root.two}.
\end{proof}


Given that the locator function has exactly two roots, $s_{\ast}  < s^{\ast}$, and is negative at both boundaries, Theorem \ref{thm:FP.characterization} implies that either $\Gamma(s,\delta) < s$ for all $s \in \mathbb{S}^{o}$, or $\Gamma$ eventually crosses the \SI{45}{\degree} line. In the first case, all trajectories converge to the lower boundary $\underbar{s}$, which is then the unique globally stable steady state. In the second case, moving from right to left, the first crossing from above must occur at $s^{\ast}$ , which must then correspond to a locally stable steady state. There are three possible sub-cases for the behavior of $\Gamma(\cdot,\delta)$ on $(\underbar{s}, s^{\ast})$: (i) it remains above the \SI{45}{\degree} line; (ii) it jumps below it at a Skiba point; or (iii) it crosses it, in which case Theorem \ref{thm:FP.characterization} implies the crossing occurs at $s_{\ast}$. In all three cases, $s^{\ast}$ remains the only locally stable interior steady state.

Finally, the added condition that $s \mapsto \pi(s,s')$ is strictly concave and satisfies $\pi_2(s,s)=0$ at some interior point ensures that the myopic planner ($\delta=0$) has an interior steady state. But since $\delta \mapsto \Gamma(s,\delta)$ is non-decreasing (see Lemma \ref{lem:PF.prop.text}), this implies that $\Gamma(s,\delta) \geq s$ somewhere, contradicting the possibility of $\Gamma$ being strictly below the \SI{45}{\degree} line throughout. This rules out global convergence to $\underbar{s}$, and since $\bar{s}$ is not a steady state, there has to be an interior locally stable steady state, which must be $s^{\ast}$.

\begin{figure}[!h]
\makebox[\textwidth][c]{
\begin{minipage}{1.0\textwidth}
	\centering
	\caption{Neoclassical growth model with inverted U-shaped locator function}
	\label{fig:ncg}
	\includegraphics[width=\textwidth]{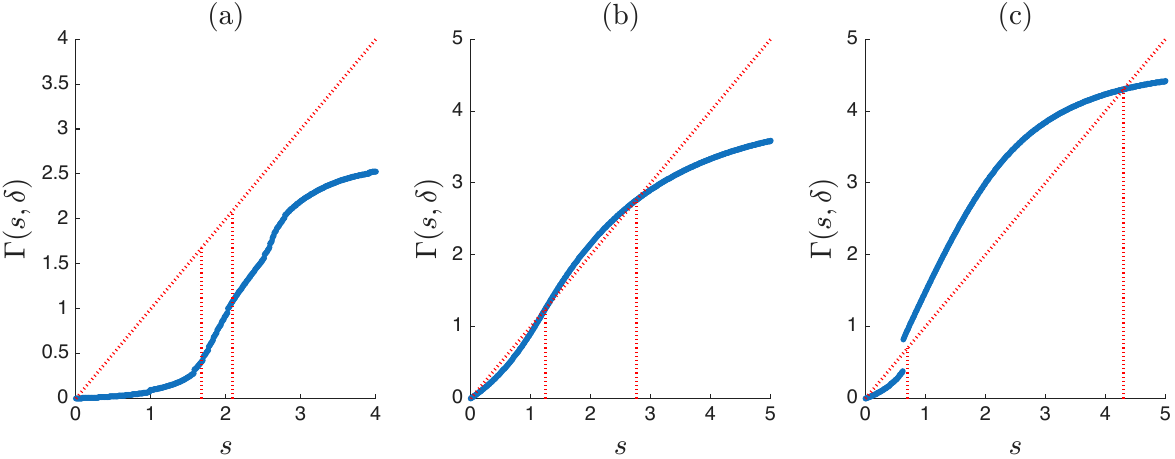}
\end{minipage}
}
\par\vspace{0.5em}
\noindent\small Note: (a) with $ a=0.266, b=1,c=0.5, \delta=0.7$, (b) with $ a=0.25, b=1,c=4.7,\delta=0.18$, (c) with $ a=0.2, b=1,c=1.4,\delta=0.5.$ Relative risk aversion coefficient: $\gamma=0.3$. Vertical dashed lines show the roots of their corresponding locator functions. 
\end{figure}

\paragraph{Application to the generalized NCG model.} 
Proposition \ref{pro:root.two} can be applied to study case (2) in the 
generalized NCG model, in which $\delta f'(0) -1 < 0 < \delta f'(s_{I}) - 1$. 
Since $s \mapsto f'(s)$ is increasing in $[0,s_{I})$ and decreasing in $(s_{I},s_{max}]$, the locator function is indeed inverted U-shaped, and since
$\delta f'(0) < 1$ and $\delta f'(s_{max}) < \delta f'(s^{\ast}) = 1$, then $\mathcal{L}(0,\delta)  < 0$ and $\mathcal{L}(s_{max},\delta)  < 0$. Thus,  by Proposition \ref{pro:root.two} either 0 is the unique globally stable steady state or $s^{\ast}$ is the only locally stable interior steady state. As it turns out, both cases are possible, and depend on the parametrization. To see this, assume CRRA preferences, $c \mapsto u(c) = \frac{c^{\gamma-1} }{\gamma-1}$, and a convex-concave production function given by $s \mapsto f(s) = -\frac{a}{3}s^3 + \frac{b}{2}s^2 + c s$. Varying the values of the different parameters, Figure \ref{fig:ncg}(a) shows a case in which $\underline{s}$ is the unique globally stable steady state, whereas Figures \ref{fig:ncg}(b) and (c) show that $r_{2}$ is the only locally stable interior steady state.


Interestingly, Figure \ref{fig:ncg}(c) illustrates a setting in which the locator function has two roots, but only one corresponds to a steady state. The other root is not a steady state and instead arises from the presence of the Skiba point. This example underscores that Theorem \ref{thm:FP.characterization} is, in a sense, sharp: the inclusion of steady states among the roots of $\mathcal{L}$ cannot generally be strengthened—except in special cases, such as those covered in Propositions \ref{pro:root.unique} and \ref{pro:SCVA.FP.characterization} below.

\subsection{The payoff function is strictly concave}
\label{sec:pi.SCVA}

In this section we discuss the particular case where $\pi$ is strictly concave. This case is the standard case studied in the literature (cf. \cite{stokey1989recursive}) and it requires $\pi_{11} \pi_{22} > (\pi_{12})^{2}$, implying that the externalities are "small" in magnitude relative to the curvature of $\pi$ over $s$ and $s'$.\footnote{The condition $\pi_{11} \pi_{22} \geq (\pi_{12})^{2}$ follows from the fact that the Hessian has to have non-negative determinant.}

The next result shows that in this setting, under mild additional conditions, the locator function not only characterizes stable fixed points, but also their basin of attraction. To state the result we need the following definition of a basin of attraction of a point $s \in \mathbb{S}$ under a function $F : \mathbb{S} \rightarrow \mathbb{R}$,
\begin{align*}
	\mathcal{B}[F](s) & : = ~ \mathcal{B}^{-}[F](e) \cup ~ \mathcal{B}^{+}[F](e), \\
	~\text{with} ~ &\mathcal{B}^{+}[F](e) :  = \{  s \in \mathbb{S} \colon s<e~\text{and}~\forall s' \in (s,e)~F(s') > 0   \} \cup \{e\} \\
	\text{and}~&\mathcal{B}^{-}[F](e) :  =  \{  s \in \mathbb{S} \colon s>e~\text{and}~\forall s' \in (e,s)~F(s') < 0   \} \cup \{e\}.
\end{align*}
That is, $	\mathcal{B}^{+}[F](e) $  is the set of all points $s \in \mathbb{S}$ such that for any point between $s$ and $e$, the function $s \mapsto F(s)$ is positive --- this set includes $e$ as a convention that simplifies the exposition. We refer to $\mathcal{B}[F](e)=\mathcal{B}^{-}[F](e) \cup 	\mathcal{B}^{+}[F](e)$ as the \emph{Basin of attraction (BoA) for $F$} --- this terminology is justified by the following result.

\begin{proposition}\label{pro:SCVA.FP.characterization}
	Suppose $\pi$ is strictly concave and satisfies
		\begin{align}\label{eqn:SCVA.cond-t}
		\max_{a \in \Upsilon(s) \colon a > s} sign \left\{  \pi_{2}(s,a) + \delta \pi_{1}( s ,a  )  \right\}  \leq 0 \leq \min_{a \in \Upsilon(s) \colon a < s} sign \left\{  \pi_{2}(s,a) + \delta \pi_{1}( s ,a  )  \right\}
	\end{align}
	for all $s \in \mathcal{R}^{o}[\mathcal{L}]$.\footnote{The function $sign$ is such that $x \mapsto sign\{ x \} = 1$ if $x>0$ and $-1$ if $x<0$ and $0$ if $x=0$.}
Then  $ \mathcal{R}^{o}[\bar{\Gamma}] =  \mathcal{R}^{o}[\mathcal{L}] $ and 
	\begin{align}
		\mathcal{B}(s,\delta) = \mathcal{B}[\bar{\Gamma}](s) =  \mathcal{B}[\mathcal{L}](s),~\forall s \in \mathcal{R}^{o}[\bar{\Gamma}].
	\end{align}
	
\end{proposition}

\begin{proof}
	See Appendix \ref{app:SCVA.FP.characterization}.
\end{proof}

This result implies that when $\pi$ is strictly concave (and satisfies condition \ref{eqn:SCVA.cond-t}) the locator function identifies not only the steady states but also their basins of attraction thereby giving an easy-to-use tool for understanding the dynamics of optimal paths. This is summarized in Table \ref{tab:SCVA}

\begin{table}[h!]
	\centering
	\begin{footnotesize}
		
		\begin{tabular}{|l|l|l|l|l|}
			\hline
			Concepts & Relationship & $\Gamma$ & Relationship &  $\mathcal{L}$ \\ \hline
			Steady State          &   $=$ (Proposition \ref{pro:equilibria.chactarization})  &   $\mathcal{R}[\bar{\Gamma}]$   & $=$ (Proposition \ref{pro:SCVA.FP.characterization})  & $\mathcal{R}[\mathcal{L}]$       \\ \hline
			Basin of Attraction ($\mathcal{B}(\cdot,\delta)$) &  $=$ (Proposition \ref{pro:Basins.chactarization})          &      $\mathcal{B}[\bar{\Gamma}](\cdot)$    & $=$ (Proposition \ref{pro:SCVA.FP.characterization}) &  $\mathcal{B}[\mathcal{L}](\cdot)$            \\ \hline
		\end{tabular}
		\caption{Equivalences under strict concavity of $\pi$ \label{tab:SCVA}}
	\end{footnotesize}
\end{table}

Condition \ref{eqn:SCVA.cond-t} allows us to link the first order condition (FOC), which is sufficient, with the locator function. To illustrate its role in the theorem, take an $s \in \mathbb{S}^0$ such that $\Gamma(s, \delta) > s$. Since $s' \mapsto V_{1}(s',\delta)$ is decreasing (by strict concavity of the value function), the envelope condition implies $\pi_{1}(s,\Gamma(s, \delta)) > \pi_{1}(\Gamma(s, \delta),\Gamma^2(s, \delta))$, and hence 
 $$\pi_{2}(s,\Gamma(s, \delta) ) + \delta \pi_{1}(s,\Gamma(s, \delta))  > \pi_{2}(s,\Gamma(s, \delta) ) + \delta \pi_{1}(\Gamma(s, \delta),\Gamma^2(s, \delta))  = 0.$$
 From the first inequality in condition \ref{eqn:SCVA.cond-t} we then conclude that $\mathcal{L}(s,\delta) > 0$. An analogous result holds for the case $\Gamma(s, \delta) < s$, thereby showing that for any $s$ such that $\Gamma(s, \delta)  \ne s$, $\mathcal{L}(s,\delta) \ne 0$. Thus, the locator function will never deliver "false zeros". 

Condition \ref{eqn:SCVA.cond-t}  appears somewhat cumbersome, but is easy to verify. For instance, it holds if $sign \{ \mathcal{L}(s,\delta)  \} = sign \{ \pi_{2}(s,a) + \delta \pi_{1}( s ,a  )  \}$ for any $a \in \Upsilon(s)$. In the \ref{exa:NCG}, $ \pi_{2}(s,a) + \delta \pi_{1}( s ,a  )  = U'(a-f(s)) ( \delta f'(s) -1 )$ and since $U'>0$, the sign of this function is determined by the sign of $\delta f'(s) -1 $, which precisely determines the sign of the locator function. Thus Condition \ref{eqn:SCVA.cond-t} holds. Additional sufficient conditions are provided by  Lemma \ref{lem:suff.cond.SSiffRoot} in the Online Appendix \ref{app:SCVA.cond}  --- for instance it shows that Condition \ref{eqn:SCVA.cond-t}  is implied by $\pi_{22} + \delta \pi_{12} \leq 0$.


\subsection{Using the Locator Function for Comparative Statics}

In this section we show how to use the locator function for deriving comparative statics results for steady states. To do this, suppose the per-period payoff is parameterized by an index $\xi \in \mathbb{R}$, against which we would like to perform comparative statics of steady states.\footnote{The results here easily extend to $\xi \in \mathbb{R}^q$ for $q>1$.} We focus on \emph{stable} steady states, as unstable ones will never be attained (unless in trivial cases where the initial state is the steady state itself). 

The next result shows that the locator function can be used to perform comparative statics of steady states against the parameter $\xi$. To show this, we slightly change notation and allow the locator function to depend explicit on $\xi$ by using $ \mathcal{L}(\cdot,\delta,\xi)$. Suppose $(s,\xi) \mapsto \mathcal{L}(s,\delta,\xi)$ is continuously differentiable. By Theorem \ref{thm:FP.characterization}, for any $ \xi$ and any stable interior steady state, $s(\xi)$, $\mathcal{L}(s( \xi), \delta ,  \xi)= 0$ and $\mathcal{L}_{1} (s(\xi), \delta ,  \xi) < 0$. Thus, by the IFT, the mapping $\xi \mapsto s( \xi)$ is continuously differentiable with derivative given by 
	\begin{align*}
		 s'(\xi)  = - \frac{  \mathcal{L}_{3}(s(\xi),\delta ,  \xi)  }{\mathcal{L}_{1}(s(\xi),\delta , \xi)},
	\end{align*}
with the sign of  $s'( \xi)  $ equal to the sign of $ \mathcal{L}_{3}(s( \xi),\delta,  \xi) $. 

This logic implies the following result
\begin{proposition}\label{pro:CS}
	Suppose $(s,\xi) \mapsto \mathcal{L}(s,\delta, \xi)$ is continuously differentiable. Then for any $ \xi \in \mathbb{R}$, any stable interior steady state, $s( \xi)$, is increasing (decreasing) in $ \xi$ iff $ \mathcal{L}_{3}(s(\xi),\delta ,  \xi) $ is positive (negative).
\end{proposition}

 \section{Applications}
\label{sec:applications}
In this section we introduce two applications to different economies that are encompassed by our framework. 


 \subsection{A Model of Rational (Un)fitness}
\label{exa:fit}

Inspired by the \textit{Rational Addiction} framework of \cite{becker1988theory}, this application examines agents who choose whether and how much to exercise, recognizing its impact on fitness and, consequently, future utility. We show that this setup can lead to multiple steady states with varying fitness levels, consistent with experimental findings that incentivizing gym attendance results in long-term fitness changes (e.g., \cite{charness2009incentives}).  The key driver of these dynamics is a direct utility boost from exercising, which increases with fitness. We refer to this as the "endorphins effect". 

\paragraph{Setup.} The fitness level, denoted by $s$, evolves according to $s' = (1-d) s + x$ where $d$ is the rate of depreciation and $x \in [0,1]$ represents exercise time. The agent's reward given fitness level $s$ and exercise time $x$ is $R(s,x) : =  s^{\alpha} + b s^{\beta} x $, where $0<\alpha<1, \beta >0$, and $b \geq 0$ 
determines the strength of
the endorphins effect
relative to the direct benefit from fitness. The agent also has a cost of exercising given by $C(x)$, with $C'\geq0$, $C''>0$. Hence, the per-period payoff and constraint correspondence are given by
\[ 
\pi(s,s')= R(s,s'-(1-d)s)  -  C(s^{\prime}-(1-d)s)~\text{and}~\Upsilon(s) = [(1-d)s,(1-d)s+1].
\]  

Contrary to the original rational addiction literature, our approach does not impose strict concavity of $\pi$. While often treated as a technical condition, imposing strict concavity would restrict the strength of the endorphins effects --- i.e., the  complementarities between fitness and exercise.\footnote{In our notation, the concavity of $\pi$ constrains the cross derivative $\pi_{12}$ relative to the second derivatives $\pi_{11}$ and $\pi_{22}$.} As we show below, this restriction has important 
implications as the dynamics of the optimal paths under a weak endorphins effect can be qualitatively different from those with a strong endorphins effect.

\paragraph{Verification of Assumptions \ref{ass:pi.properties}-\ref{ass:C.properties}.} Given the constant depreciation rate $d$, the maximum sustainable fitness level is $1/d$, so we define the state space as $\mathbb{S} = [0, k/d]$ for some scaling factor $k\geq1$. It is straightforward to show the validity of Assumption \ref{ass:pi.properties}(ii)(iii), but Assumption \ref{ass:pi.properties}(i) is less straightforward, as $\pi_{1}$ may not be strictly positive due to the endorphins effect ($R_{2}>0$). The next lemma shows a condition on parameters under which Assumption \ref{ass:pi.properties} holds.

\begin{lemma}\label{lem:fit.sufficient}
	Suppose $b (1-d)  \leq  \alpha (d/k)^{1+\beta-\alpha}$ and $1+\beta-\alpha \geq0$. 	Then, Assumption \ref{ass:pi.properties} is satisfied.
\end{lemma}

\begin{proof}
	See Appendix \ref{app:fit}.
\end{proof}

Since $s \mapsto \Upsilon(s) = [(1-d)s,(1-d)s+1]$, parts (i) and (iii) of Assumption \ref{ass:C.properties} are readily satisfied, but part (ii) is not. However, Assumption \ref{ass:C.properties}(ii) is only needed to establish monotonicity of the value function. Lemma \ref{lem:VF.incr.v2} and Remark \ref{rmk:VF.incr.v2}(3) in the Online Appendix \ref{app:VF.properties} shows that this monotonicity result holds in this (un)fitness model without Assumption \ref{ass:C.properties}(ii), so this assumption is not needed in this example.

\paragraph{The Locator function and optimal path dynamics.}  The locator function is given by 
\begin{align*}
	s \mapsto \mathcal{L}(s,\delta) = H(s) - C'(ds)(1-\delta(1-d)),   
\end{align*}
where $H(s) : = (1-\delta(1-d) + \delta \beta d) b s^{\beta} + \delta \alpha s^{\alpha-1}$ captures the 
direct payoff from higher fitness as well as the endorphins effect, and $s \mapsto C'(ds)(1-\delta(1-d))$ captures the marginal cost of exercise in steady state.  This decomposition of the locator function shows that its roots are given by the points such that $H(s) = C'(ds) ( 1 - \delta (1-d)) $. The RHS is increasing by assumption, while monotonicity of $H$ depends on parameters $\alpha,\beta,b,d$. 

Consider first the extreme case in which there is no “endorphins” effect—that is, $b = 0$. In this case, the function $H$ is decreasing (since $\alpha \in [0,1)$), and the locator function is single crossing from above. By Proposition \ref{pro:root.unique}, the unique interior steady state—given by the root of the locator function—is globally stable. This result suggests that, in the absence of the “endorphins” effect, individuals with low initial fitness find it optimal to increase their fitness, while those with high initial fitness allow it to depreciate to some extent, with everyone eventually converging to the same steady state.

\begin{figure}[!h]
\makebox[\textwidth][c]{
\begin{minipage}{1.0\textwidth}
	\centering
	\caption{(Un)fitness model policy and locator functions}
	\label{fig:fit}
	\includegraphics[width=\textwidth]{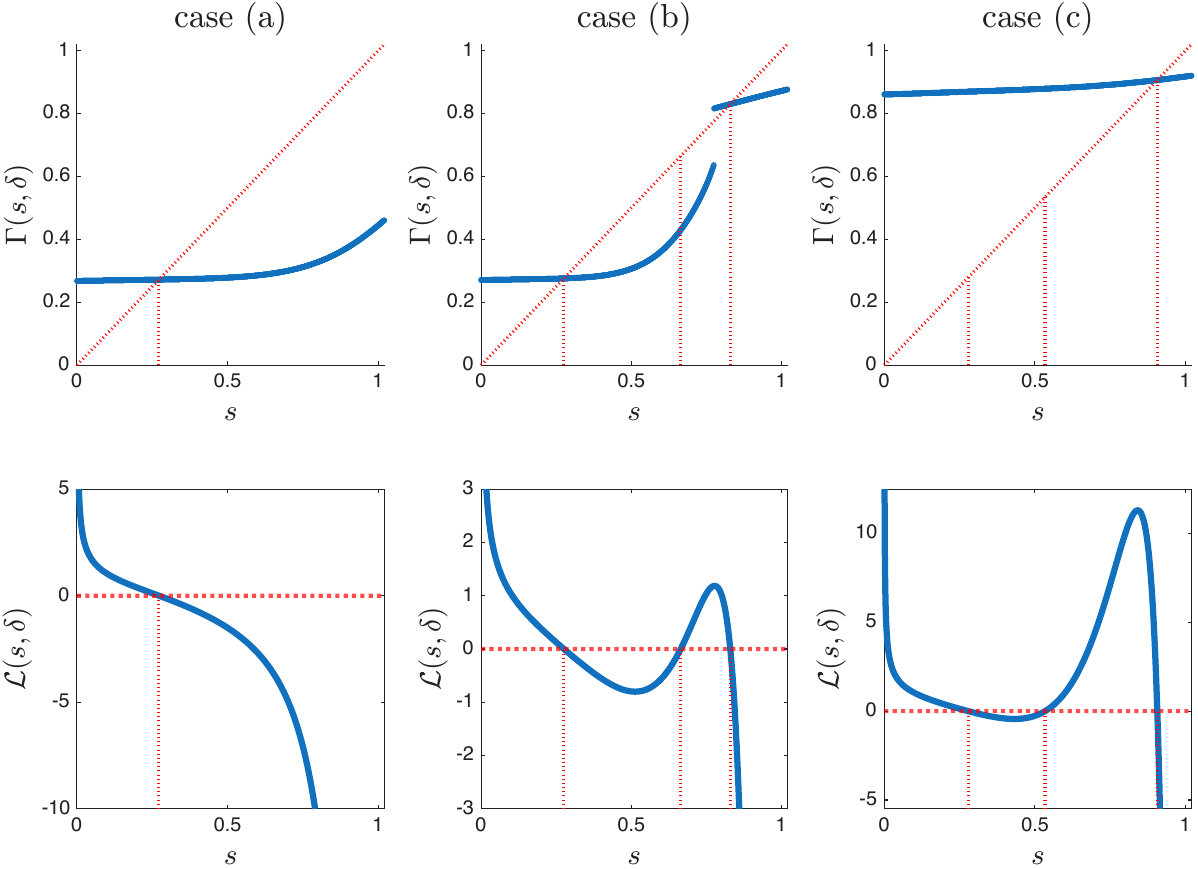}
\end{minipage}
}
\par\vspace{0.5em}
\noindent\small Note: Case (a) with "moderate" endorphin effects: $b=1$; Case (b) with high endorphin effects: $b=9$; Case (c) with very high endorphin effects: $b=15$. Other parameters are constant in all 3 cases: $\alpha=0.5,\beta=6,a=1,c=0.6,d=0.98,\delta=0.8$.
Vertical dashed lines show the roots of corresponding locator functions. 
\end{figure}


A positive endorphins effect introduces more nuanced dynamics, as the payoff function from fitness and exercise, $H$, can now be increasing—potentially giving rise to multiple steady states. To illustrate the range of possibilities, we adopt the following functional form for the cost function: $x \mapsto C(x) = \frac{a}{2}x^2 + c\left( \frac{1}{1 - x} + x \right)$, with $\mathbb{S} = [0, \frac{1}{d}]$. This specification ensures that the planner always chooses an interior level of exercise, i.e., $x \in (0,1)$. In Figure \ref{fig:fit}, we explore different parameter values, numerically solve for the policy correspondence (upper panel), and examine what the locator function reveals about steady states (lower panel).

Figure \ref{fig:fit}(a) shows a case with a moderate endorphins effect, resulting in a unique and globally stable interior steady state at a low fitness level. In Figure \ref{fig:fit}(b), a stronger endorphins effect gives rise to two interior locally stable steady states --- one at a low fitness level, which we call a "couch potato" steady state, and one at a high level, which we call, for obvious reasons, a "Julian Alvarez" steady state --- and an intermediate Skiba point. In contrast, Figure \ref{fig:fit}(c) depicts a very strong endorphins effect, again resulting in a unique and globally stable interior steady state—this time at a high fitness level. In case (a), the locator function cleanly identifies the steady state: since it is single crossing decreasing, Proposition \ref{pro:root.unique} applies. Cases (b) and (c) illustrate both the strengths and limitations of the locator function approach. In both, the locator function has two roots with negative slope, signaling candidates for locally stable interior steady states. This prediction is accurate in case (b), as confirmed by the policy correspondence in the upper panel. In case (c), however, the lower root is a “false positive”—a root that does not correspond to an actual steady state.

Based on our results and supporting experimental evidence (e.g., \cite{charness2009incentives}), a key takeaway is that imposing strict concavity is not merely a technical assumption. Rather, it rules out economically plausible features of the utility function and limits the model’s ability to reflect observed behavior. Even without concavity, sharp predictions can still be obtained when the locator function is single crossing decreasing, and even without that, the high-fitness locally stable steady state is correctly identified. The main limitation is that the low-fitness root of the locator function may be a false positive---that is, it may not correspond to an actual steady state. Nonetheless, even in this case, the locator function supports meaningful comparative statics. For example, since it is increasing in $b$, Proposition~\ref{pro:CS} implies that interior locally stable steady-state fitness levels rise as the endorphin effect becomes stronger.





\subsection{An Economy with Intertemporal Economies of Scale}\label{exa:intertemporal}

Our second application considers a planner’s problem in a two-sector economy where only one sector exhibits learning-by-doing, generating a trade-off between present consumption and future productivity gains. Depending on the strength of learning by doing, the planner's optimal path may have a globally stable interior steady state, or instead exhibit path dependence with multiple steady states. 

This application is motivated by a large literature, including \citet{krugman1987}, \citet{lucas1988}, \citet{young1991}, and \citet{redding1999}, which study how learning-by-doing externalities can give rise to multiple steady states and inefficient specialization in market economies absent industrial policies. Closer to our analysis, \citet{bardhan1971} and \citet{melitz2005} derive necessary conditions for optimality, but restrict the analysis to settings with strict concavity and do not explore the possibility of multiple steady states. 

We study an open economy, Home, with two goods, and a unit of labor
supplied inelastically. We use $s$ and $s'$ to denote employment in production
of good $1$ in the previous and current periods. There is learning by doing in the production
of good 1, with this period's productivity shifter a function of last
period's employment. 
 Production of good 1 in the current period is thus given by $H(s)F(s')$, where  $s \mapsto H(s) : = s^{\theta}$ and $s' \mapsto F(s') : = (s')^{\alpha}$ with $\theta,\alpha \in (0,1)$.
Production of good $0$ in the current period is given
by $G(1-s')$, with $G:\mathbb{R}_{+}\rightarrow\mathbb{R}$ smooth,
increasing and concave. 

To simplify the analysis, we further assume that Home's per-period
utility function $U:\mathbb{R}_{+}^{2}\rightarrow\mathbb{R}$ is quasilinear,
$U(c_0, c_1)=c_0 + u\left(c_{1}\right)$, with $u:\mathbb{R}_{+}\rightarrow\mathbb{R}$
smooth, increasing and strictly concave. Home is the only producer
of good $1$, with 
the rest
of the world's inverse demand curve for Home's exports of good $1$ given by $e \mapsto p(e)=b e^{1/\varepsilon}$.
Using good $0$ as numeraire, and suppressing subindex $1$ for good $1$ consumption, the per-period payoff
function of Home's planner is then
\begin{align*}
	(s,s')\mapsto\pi(s,s'):=\max_{c}u(c)+G(1-s' )+p\left(H(s)F(s')-c\right)\cdot\left(H(s)F(s') - c\right).
\end{align*}
The only restriction on the current period's employment in production of good 1 is that it respect the resource constraint on labor, hence $s' \in \Upsilon(s)$ with $s\mapsto \Upsilon(s):=[0,1].$ Home's production of good $1$ is
$H(s) F(s')$, of which $e(s,s') := H(s)F(s')-c(s,s')$
units are exported at price  $p\left( e(s,s') \right)$, with $c(s,s')$ denoting the solution of the above optimization problem for a given $(s,s')$ pair.
Export revenues fund imports of good $0$, complementing Home's own
production of that good for consumption.\footnote{We can generalize several of these assumptions. First, we can allow for there to be a foreign variety of good $1$ (as in the standard
	Armington trade model) and have Home's utility given by $U(c_{0},c_1,c_{1}^{*})=c_0 + u\left(c_{1}\right)+u^{*}\left(c_{1}^{*}\right)$,
	where $c_{1}^{*}$ is Home's consumption of the foreign variety of
	good $1$ and $u^{*}:\mathbb{R}_{+}\rightarrow\mathbb{R}$ is smooth,
	increasing and strictly concave. Assuming that the relative price
	between the foreign variety of good $0$ and good $1$ is fixed, the
	results below carry through without modification. Alternatively, we
	can remove the separability assumption between goods $0$ and $1$
	and assume that Home is a small open economy (i.e., the relative price
	at which it can trade these two goods is fixed), or that it is a closed
	economy with $U(c_{0},c_{1})=\frac{\gamma}{\gamma-1}\left(c_{0}^{\frac{\sigma-1}{\sigma}}+c_{1}^{\frac{\sigma-1}{\sigma}}\right)^{\frac{\gamma-1}{\gamma}\frac{\sigma}{\sigma-1}}$
	, $1<\gamma<\sigma$. These results are discussed in the Online Appendix \ref{app:ie.extension}.
	What is difficult is to simultaneously allow for non-separability
	in preferences with Home open to trade with endogenous foreign prices.
	This is because then $\pi_{12}(s,s')>0$ is hard to verify for the whole
	state space.  } 

\paragraph{Verification of Assumptions \ref{ass:pi.properties}-\ref{ass:C.properties}.} 

Since the planner behaves as a monopolist for good $1$ in the foreign market, we need to assume the elasticity of foreign demand for Home's exports of good $1$, $\varepsilon =-\left(\frac{d\ln p\left(e\right)}{d\ln e}\right)^{-1} $, is larger than 1 for there to be an interior solution. In the following lemma, we show that as long as the elasticity of marginal utility of good 1, $c \mapsto \gamma(c):=-\left(\frac{d\ln u'(c)}{d\ln c}\right)^{-1}$, is always larger than 1, Assumption \ref{ass:pi.properties} holds.

\begin{lemma}\label{lem:ie.pi12}
If $\varepsilon>1$ and $\gamma(\cdot)>1$, then Assumption \ref{ass:pi.properties} is satisfied.
\end{lemma}

\begin{proof}
See Appendix \ref{app:intertemporal}.
\end{proof}

To gain some intuition for this condition, note that by the Envelope
Theorem we have $\pi_{1}(s,s')=u'(c(s,s'))H'(s)F(s')$.
An increase in $s'$ implies a higher $F(s')$ and hence a
higher $\pi_{1}(s,s')$. This positive direct effect is counteracted by
a negative indirect effect arising from the fact that a higher $F(s')$
implies more consumption of good $1$ and hence a lower marginal utility
$u'(c)$. In a closed economy $\gamma>1$ is necessary and sufficient
to ensure that the positive direct effect dominates the indirect negative
effect. In an open economy this is relaxed because part of the extra
output is exported, hence marginal utility declines less. The proof of the lemma shows
\[
\pi_{12}(s,s')=u'(c(s,s')) H'(s) F'(s') \frac{\gamma(c(s,s') )-1+\left(\varepsilon-1\right)e(s,s')/c(s,s')}{1+\left(\varepsilon/\gamma(c(s,s') )\right)e(s,s')/c(s,s')},
\]
hence $\pi_{12}(s,s') > 0$ if $\gamma(\cdot)>1$.\footnote{Note that $\gamma(\cdot)>1$ is a not a necessary condition. If exports are high relative to consumption and if foreign demand is highly elastic then most of the extra output associated with a higher $s'$ will be exported, so we can have $\pi_{12}(s,s')>0$ even with $\gamma(\cdot)$ lower than 1.}  Assumption \ref{ass:C.properties} is trivially satisfied. 

\paragraph{The locator function and optimal path dynamics.} We focus on the case with $\gamma(\cdot)= \varepsilon = : \gamma$, which significantly simplifies  the analysis because it implies that a constant share of production is exported (with the rest consumed). The locator function is then simply
\begin{align}  \label{eq: ie locator} 
s \mapsto \mathcal{L}(s, \delta) =\left(1+ \left(b\left(1-1/\gamma \right)\right)^{\gamma} \right) ^ {1/\gamma} 
(\alpha+\delta \theta)s^{ (\theta+\alpha) (1-1/\gamma ) -1 } - G'(1-s).
\end{align}
Further imposing a restriction on $G$, we can use this locator function to find an intuitive condition on the strength of learning externalities determining whether we have a single or multiple steady states:
\begin{lemma}\label{lem:ie.L}
Assuming that $ \varepsilon  = \gamma(c) = : \gamma $ for all $c$ and $G(1-s')=1-s'$, the following are true:
\begin{enumerate}
	\item If $\theta < \frac{1}{\gamma - 1} + 1 - \alpha$, then the locator function is single crossing from above, and its interior root is the globally stable steady state. 
	\item If $\theta > \frac{1}{\gamma - 1} + 1 - \alpha$ and $\mathcal{L}(\bar{s},\delta)>0$ then the locator function is single crossing from below, and does not have an interior stable steady state.
\end{enumerate}
\end{lemma}

\begin{proof}
See Appendix \ref{app:intertemporal}.
\end{proof}

The mechanism that can lead to multiple steady states is learning by doing, captured by $H(s) = s^{\theta}$. When the learning elasticity $\theta$ is sufficiently high, the future productivity gains from higher current employment in sector 1 can outweigh the effects of diminishing returns—specifically, the diminishing marginal utility of consumption (or declining export prices), represented by 
$\frac{1}{\gamma - 1}$, and the decreasing marginal product of labor, given by $1 - \alpha$. To see how these forces interact, consider a simplified setting: a static, closed economy in which the planner chooses employment in sector 1 to maximize utility, i.e., $\max_x \left(x^{\theta}x^{\alpha}\right)^{1 - 1/\gamma} + 1 - x$. In this case, the condition for an interior solution is precisely $\theta < \frac{1}{\gamma - 1} + 1 - \alpha$.

\begin{figure}[!h]
\makebox[\textwidth][c]{
\begin{minipage}{1.0\textwidth}
	\centering
	\caption{Intertemporal externality model}
	\label{fig:ie}
	\includegraphics[width=\textwidth]{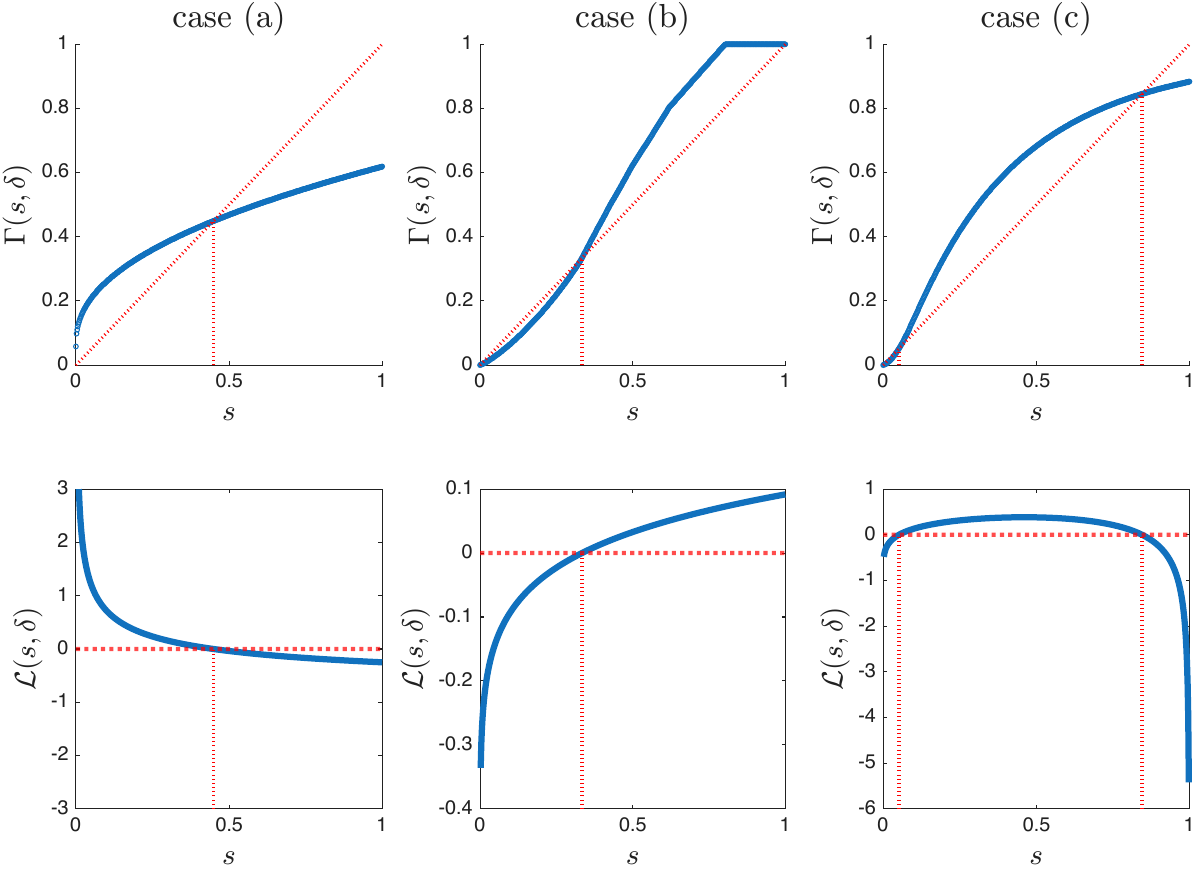}
\end{minipage}
}
\par\vspace{0.5em}
\noindent\small Note: Case (a) with $\theta =0.5,\gamma=\varepsilon=5, \alpha=0.3,\beta=1,b=2,\delta=0.32.$ Case (b) with $\theta =1,\gamma=\varepsilon=5, \alpha=0.35,\beta=1,b=2,\delta=0.32.$ Case (c) with $\theta =1,\gamma=\varepsilon=8, \alpha=0.38,b=2.5,\delta=0.12,\beta = 0.68.$ $s'\mapsto \pi(s,s')$ is concave and there exists $s$ in $\mathbb{S}^o$ such that $\pi_2(s,s)=0$ in case (c). Vertical dashed lines show the roots of corresponding locator functions.  
\end{figure}

As in the previous application, we use a figure to illustrate the range of possibilities for the optimal path. For each case, the upper panel displays the numerically computed policy correspondence, and the lower panel shows the corresponding locator function. Cases (a) and (b) in Figure \ref{fig:ie} assume a linear payoff in sector 0, $G(s') = 1 - s'$ (as in Proposition \ref{lem:ie.L}), and differ only in the value of the learning elasticity: $\theta < \frac{1}{\gamma - 1} + 1 - \alpha$ in case (a), and $\theta > \frac{1}{\gamma - 1} + 1 - \alpha$ in case (b). Case (c) also assumes a high learning elasticity but introduces diminishing returns in sector 0, with $G(1 - s') = \frac{(1 - s')^\beta}{\beta}$ for $\beta \in (0,1]$.

In case (a), a low learning elasticity leads to a unique, globally stable interior steady state. In case (b), the interior steady state becomes unstable, and the dynamics exhibit threshold behavior: if the initial productivity of sector 1 is below the unstable steady state, the planner shifts employment fully to sector 0; if it is above, full specialization in sector 1 emerges. In case (c), the introduction of diminishing returns in sector 0 shifts the upper steady state into the interior, breaking the corner solution seen in case (b).

The locator function in the lower panels successfully captures these dynamics. In case (a), it has a single interior root with negative slope, corresponding to a globally stable interior steady state, as implied by Proposition \ref{pro:root.unique} and Lemma \ref{lem:ie.L}. In case (b), the locator function has a single interior root with positive slope, ruling out interior locally stable steady states and confirming that the only stable steady states lie at the boundaries. In case (c), the locator function exhibits an inverse-U shape with negative values at both extremes. By Proposition \ref{pro:root.two}, this implies either a single interior locally stable steady state at the higher root or global stability at the lower boundary. Moreover, since $\pi_{22}(s,s')<0$ and the myopic policy function has an interior fixed point, we can invoke the latter part of Proposition \ref{pro:root.two} to conclude that the higher root of the locator function must be a locally stable steady state.

When an interior locally stable steady state exists (as in cases a and c in Figure \ref{fig:ie} ), we can use Proposition \ref{pro:CS} to study how various shocks affect the steady state allocation, $s$. We see from \eqref{eq: ie locator} that the locator function is increasing in $b$, which governs foreign demand for Home's exports, and the discount rate $\delta$. Thus, Proposition \ref{pro:CS} implies that steady state $s$ increases with $b$ and $\delta$: not surprisingly, higher export demand for good $1$ leads to higher employment in that sector, while lower discounting leads to higher employment in the sector that exhibits learning by doing.

  \section{Remarks}
\label{sec:remarks}

In this section, we review existing approaches for analyzing steady states and their stability, discuss how our proposed method relates to the most widely used of these approaches, and conclude with a discussion of the role played by Assumption \ref{ass:pi.properties}.

\paragraph{Taxonomy of the existing literature.} While the stability of steady states can always be analyzed on a case-by-case basis, the general analytical approaches used in the literature fall roughly into two categories: the Lyapunov method and linear approximation methods. The Lyapunov approach (e.g., \cite{brock1976global}; \cite{stokey1989recursive}, Section 6.2) is global in nature but there is almost no guidance on how to establish existence or construct the Lyapounov function, thereby limiting its applicability in many settings.\footnote{See the discussion in \cite{stokey1989recursive} p. 139-140. }

Linear approximation methods require additional assumptions, such as strict concavity of the per-period payoff, and are both local in nature and uninformative about the precise shape of the basin of attraction.\footnote{To our knowledge, global results based on linear approximation have been obtained only for the Cass--Koopmans growth model; see \cite{nishimura2004discrete} and references therein.} To fix ideas, we consider a version of the approach presented in Sections 6.2 and 6.3 of \cite{stokey1989recursive}, and summarized in Theorem 6.9.  
We formally discuss this theorem and its connection to our results below, but the main takeaway is that Theorem 6.9 assumes both strict concavity of the per-period payoff and uniqueness of the steady state---assumptions not required by our method. More importantly,  
a key limitation of this result is that it guarantees the existence of a local neighborhood around the steady state in which convergence occurs, but does not characterize this neighborhood. As a result, 
given an initial condition, one cannot determine whether it lies within this region, and thus whether convergence to the steady state will in fact occur. We also point out how, for cases in which $\pi$ is \emph{not} strictly concave
the aforementioned method cannot be implemented. Our method, on the other hand, can be applied, albeit delivering conservative estimates of the basin of attractions. 

\paragraph{Relationship with local approximation results.} 
We now present a more thorough discussion of the similarities and differences between Proposition \ref{pro:SCVA.FP.characterization} and the local approximation method, represented by
Theorem 6 in \cite{stokey1989recursive}.
The approach in \cite{stokey1989recursive} considers the dynamical system $\zeta_{t+1} = \Xi(\zeta_{t}) : = (  \zeta_{t}^{(2)}, G(\zeta_{t}) )$, where $G : \mathbb{S}^{2} \rightarrow \mathbb{S}$ is defined implicitly by $0 = \pi_{2}(\zeta^{(1)},\zeta^{(2)}) + \delta \pi_{1}(\zeta^{(2)},G(\zeta))$ for any $\zeta = (\zeta^{(1)},\zeta^{(2)}) \in \mathbb{S}^{2}$. 
It is clear that a fixed point of $\Xi$ will be a fixed point of $\Gamma(\cdot, \delta)$.  Moreover there exists an equivalence among the respective flows:  A flow of $\Xi$, $(\zeta_{t})_{t}$ can be seen as $\zeta_{t} =  (s_{t},s_{t+1})$ where $(s_{t})_{t}$ is a flow of $\Gamma(\cdot, \delta)$. Hence, it suffices to analyze the stability properties of fixed points $\Xi$. 

 \citet{stokey1989recursive} study the Jacobian of $\Xi$ at a fixed point $\zeta = (s,s)$, 
\begin{align*}
	J_{\Xi}(\zeta) = \left[
	\begin{array}{cc}
		0 &    1 \\
		G_1(\zeta) & G_2(\zeta) \\
	\end{array}
	\right],
\end{align*}
and claim that local stability of a fixed point $\zeta$ suffices to show $J_{\Xi}$ has a unique characteristic root with absolute value less than one (cf. Theorem 6.9). $J_{\Xi}$'s characteristic polynomial is given by $\lambda^{2} - \lambda G_{2}(\zeta) - G_{1}(\zeta) = 0$. By the IFT the derivatives of $G$ are given by $G_1(\zeta)=-\pi_{21}(\zeta^{(1)}, \zeta^{(2)}) / (\delta \pi_{12}(\zeta^{(2)}, G(\zeta))) $ and $G_2(\zeta) = - ( \pi_{22}(\zeta^{(1)}, \zeta^{(2)}) + \delta \pi_{11}(\zeta^{(2)}, G(\zeta))  ) / (\delta \pi_{12}(\zeta^{(2)}, G(\zeta))) $ respectively. Evaluating these expressions at $\zeta =(s,s)$ and some straightforward algebra implies the following characterization of the characteristic polynomial:
\begin{align*}
	\lambda^{2} + \lambda (\pi_{22}(s,s) + \delta \pi_{11}(s,s))/(\delta \pi_{21}(s,s) )  + 1/\delta = 0.
\end{align*}
Since the LHS is a quadratic function and coefficient of $\lambda$ is negative, by Vieta's formula the two roots should both be positive. Then note that LHS at $\lambda = 0$ is positive, so existence of a positive root with absolute value less than one is implied by the LHS being negative at $\lambda =1$; i.e., $\delta \pi_{21}(s,s)  + \pi_{22}(s,s) + \delta \pi_{11}(s,s) + \pi_{21}(s,s) < 0$. This is \emph{exactly} the condition of $\mathcal{L}_{1}(s,\delta) < 0$ which we derived in Theorem \ref{thm:FP.characterization}. That is, the linear approximation method uses the \emph{same} mathematical quantity--the sign of $\mathcal{L}_{1}(s,\delta)$-- to assess whether a fixed point is stable. The difference, however, lies in how each approach arrives at this result. 

Instead of using a linear approximation, our approach relies on the locator function, which is linked to the dynamical behavior under "order conditions" (Assumption \ref{ass:pi.properties}), and on drawing insights from the Samuelson principle and comparative statistics. This discrepancy has important implications in terms of the characterization of the basins of attraction. Theorem 6.9 in \cite{stokey1989recursive} only proves an open neighborhood of initial conditions leading to convergence but does not specify it, limiting its applicability since one cannot determine if a given initial condition will converge. Proposition \ref{pro:SCVA.FP.characterization} on the other hand, provides, under an easy-to-check additional condition, a complete characterization of the basin of attractions and of steady states.

\paragraph{On the role of Assumption \ref{ass:pi.properties}.} 
As noted above, our approach replaces the strict concavity assumption common in the literature with the monotonicity conditions outlined in Assumption \ref{ass:pi.properties}. 
These conditions are essential for establishing the comparative statics properties of $\Gamma$ that underpin Theorem \ref{thm:FP.characterization}.

Formally, the key comparative statics result states that $\delta \mapsto \Gamma(s, \delta)$ is monotonic in the sense that $\max \Gamma(s, \delta) \leq \min \Gamma(s, \delta')$ for any $\delta'> \delta$. To obtain this result we invoke the Milgrom-Shannon theorem combined with 	
\begin{align}\label{eqn:Gamma.CS-1}
	s' \mapsto  V(s', \delta) + \delta 	\frac{ d V(s', \delta)}{d \delta} 
\end{align}
being 
non-decreasing for some $s'' \in  \Gamma(s', \delta)$. 
This expression captures the marginal effect of the discount factor on the payoff and consists of two components: (i) a direct effect, reflecting how the discount factor influences the valuation of the "next period" payoff (the first term on the right-hand side), and (ii) an indirect effect, reflecting how it affects the valuation of future payoffs beyond the next period (the second term). To show that the overall expression is non-decreasing in $s'$, it suffices to verify that both components are themselves non-decreasing. For the direct effect (i), this requires the value function to be non-decreasing. For the indirect effect (ii), given the monotonicity of the value function, we need that higher $s'$ leads to higher $s''$, which in turn requires that $\Gamma(\cdot, \delta)$ is non-decreasing. The main role of  Assumption \ref{ass:pi.properties} is precisely to establish that both the value function and policy 
correspondence are non-decreasing. Indeed, the monotonicity restriction in Assumption \ref{ass:pi.properties}(i) is (only) used to establish the monotonicity of the  value function, and monotonicity restriction in Assumption \ref{ass:pi.properties}(iii) is (only) used to establish the monotonicity of the  policy correspondence.

We now discuss extensions (and limitations) of these assumptions. Consider the following versions of Assumption \ref{ass:pi.properties} and \ref{ass:C.properties} where, essentially, the order conditions are flipped.

\setcounter{assumption}{0}
\renewcommand{\theassumption}{ALT.\arabic{assumption}}

\begin{assumption}
	\label{ass:pi.properties.ALT} $\pi$ is continuous and (i) $s \mapsto \pi(s,s')$ is decreasing, uniformly smooth a.e. with $\partial^{+}_{1} \pi \leq \partial^{-}_{1} \pi$;
	(ii) $s' \mapsto \pi(s,s')$ is smooth with derivative denoted as $\pi_{2}$; (iii) $s \mapsto \pi_{2}(s,s')$ is decreasing. 
\end{assumption}

\begin{assumption}\label{ass:C.properties.ALT}
	(i) $s \mapsto \Upsilon(s)$ is continuous, compact- and nonempty interior-valued; (ii) $s \mapsto \Upsilon(s)$ is non-increasing in the inclusion sense; (iii) $s \mapsto \Upsilon(s)$ is non-increasing in the strong set order sense.
\end{assumption}

By inspection of the proofs of Lemmas \ref{lem:VF.prop.text} and \ref{lem:PF.prop.text}, it is easy to conclude that, under these assumptions, the value function and policy correspondence are \emph{non-increasing}. From the discussion above, a comparative statics result still holds under this case, but now $\delta \mapsto \Gamma(s, \delta)$ is monotonically \emph{decreasing} in the sense that $\min \Gamma(s, \delta) \geq \max \Gamma(s, \delta')$ for any $\delta'> \delta$.
In turn, this result implies 
a flipped
version of Lemma \ref{lem:FP.IFT}, with the mapping $q$ now \emph{non-increasing} in $\delta$ when evaluated at a stable steady state and \emph{non-decreasing} in $\delta$  when evaluated at an unstable steady state.

From this observation and since $\pi_{1} < 0$ under Assumption \ref{ass:pi.properties.ALT}, we conjecture the following alternative version of Theorem \ref{thm:FP.characterization} holds: if Assumptions \ref{ass:pi.properties.ALT}, \ref{ass:C.properties.ALT}, and \ref{ass:Pi.roots} hold, then for any interior steady state $s$ we have $s \in \mathcal{R}^{o}[\mathcal{L}]$, with $s$ stable if $\mathcal{L}_{1}(s,\delta) < 0$ and unstable if $\mathcal{L}_{1}(s,\delta) > 0$.

Unfortunately, our analysis does not allow for having $\pi_{1}$ and $\pi_{12}$ of opposite signs.
This is because a situation in which $V$ is increasing but $\Gamma$ is decreasing, or vice-versa, would 
no longer ensure monotonicity in expression \ref{eqn:Gamma.CS-1}, 
implying that we no longer obtain useful comparative statics with respect to the discount rate.

%
%
%
%

We conclude this discussion by pointing out that in situations where $s \mapsto \Gamma(s,\delta)$ is non-increasing (the case under $\pi_{12}<0$) fixed points may not be the only steady states as there could be cycles or other chaotic behavior. Such possibilities have been pointed out even in simple growth models (e.g. Chapters 4, 7 and 8 in \cite{stachurski2012nonlinear}), but to our knowledge there is no general theory comparable to the one for fixed points. 

 \paragraph{State Space.} The state space, $\mathbb{S}$, is assumed to be bounded, convex, and uni-dimensional. We now discuss the role of each of these assumptions and whether it is possible to
  relax them.
 
The convexity assumption is technical and for convenience as it allow us to perform differentiation of the various functions.  The bounded assumption can be relaxed, provided that one restricts the behavior of the tails of $\pi$, as we show in the Online Appendix \ref{app:state.unbounded}. As discussed there, an unbouded state space implies that 
some optimal paths can drift to plus or minus infinity, however our results are not affected.  Finally, the assumption of uni-dimensionality is central to the tractability of our results, as it renders the state space totally ordered, and also allows us to use simple scalar derivatives to identify steady states and characterize their local stability. In higher dimensions, extending Theorem \ref{thm:FP.characterization} (and its refinements) is in principle possible under appropriate regularity and monotonicity conditions on the gradient and cross-derivative matrix of the payoff function $\pi$. 
We believe that extending our framework to the multidimensional case is a fruitful open question for future research as it will greatly extend the scope of the current theory. However, such an extension involves substantial technical complications and is beyond the scope of this paper. Moreover, in the multidimensional setting, it is no longer immediate that the asymptotic behavior of optimal paths is fully described by fixed points (Proposition \ref{pro:equilibria.chactarization}); the possibility of cycles or more complex invariant sets may not be ruled out \textit{a priori}.

  \section{Conclusion}
\label{sec:conclusion}

We have introduced a monotonicity-based approach for characterizing optimal paths in dynamic optimization problems. By centering the analysis on a simple locator function constructed from model primitives, we can identify steady states, assess their stability, and in some cases describe their basins of attraction, all without solving the full dynamic program. Applications to a generalized neoclassical growth model, a rational (un)fitness model, and an economy with learning-by-doing illustrate both the tractability and the breadth of the framework, and point to its usefulness for applied work that seeks qualitative insights rather than exact solutions.

The main limitation of the present analysis is its restriction to a one-dimensional state space. Extending the framework to multiple states—where cycles, richer path dependence, and interactions across dimensions may arise—remains an important avenue for future research. Developing higher-dimensional analogues of the locator function would significantly expand the scope of monotonicity-based methods for analyzing optimal paths.

\bibliographystyle{apalike}
\bibliography{references}
	
	\pagebreak
	
	\appendix
	\setcounter{page}{1}

	\section{Appendix: Proofs and Results in the Text}

\subsection{Appendix for Section \ref{sec:results.VF+PF}} 
\label{app:results.VF+PF}

\begin{proof}[Proof of Lemma \ref{lem:VF.prop.text}]
	$(\delta,s) \mapsto V(s,\delta)$ is unique, continuous and increasing by Lemmas \ref{lem:VF.contraction} and \ref{lem:VF.incr} in Appendix \ref{app:VF.properties}.
\end{proof}


\begin{proof}[Proof of Lemma \ref{lem:PF.prop.text}]
	$(s,\delta) \mapsto \Gamma(s, \delta)$ is non-empty, compact-valued, and upper hemicontinuous, also \(s \mapsto \Gamma(s, \delta)\) is function-like and non-decreasing by Lemmas \ref{lem:PF.cont} and \ref{lem:PF.incr} in Appendix \ref{app:PF.properties}. 
	
	We now show that $\delta \mapsto \Gamma(s,\delta)$ is non-decreasing in the sense specified in the lemma. This follows by Lemma \ref{lem:PF.delta.incr} in Appendix \ref{app:PF.properties} --- the condition in the Lemma is satisfied by Lemma \ref{lem:VF.prop.text}.
	
\end{proof}



\begin{proof}[Proof of Proposition \ref{pro:VF.diff}]
	We first claim that there exists a $\Delta > 0$ small enough such that there exists a $y_{\Delta} \in \Gamma(s+\Delta, \delta)$ such that $y_{\Delta} \in \Upsilon(s)$. To show this, suppose all $s' \in \Gamma(s, \delta)$ are such that $s' < \max \Upsilon(s)$. Then let $V' : = ( \min \Upsilon(s) - \gamma,\max \Upsilon(s)  ) $ for some $\gamma>0$. Since $\Upsilon(s)$ is convex-valued, $\Gamma(s,\delta) \subset  V' $, so by UHC of $\Gamma$ (Lemma \ref{lem:PF.prop.text}) there exists a $\Delta>0$ such that $\Gamma(s,\delta) \subset  V'$ in $s \in (s-2 \Delta, s + 2 \Delta)$. In particular, $\Gamma(s + \Delta, \delta) \subset (\min \Upsilon(s) - \gamma,\max \Upsilon(s) )$. Moreover, by Lemma \ref{lem:PF.prop.text}, $ \min \Gamma(s + \Delta, \delta) \geq s' \geq \min \Upsilon(s)$, consequently $\Gamma(s + \Delta, \delta) \subset [\min \Upsilon(s),\max \Upsilon(s) ) \subseteq \Upsilon(s)$. So the claim follows assuming all $s' \in \Gamma(s, \delta)$ are such that $s' < \max \Upsilon(s)$, but this follows from the fact that $\Gamma(s, \delta) \subseteq \Upsilon^{o}(s)$.

	From this claim it follows that 
	\begin{align*}
		\frac{ V(s + \Delta ,\delta) - V(s,\delta) }{\Delta} \leq  \frac{ \pi(s+\Delta, y_{\Delta})  -  \pi(s, y_{\Delta}) }{\Delta}.
	\end{align*}
	Since $s \mapsto \Gamma(s,\delta)$ is UHC (Lemma \ref{lem:PF.prop.text}), by possibly going to a subsequence, it follows that $y = \lim_{\Delta \rightarrow 0} y_{\Delta} \in \Gamma(s, \delta)$. By Assumption \ref{ass:pi.properties},  $\lim_{\Delta \rightarrow 0} \frac{ \pi(s+\Delta, y_{\Delta})  -  \pi(s, y_{\Delta}) }{\Delta} = \partial^{+}_{1} \pi(s,y) \leq \max_{y \in \Gamma(s, \delta)}  \partial^{+}_{1} \pi(s,y) $.  Thus,
	\begin{align*}
		\limsup_{\Delta \downarrow 0}  \frac{ V(s + \Delta ,\delta ) - V(s,\delta) }{\Delta} \leq  \max_{y \in \Gamma(s, \delta)}  \partial^{+}_{1} \pi(s,y).
	\end{align*}

	We now claim that there exists a $\Delta > 0$ small enough such that $\Gamma(s, \delta) \subseteq \Upsilon(s + \Delta)$. This follows by the fact that  $\Gamma(s, \delta) \subseteq \Upsilon^{o}(s)$ and  continuity of $\Upsilon$ (Assumption \ref{ass:C.properties}). Hence, for all $y \in \Gamma(s,\delta)$,
	\begin{align*}
		\frac{ V(s + \Delta ,\delta ) - V(s,\delta) }{\Delta} \geq  \frac{ \pi(s+\Delta, y)  -  \pi(s, y) }{\Delta}.
	\end{align*}	
	Thus, 
	\begin{align*}
		\liminf_{\Delta \downarrow 0}  \frac{ V(s + \Delta  ,\delta) - V(s,\delta) }{\Delta} \geq  \partial^{+}_{1} \pi(s,y).
	\end{align*}
	Since this holds for all $y \in \Gamma(s, \delta)$, it holds for the maximal element, and thus we showed that
	\begin{align*}
		\partial^{+} V(s,\delta)  = \lim_{\Delta \downarrow 0}  \frac{ V(s + \Delta ,\delta ) - V(s,\delta) }{\Delta} = \max_{y \in \Gamma(s, \delta)}  \partial^{+}_{1} \pi(s,y).
	\end{align*}
	
	We now analyze $\partial^{-} V(s,\delta)$, We first claim that there exists a $\Delta < 0$ small enough such that there exists a $y_{\Delta} \in \Gamma(s+\Delta, \delta)$ such that $y_{\Delta} \in \Upsilon(s)$. To show this, suppose all $s' \in \Gamma(s, \delta)$ are such that $s' > \min \Upsilon(s)$. Then let $V' : = ( \min \Upsilon(s) ,\max \Upsilon(s) + \gamma ) $ for some $\gamma>0$. Since $\Upsilon(s)$ is convex-valued, $\Gamma(s,\delta) \subset  V' $, so by UHC of $\Gamma$ (Lemma \ref{lem:PF.prop.text}) there exists a $\Delta<0$ such that $\Gamma(s,\delta) \subset  V'$ in $s \in (s+2 \Delta, s - 2 \Delta)$. In particular, $\Gamma(s + \Delta, \delta) \subset (\min \Upsilon(s) ,\max \Upsilon(s) + \gamma )$. Moreover, by Lemma \ref{lem:PF.prop.text}, $ \max \Gamma(s + \Delta, \delta) \leq s' \leq \min \Upsilon(s)$, consequently $\Gamma(s + \Delta, \delta) \subset (\min \Upsilon(s),\max \Upsilon(s) ] \subseteq \Upsilon(s)$.  So the claim follows assuming all $s' \in \Gamma(s, \delta)$ are such that $s' > \min \Upsilon(s)$, but this follows from the fact that $\Gamma(s, \delta) \subseteq \Upsilon^{o}(s)$.

	Hence, for any $\Delta<0$ and $y_{\Delta} \in \Gamma(s+\Delta, \delta) \cap \Upsilon(s)$,
	\begin{align*}
		\frac{ V(s + \Delta  ,\delta) - V(s,\delta) }{\Delta} \geq  \frac{ \pi(s+\Delta, y_{\Delta})  -  \pi(s, y_{\Delta}) }{\Delta}.
	\end{align*}
	And by taking limits it follows that for some $y \in \Gamma(s, \delta)$, $\liminf_{\Delta \uparrow 0}  \frac{ V(s + \Delta, \delta ) - V(s,\delta) }{\Delta} \geq   \partial^{-}_{1} \pi(s,y) \geq \min_{y \in \Gamma(s, \delta)}  \partial^{-}_{1} \pi(s,y)$. Analogous calculations to those above imply $\limsup_{\Delta \uparrow 0}  \frac{ V(s + \Delta, \delta ) - V(s,\delta) }{\Delta} \leq  \partial^{-}_{1} \pi(s,y)$ for all $y \in \Gamma(s, \delta)$. And thus, 
	\begin{align*}
		\partial^{-} V(s,\delta)  = \lim_{\Delta \uparrow 0}  \frac{ V(s + \Delta  ,\delta) - V(s,\delta) }{\Delta} = \min_{y \in \Gamma(s, \delta) }  \partial^{-}_{1} \pi(s,y).
	\end{align*}
\end{proof}

\paragraph{Proof of Corollary \ref{cor:VF.diff.eqn}.} Lemma \ref{lem:VF.diff.cont} in Appendix \ref{app:VF.properties}. $\square$

\subsection{Appendix for Section \ref{sec:planner}} 
\label{app:app_planner}

%

Here we prove a more general version of Proposition \ref{pro:equilibria.chactarization} under unbounded state space. It is clear this proposition implies Proposition \ref{pro:equilibria.chactarization} under the assumption of bounded state space.

  \begin{proposition}\label{pro:equilibria.chactarization.app}
	For any $s_0 \in \mathbb{S}$ and any $\phi(.,s_0) \in \Phi(s_0)$, $\lim_{t \rightarrow \infty} \phi(t,s_{0}) \in  \mathcal{R}[\bar{\Gamma}] \cup \{\pm \infty \}$.  
\end{proposition}

\begin{proof}[Proof of Proposition \ref{pro:equilibria.chactarization.app}]
	For any $s_{0}$ either $\phi_{\delta}(1,s_{0}) > s_{0}$, $\phi_{\delta}(1,s_{0}) < s_{0}$ or $\phi_{\delta}(1,s_{0}) = s_{0}$ . In the third case the result trivially holds so we focus on the other two. Suppose $\phi_{\delta}(1,s_{0}) < s_{0}$. Since $\Gamma(\cdot, \delta)$ is non-decreasing (see Lemma \ref{lem:PF.prop.text}) it follows that $\Gamma(\phi_{\delta}(1,s_{0}), \delta) \leq \Gamma(s_{0}, \delta)$ which implies that $\phi_{\delta}(2,s_{0}) \leq \phi_{\delta}(1,s_{0}) < s_{0}$. By applying this logic recursively we obtain that $(\phi_{\delta}(t,s_{0}))_{t}$ is a non-increasing sequence. Hence, the limit exists and is given by $r = \inf_{t} \phi_{\delta}(t,s_{0})$. Observe that $r$ is either finite or infinite, if it is finite, since $\Gamma(\cdot, \delta)$ is UHC (see Lemma \ref{lem:PF.prop.text}), $r \in \Gamma(r, \delta)$. By Lemma \ref{lem:PF.FP.single} in the Online Appendix \ref{app:PF.properties} this implies that $r \in  \mathcal{R}[\bar{\Gamma}]$ as desired. 
	
	Suppose $\phi_{\delta}(1,s_{0}) > s_{0}$. By analogous arguments to those above, $(\phi_{\delta}(t,s_{0}))_{t}$ is a non-decreasing sequence, hence the limit exists and is given by $s : = \sup_{t} \phi_{\delta}(t,s_{0})$.  Observe that $s$ is either finite or infinite, if it is finite, since $\Gamma(\cdot, \delta)$ is UHC (see Lemma \ref{lem:PF.prop.text}), $s \in \Gamma(s, \delta)$. By Lemma \ref{lem:PF.FP.single} this implies that $s \in  \mathcal{R}[\bar{\Gamma}]$ as desired. 
\end{proof}

  \begin{remark}
	The result allows for $\mp\infty$ to be a limit point. This is natural as in this proposition  $\mathbb{S}$ can be unbounded, so some paths of $\Gamma$ may drift to infinity. A sufficient condition to rule out these case is $\limsup_{s \rightarrow \infty}  \max \Upsilon(s)/s < 1$ and $\liminf_{s \rightarrow -\infty}  \min \Upsilon(s)/s > 1$; i.e., it is not feasible to maintain arbitrary high  or low levels of $s$. In this paper, however, we don't impose this condition and allow limits of paths to take the value $+\infty$. $\triangle$
\end{remark}

\subsection{Appendix for Section \ref{sec:refinements}}
\label{app:refinements} 

The proof of the propositions in Section \ref{sec:refinements} rely on the following lemma which provides information for the behavior of the optimal policy correspondence at the boundary.

\begin{lemma}\label{lem:locator.boundary}
	Suppose $\mathbb{S} = [\underbar{s},\bar{s}]$ and $s \in \Upsilon^{o}(s)$ for all $s \in \mathbb{S}^{o}$, and suppose $\pi_{1}$ and $\pi_{2}$ are continuous. The following are true: 
	\begin{enumerate}
		\item 	If $\mathcal{L}(\underline{s},\delta) > 0$ then either $\Gamma(\underbar{s},\delta) > \underbar{s}$ or  there exists a $\varepsilon>0$ such that  $\Gamma(s, \delta) > s$ for all $s \in (\underbar{s},\underbar{s}+\varepsilon)$. 
		\item 	If $\mathcal{L}(\bar{s},\delta) < 0$ then either $\Gamma(\bar{s},\delta) < \bar{s}$ or  there exists a $\varepsilon>0$ such that  $\Gamma(s, \delta) < s$ for all $s \in (\bar{s} - \varepsilon,\bar{s})$. 
		
		\item 	If $\mathcal{L}(\underline{s},\delta) < 0$ and $\Gamma(\underline{s},\delta) = \underbar{s}$, then  there exists a $\varepsilon>0$ such that  $\Gamma(s, \delta) < s$ for all $s \in (\underbar{s},\underbar{s}+\varepsilon)$.
		
		\item 	If $\mathcal{L}(\bar{s},\delta) > 0$ and $\Gamma(\bar{s},\delta) = \underbar{s}$, then  there exists a $\varepsilon>0$ such that  $\Gamma(s, \delta) > s$ for all $s \in (\bar{s} - \varepsilon,\bar{s})$.  
	\end{enumerate}

\end{lemma}

\begin{proof}
	See Online Appendix \ref{OA:lemmas.aux}
\end{proof}

\subsubsection{Proof of Proposition \ref{pro:root.unique}}
\label{app:root.unique}

\begin{lemma}\label{lem:root.unique}
	
	Suppose $\mathbb{S} = [\underbar{s},\bar{s}]$ and $s \in \Upsilon^{o}(s)$ for all $s \in \mathbb{S}^{o}$, and suppose $\pi_{1}$ and $\pi_{2}$ are continuous. If $\mathcal{R}[\mathcal{L}] = \{e\}$, then the following are true:
	\begin{enumerate}
		\item If  $\mathcal{L}(\underline{s},\delta) > 0$, then there exists a $c \in \mathbb{S}$ such that $(\underline{s},c) \subseteq \mathcal{B}(e,\delta)$ and $(c,\bar{s}]  \subseteq \mathcal{B}(\bar{s},\delta)$.
		
		\item If $\mathcal{L}(\bar{s},\delta) < 0$, then there exists a $c \in \mathbb{S}$ such that $[\underline{s},c) \subseteq \mathcal{B}(\underline{s},\delta)$ and $(c,\bar{s})  \subseteq \mathcal{B}(e,\delta)$.
	\end{enumerate}
	
\end{lemma}

\begin{proof}
	See Online Appendix \ref{OA:lemmas.aux}.
\end{proof}

\begin{proof}[Proof of Proposition \ref{pro:root.unique}]
	Under our assumptions, the conditions for Lemma \ref{lem:root.unique} hold. Moreover, since the locator function satisfies single crossing from above, both parts of Lemma \ref{lem:root.unique} hold. Observe that $c$ in Part 1, which we denote as $c_{1}$, must equal $\bar{s}$. Otherwise, would contradict the claim about $\mathcal{B}(e,\delta)$ in part 2. Similarly, the $c$ in Part 2, which we denote as $c_{2}$, equals $\underline{s}$; otherwise it would contradict part 1.

	This shows that the interior unique root of $s \mapsto \mathcal{L}(s,\delta)$ is globally stable over $\mathbb{S}^{o}$. We now show that such root must be a steady state. To show this, observe that by the proof of Lemma \ref{lem:root.unique}, $\Gamma(s, \delta) > s$ for some $s$ arbitrarily close to $\underline{s}$  and $\Gamma(s,  \delta) < s$ for some $s$ arbitrarily close to $\overline{s}$. Since $\Gamma(\cdot, \delta)$ is non-decreasing, these inequalities imply that $s \mapsto \Gamma(s, \delta) - s$ must cross zero. By Lemma \ref{lem:PF.FP.single}, it must do so at a fixed point and by Theorem \ref{thm:FP.characterization}, it thus must be a root. Thus the desired result holds. 
\end{proof}

\subsubsection{Proof of Proposition \ref{pro:root.two}}
\label{app:root.two}

\begin{proof}[Proof of Proposition \ref{pro:root.two}]
	Since $\mathcal{L}(\bar{s},\delta)  < 0$, by Lemma \ref{lem:locator.boundary}(2), 
	$\Gamma(s,\delta) < s$ for all $s$ in an open ball to the left of  $\bar{s}$. 
	Therefore, one possibility is $\Gamma(s, \delta) <s$ for all $s \in \mathbb{S}^{o}$. In this case no interior steady states exist and $\underbar{s}$ is the only (globally stable) steady state.
	
	Another possibility, however, is for $\Gamma(\cdot, \delta)$ to intersect the \SI{45}{\degree} line for the first time (coming "from the right") in some interior point, denoted by $e$. By Theorem \ref{thm:FP.characterization} $e$ must be one of the roots. But since $s \mapsto \bar{\Gamma}(s, \delta)$ is decreasing at such point, $e$ is a stable steady state (see Figure \ref{fig:BoA.Gamma}). Thus, by Theorem \ref{thm:FP.characterization}, $ e=s^{\ast}$. 
	
	From $s^{\ast}$ and moving "to the left'', we claim $\Gamma(\cdot, \delta)$ can intersect the \SI{45}{\degree} at most one time over $( \underbar{s} ,s^{\ast})$. This claim follows directly from Theorem \ref{thm:FP.characterization} and the assumption that the locator function has only two roots. Also by the same theorem, we can conclude that the only possible fixed point over  $( \underbar{s} ,s^{\ast})$ would be located at $s_{\ast}$ and  would be unstable. The other two possibilities are for $\Gamma(\cdot,\delta)$ to "jump" below the \SI{45}{\degree} line and remain below the \SI{45}{\degree} line until we reach $\underbar{s}$ or to remain above the \SI{45}{\degree} line until we reach $\underbar{s}$.
	
	Hence, we have shown that if the locator function has two roots and $\mathcal{L}(\underbar{s},\delta), \mathcal{L}(\bar{s},\delta)  < 0$, then either $\underbar{s}$ is a globally stable steady state, or $s^{\ast}$ is the only interior (locally) stable steady state.
	
	Now suppose  there exists $s_0 \in \mathbb{S}^{o}$ such that  $\pi_{2}(s_0,s_0) = 0$. Since $s' \mapsto \pi(s_0,s')$ is strictly concave by assumption, this implies that $\Gamma(s_0,0)= s_0$. By Lemma \ref{lem:PF.prop.text}, $\Gamma(s_0,\delta) \geq \Gamma(s_0,0) = s_0$, and hence $\Gamma(s, \delta) <s$ for all $s \in \mathbb{S}^{o}$ cannot hold. So case one in the lemma is ruled out. 	
\end{proof}

\subsubsection{Proof of Proposition \ref{pro:SCVA.FP.characterization}}
\label{app:SCVA.FP.characterization}

To establish Proposition  \ref{pro:SCVA.FP.characterization} we employ the following lemmas.

\begin{lemma}\label{lem:piSCVA.SSiffRoot}
	Suppose $\pi$ is strictly concave,  $s \in \Upsilon(s)$ for all $s \in \mathbb{S}^{o}$, and for all $s \in \mathcal{R}^{o}[\mathcal{L}]$,
	\begin{align}\label{eqn:SCVA.cond-0}
		\max_{a \in \Upsilon(s) \colon a > s} sign \left\{  \pi_{2}(s,a) + \delta \pi_{1}( s ,a  )  \right\}  \leq	sign \left\{ \mathcal{L}(s,\delta) \right\} \leq \min_{a \in \Upsilon(s) \colon a < s} sign \left\{  \pi_{2}(s,a) + \delta \pi_{1}( s ,a  )  \right\} 
	\end{align}
	Then $s$ is an interior steady state if and only if is an interior root of the locator function.
\end{lemma}

\begin{proof}
	See Online Appendix \ref{OA:lemmas.aux}.
\end{proof}

\begin{lemma}\label{lem:Basin.characterization}
	Suppose $s \mapsto \Gamma(s, \delta)$ is a function. Then, for any interior steady state $e$, 
	\begin{align*}
		\mathcal{B}(e,\delta) \supseteq 	\mathcal{B}[\mathcal{L}](e).
	\end{align*}
\end{lemma}

\begin{proof}
	See Online Appendix \ref{OA:lemmas.aux}.
\end{proof}

\begin{proof}[Proof of Proposition \ref{pro:SCVA.FP.characterization}]
	Since $\pi$ is strictly concave, so is $V(\cdot, \delta)$ and thus $\Gamma(\cdot, \delta)$ is a function. Thus, by Lemma \ref{lem:Basin.characterization}, $\mathcal{B}[\mathcal{L}](s) \subseteq \mathcal{B}(s,\delta)$ for any interior steady state $s$. 
	
	We now show $\mathcal{B}[\mathcal{L}](s) \supseteq \mathcal{B}(s,\delta)$ for any interior steady state $s$. If $s$ is unstable this statement is trivially true, so we focus on $s$ stable.
	
	Consider any $a \in \mathcal{B}(s,\delta)$ such that $a < s$ --- the proof for the case $a>s$ is analogous and thus omitted. Suppose $ a \notin \mathcal{B}[\mathcal{L}(\cdot,\delta)](s)$. The only way $ a \notin \mathcal{B}[\mathcal{L}(\cdot,\delta)](s)$ is that there  exists a $b \in (a,s)$ that is a root of $\mathcal{L}(\cdot,\delta)$, but by Lemma \ref{lem:piSCVA.SSiffRoot} $b$ is a steady state, and Proposition \ref{pro:equilibria.chactarization}, $b$ is a fixed point of $\Gamma(\cdot, \delta)$, which implies that $b \notin \mathcal{B}(s,\delta)$. This yields a contradiction because  $ \mathcal{B}(s,\delta) = \mathcal{B}[\bar{\Gamma}(\cdot, \delta)](s)$ is convex.
\end{proof}

%
%
%

\subsection{Appendix for Proof of Theorem \ref{thm:FP.characterization}}
\label{app:app_proofs.main}

\begin{proof}[Proof of Lemma \ref{lem:eqn.roots}]
	For any $s \in  \mathcal{R}^{o}[\bar{\Gamma}(\cdot, \delta)]$ it clearly follows that $s \in Range(\Gamma(\cdot, \delta))$. Thus, optimality and Proposition \ref{pro:equilibria.chactarization} imply 	\begin{align*}
		0 = 	\pi_{2}(s,s) + \delta \pi_{1}(s,y) 
	\end{align*}
	for any $y \in \Gamma(s, \delta)$. But since $s$ is a fixed point, $y=s$. Hence, $\mathcal{L}(s,\delta) = 0$ and thereby the first part of Theorem \ref{thm:FP.characterization} is proven.
\end{proof}	

\begin{proof}[Proof of Lemma \ref{lem:FP.IFT}]

We rely on Lemma \ref{lem:PF.prop.text} that implies that for any $s \in \mathbb{S}$ for which $\Gamma(s, \delta) \in  \Upsilon^{o}(s)$, and for any $\delta ' >\delta$,\footnote{For a set $S$, $\min S : = \min\{ s \colon s \in S\}$ and analogously with $\max S$.}
		\begin{align}\label{eqn:PF.delta.incr.text}
			\min \Gamma(s, \delta') \geq \max \Gamma(s, \delta).
		\end{align}
	
	We first show the case of stable steady state. We do this by contradiction, i.e., suppose there exists $(s, \delta) \in Graph \mathcal{E}^{o}$ with $s$ stable such that for any open neighborhood of $\delta$ either (a) there exists a $\delta' >\delta$ in this neighborhood such that  $q(\delta') < q(\delta) = s$ or (b) there exists a $\delta' < \delta $ in this neighborhood such that  $q(\delta') > q(\delta) = s$.  
	
	Suppose (a) holds (the proof for (b) is analogous and thus omitted). Since $s$ is a stable steady state, by Proposition \ref{pro:Basins.chactarization} in the Online Appendix \ref{app:PF.properties}, there exists a  $\underline{s} \in \mathbb{S}$ such that for any $s \in U(s): = (\underline{s},s)$, $\Gamma(s, \delta) > s$. By expression \ref{eqn:PF.delta.incr.text} this implies that 
	\begin{align*}
		\Gamma(s, \delta') \geq \Gamma(s, \delta) > s 
	\end{align*} 
	for any $\delta' >\delta$. 
	
	By our assumption and continuity of $q$, there exists a $\delta'>\delta$ such that $q(\delta') \in U(s)$. So, the previous display implies that $\Gamma(q(\delta'),\delta') > q(\delta')$, but this is a contradiction to the fact that $q(\delta') \in \mathcal{R}^{o}[\bar{\Gamma}(\cdot, \delta')]$. Hence, $q(\delta)$ is increasing in $\delta$ with $q(\delta)$ stable. 
	
	The proof for the case of unstable steady state is analogous. I.e., suppose there exists a  $(u,\delta) \in Graph \mathcal{E}^{o}$ with $u$ unstable, such that for any open neighborhood of $\delta$ either (a) there exists a $\delta' >\delta$ in this neighborhood such that  $q(\delta') > q(\delta) = u$ or (b) there exists a $\delta' < \delta $ in this neighborhood such that  $q(\delta') < q(\delta) = u$.  Proposition \ref{pro:Basins.chactarization} implies there existence of  $\underline{u}, \overline{u} \in \mathbb{S}$ such that for any $s \in U_{-}(u): = (\underline{u},u)$, $\Gamma(s, \delta) < s$ and for any $s \in U_{+}(u): = (u,\overline{u})$, $\Gamma(s, \delta) > s$. This result and expression \ref{eqn:PF.delta.incr.text} imply that for any $\delta'>\delta$, $\Gamma(s, \delta') > s$ for any $s \in U_{+}(u)$;  and for any $\delta'<\delta$, $\Gamma(s, \delta') < s$ for any $s \in U_{-}(u)$. But these expressions contradict (a) and (b) respectively. 
\end{proof}

\begin{remark}\label{rem:q.existence}
	Existence of the neighborhood $U_{q}(\delta)$ and the mapping $q$ is ensured by our assumption of genericity. For non-generic fixed points, $q(\delta)$ exist, but there could exist a sequence $(\delta'_{n})_{n}$ converging to $\delta$ for which either $q(\delta'_{n})$ does not exist or $q(\delta'_{n})$ is multi-valued (not a function). The first case could occur because $\bar{\Gamma}(\cdot,\delta'_{n})$ does not intersect zero, the second case could occur because $\bar{\Gamma}(\cdot,\delta)$ does intersect zero but it does so tangentially. $\triangle$
\end{remark}

%


\subsection{Appendix for Section \ref{exa:fit}}\label{app:fit}

\begin{proof}[Proof of Lemma \ref{lem:fit.sufficient}]
Assumption \ref{ass:pi.properties}(ii) holds as $R$ and $C$ are smooth. Assumption \ref{ass:pi.properties}(iii) also holds as $\pi_{12}(s,s') =   b \beta  s^{\beta-1}  + (1-d) C''(s'-(1-d)s) >0  $. 

Regarding  Assumption \ref{ass:pi.properties}(i) observe that 
\begin{align*}
 	\pi_{1}(s,s')\geq \alpha s^{\alpha-1}-b(1-d) s^{\beta}.
\end{align*}
 	If the state space is $[0,k/d]$ for some $k \geq 1$, choosing $1+\beta-\alpha>0$ and $b(1-d) \leq \alpha (d/k)^{1+\beta-\alpha}$ ensures $\pi_{1}(s,s')>0$ for all $s,s'\in \mathbb{S}$, satisfying Assumption \ref{ass:pi.properties}(i).  
\end{proof}

\subsection{Appendix for Section \ref{exa:intertemporal}}\label{app:intertemporal}

\begin{proof}[Proof of Lemma \ref{lem:ie.pi12}]
We show Assumption \ref{ass:pi.properties} holds by showing $\pi_1(s,s')>1$ and $\pi_{12}(s,s')>0$. 
By the Envelope Theorem, we have $\pi_{1}(s,s')=u'(c(s,s'))H'(s)F(s')$, which is always positive. 

Also, 
\begin{align*}
	\pi_{12}(s,s') = & u''(c(s,s')) \frac{\partial c(s,s')}{\partial s'} H'(s) F(s) + u'(c(s,s')) H'(s)F'(s')   \\
	 = & u'(c(s,s') ) H'( s ) \left( F'(s') - \frac{ 1 }{c(s,s') \gamma(c)} \frac{\partial c(s,s') }{\partial s'}F(s') \right). 
\end{align*}
The first order condition determining $c$ given
$s$ and $ s'$ can be written as
\begin{align*}
	u'(c(s,s'))=\frac{\varepsilon -1}{\varepsilon  }p( e(s,s') ).
\end{align*}
 In words, this condition states that the marginal
utility of consuming good $1$ must equal the marginal revenue of
exporting this good, which is just the price divided by the markup
$\varepsilon/\left(\varepsilon-1\right)$. Fix $s$ and differentiate
this condition, we have $u''(c(s,s'))  d c(s,s')  = \frac{\varepsilon - 1}{\varepsilon} p'(H(s)F(s') - c(s,s')) (H(s)F'(s') d s' - dc(s,s'))$, which implies
\begin{align*}
	\frac{\partial c(s,s')}{\partial s'} =& \frac{ \frac{\varepsilon-1}{\varepsilon} p'(e(s,s'))H(s)F'(s') }{\frac{\varepsilon-1}{\varepsilon}p'(e(s,s'))+u''(c(s,s'))} \\
	=& \frac{ \frac{\varepsilon-1}{\varepsilon} p'(e(s,s'))H(s)F'(s') }{\frac{\varepsilon-1}{\varepsilon}p'(e(s,s'))- \frac{(\varepsilon - 1)p(e(s,s'))}{\varepsilon \gamma(c(s,s') ) c(s,s') } } >0.
\end{align*}
Plugging this back into our last expression of $\pi_{12}(s,s')$ yields,
after some algebra,
\begin{align*}
	\pi_{12}(s,s')= u'(c(s,s'))F'(s')H'(s) \frac{\left(\gamma(c(s,s'))-1\right)c(s,s')+\left(\varepsilon-1\right)e(s,s')}{c(s,s')+\left(\varepsilon/\gamma(c(s,s') )\right)e(s,s')}.
\end{align*}
Since $\varepsilon>1$ and $\gamma(\cdot)>1$ then $\left(\gamma(c(s,s') )-1\right)c(s,s') +\left(\varepsilon-1\right)e(s,s') >0$,
implying that $\pi_{12}(s,s')>0$ and verifying Assumption \ref{ass:pi.properties}.
\end{proof}

\begin{proof}[Proof of Lemma \ref{lem:ie.L}]
Because of  $\gamma(c) = \varepsilon $, which is denoted by $\gamma$ throughout this proof, the optimal consumption is a constant share of production, $c_1 = \mu sF(s')$, where $\mu :=  \frac{1}{1+ \left( b \frac{\gamma-1}{\gamma}  \right)^\gamma }$. Firstly, the locator function is 
\begin{align*}
	\mathcal{L}(s, \delta) = b\left(1-1/\gamma \right) \left( \frac{1}{1+\left(b\frac{\gamma-1}{ \gamma } \right)^{-\gamma }} \right) ^{-1/\gamma} (\alpha+\delta \theta)s^{ (\theta+\alpha) (1-1/\gamma ) -1 } - 1.
\end{align*}
Hence, if $ \theta < \frac{1}{\gamma-1} + 1 - \alpha $, $s\mapsto \mathcal{L}(s, \delta)$ is decreasing. Note that $\lim_{s\downarrow \underline{s}} \mathcal{L}(s) = +\infty$ and $\mathcal{L}(\bar{s}) = -1$, so $s\mapsto \mathcal{L}(s, \delta) $ is single-crossing from above. By Proposition \ref{pro:root.unique}, there is a unique interior steady state which is globally stable. On the other hand, if $ \theta > \frac{1}{\gamma-1} + 1 - \alpha$ and $\mathcal{L}(\bar{s})>0$, then $\mathcal{L}$ has unique root with positive slope. Therefore, by Theorem \ref{thm:FP.characterization} the interior root cannot be a potential stable steady state. 


\end{proof}

\pagebreak 

	\setcounter{page}{1}

\begin{center}
	{\Huge{Online Appendix}}
\end{center}

	\setcounter{section}{0}
\renewcommand{\thesection}{OA.\arabic{section}}

Throughout this Online Appendix it is convenient to make the dependence on the discount factor, $\delta$, explicit on $X[\bar{\Gamma}]$ and $X[\mathcal{L}]$ where $X$ can be wither $\mathcal{R}$ or $\mathcal{B}$. For this, we will use the notation $X[\bar{\Gamma}(\cdot,\delta)]$ and $X[\mathcal{L}(\cdot,\delta)]$.

\section{Unbounded State Space}
\label{app:state.unbounded}

It is possible to results in the text to the unbounded state space by imposing following technical assumption that is used to establish the existence of the value function. 
\setcounter{assumption}{0}
\renewcommand{\theassumption}{\thesection.\arabic{assumption}}

\begin{assumption}\label{ass:growth}
	There exists a continuous function $s \mapsto \varphi(s,\delta)$ such that $\varphi \geq c > 0$ and 
	\begin{align*}
		\sup_{s \geq 0 }	\sup_{s' \in \Upsilon(s)}  \delta \frac{\varphi(s',\delta)}{\varphi(s,\delta)} < 1~and~\sup_{s \geq 0} \sup_{s' \in \Upsilon(s)} \frac{\pi(s,s')} { \varphi(s,\delta)} < \infty.
	\end{align*}	
\end{assumption}

If $\mathbb{S}$ is bounded, this assumption is vacuous ($\varphi$ can be taken to be a constant). For unbounded state space, however, $\varphi$ controls the growth of the per-period payoff and defines the relevant function space to which the value function belongs to: All $s \mapsto f(s)$ continuous  such that $||f||_{\delta} : = \sup_{s \in \mathbb{S}} | f(s) / \varphi(s,\delta)  | < \infty$.\footnote{The concept underpinning Assumption \ref{ass:growth} is by no means new, e.g. see \cite{alvarez1998dynamic} for analogous conditions.}

For example, in the \ref{exa:NCG}, if one defines $\mathbb{S} = \mathbb{R}_{+}$, then Assumption \ref{ass:growth} is satisfied with  $s \mapsto \varphi(s,\delta) = u(f(s)) + A$, where $A$ is such that  $1 + \sup_{Y \in [0,1] } \frac{ u'(Y)  }{u(Y) + A } ( f(Y) - Y  )$ is strictly less than $1/\delta$. 
 
Under Assumption \ref{ass:growth} the results in the text go through essentially without change. The only change lies in Proposition \ref{pro:equilibria.chactarization} which continues to hold, but with the only caveat that $\mp \infty$ could be also steady states (see Proposition \ref{pro:equilibria.chactarization.app} in Appendix \ref{app:app_planner}). Theorem \ref{thm:FP.characterization}, however, remains unchanged and so its refinements.

The proofs in the appendix will be done under this, more general, assumption.  
	\section{Properties of the Value Function}
\label{app:VF.properties}

Throughout this section we use
\begin{align}
	V \mapsto	\mathbb{B}[V](s,\delta) : = \max_{s' \in \Upsilon(s)} \pi(s,s') + \delta V(s',\delta)~\forall (s,\delta) \in \mathbb{S} \times [0,1),
\end{align}
which is  the Bellman operator defining the value function $(s,\delta) \mapsto V(s,\delta)$ as its fixed point.

Let $\mathbb{V} : = \{ f \colon \mathbb{S}\times [0,1) \rightarrow \mathbb{R} \colon f~is~continuous~and~f(.,\delta) \in \mathbb{V}_{\delta}~\forall \delta \in [0,1)  \}$, where $\mathbb{V}_{\delta}$ is the space of functions over $\mathbb{S}$ such that $||f||_{\delta} : = \sup_{s \in \mathbb{S}} |f(s)/\varphi(s,\delta)| < \infty$. It is straightforward to show that $(\mathbb{V}_{\delta},||.||_{\delta})$ is a Banach space. 

Throughout, for any correspondence $s \mapsto F(s)$ let $F^{o}$ denote the interior of it, i.e., for each $s$, $F^{o}(s)$ is the interior of the set $F(s)$.

\subsection{Existence and Uniqueness of the Value Function}

\begin{lemma}\label{lem:VF.contraction}
	$(s,\delta) \mapsto V(s,\delta)$ is the unique solution to $V = \mathbb{B}[V]$ in the space $\mathbb{V}$.
\end{lemma}

\begin{proof}[Proof of Lemma \ref{lem:VF.contraction}]
	 For each $\delta \in [0,1)$, we show that the Bellman operator, $\mathbb{B}_{\delta}$ --- acting on functions of $s$, not $\delta$ --- maps $\mathbb{V}_{\delta}$ into itself and it is a contraction. By Assumption \ref{ass:growth}, $(s,s') \mapsto \pi(s,s')/\varphi(s,\delta)$ is bounded over $Graph \Upsilon$. Hence, $\mathbb{B}_{\delta}$ maps bounded function into itself. By continuity of $\pi$ and the fact that $s \mapsto \Upsilon(s)$ is continuous and compact-valued (see Assumption \ref{ass:C.properties}), the ToM implies that $\mathbb{B}_{\delta}$ maps  $\mathbb{V}_{\delta}$ into itself. The contraction property follows from standard Blackwell sufficient conditions. 
	
	Therefore, for each $\delta \in [0,1)$, there exists a unique function on $\mathbb{V}_{\delta}$, denoted as $V(\cdot, \delta)$, that is a fixed point of $\mathbb{B}_{\delta}$. Now defined $(s,\delta) \mapsto V(s,\delta) : = V(s,\delta)$. Clearly, $V \in \mathbb{V}$ and is the unique function such that $V(.,\delta) = \mathbb{B}_{\delta}[V(.,\delta)]$ for all $\delta \in [0,1)$, but this readily implies that $V = \mathbb{B}[V]$. 
\end{proof}

\subsection{Monotonicity of the Value Function}
\label{app:VF.mono}

In this section we present three approaches to establish that $s \mapsto V(s,\delta)$ is increasing, each relying on different assumptions and different proofs technique. One, relies on $s \mapsto \Upsilon(s)$ being increasing in the sense of inclusion and is standard (cf. \cite{stokey1989recursive}). However, this condition on $\Upsilon$ might be too strong in some settings. So, we also present two alternative approaches, which dispenses with the aforementioned assumption. One relies on weaker assumptions --- namely, on being able to `replicate' payoffs generated for different values of the state variables ---, the other one relies on a generalized version of the mean value theorem for a.e. smooth functions. To our knowledge, these last two approaches are novel and might be of independent interest.

Here are the results under the first approach; it is a well-known result and it is here merely for completeness.

\begin{lemma}\label{lem:VF.incr}
	Suppose $s \mapsto \Upsilon(s)$ is non-decreasing in the inclusion sense. Then $s \mapsto V(s,\delta)$ is increasing.
\end{lemma}

\begin{proof}[Proof of Lemma \ref{lem:VF.incr}]
	We first show that $\mathbb{B}_{\delta}$ maps non-decreasing functions into themselves. To do this take any $f \in \mathbb{V}_{\delta}$ that is non-decreasing. Then
	\begin{align*}
		\mathbb{B}_{\delta}[f](s_{0}) \geq \pi(s_{0},s') + \delta f(s')  > \pi(s_{1},s') + \delta f(s') 
	\end{align*}
	for any $s_{0} > s_{1}$ and any $s' \in \Upsilon(s_{0})$; the second inequality follows from the fact that $s \mapsto \pi(s,s')$ is increasing (Assumption \ref{ass:pi.properties}(i)). 
	
	By assumption, $\Upsilon(s_{0}) \supseteq \Upsilon(s_{1})$, so $s'$ can be taken to be in $\Gamma_{\delta}(s_{1})$ and thus obtain $	\mathbb{B}_{\delta}[f](s_{0}) > \mathbb{B}_{\delta}[f](s_{1})$. Thus, $s \mapsto  \mathbb{B}_{\delta}[f](s)$ is increasing, and so $\mathbb{B}_{\delta}$ maps non-decreasing function in increasing (and thus non-decreasing). Since the class of non-decreasing is closed, it follows that $V(\cdot, \delta)$ is non-decreasing. Indeed, since   $V(\cdot, \delta)$ is unique, it actually is increasing. 
\end{proof}

Here is the second approach

\begin{lemma}\label{lem:VF.incr.v2}
	Suppose for any $s_{1} > s_{0}$ there exists a pair $s_1',s_0'$ with $s_1' \in \Upsilon(s_{1})$ and $s_0' \in \Gamma(s_{0},\delta)$ such that (i) $s_1' \geq s_0'$ and  (ii) $\pi(s_{1},s_1') \geq \pi(s_{0},s_0')$ with at least one of the inequalities strict. Then $s \mapsto V(s,\delta)$ is increasing.
\end{lemma}

\begin{proof}[Proof of Lemma \ref{lem:VF.incr.v2}]
	We first show that $\mathbb{B}_{\delta}$ maps non-decreasing functions into themselves. To do this take any $f \in \mathbb{V}_{\delta}$ that is non-decreasing. Then for any $s_{1} > s_{0}$, 
	\begin{align*}
		\mathbb{B}_{\delta}[f](s_{1}) \geq \pi(s_{1},s_1') + \delta f(s_1')  \geq  \pi(s_{0},s_0') + \delta f(s_1') \geq   \pi(s_{0},s_0') + \delta f(s_0') =  	\mathbb{B}_{\delta}[f](s_{0}), 
	\end{align*}
	where the first inequality holds because $s_1' \in \Upsilon(s_{1})$; the second inequality follows from condition (ii); the third inequality follows from condition (i) and the fact that $f$ is non-decreasing; the last equality follows because $s_0' \in \Gamma(s_{0},\delta)$. 
	
	Thus, $s \mapsto  \mathbb{B}_{\delta}[f](s)$ is increasing, and so $\mathbb{B}_{\delta}$ maps non-decreasing function in increasing (and thus non-decreasing). Since the class of non-decreasing is closed, it follows that $s \mapsto V(s,\delta)$ is non-decreasing. Indeed, since one of the equalities is strictly and $s \mapsto V(s,\delta)$ is unique, it actually is increasing. 
\end{proof}

\begin{remark}\label{rmk:VF.incr.v2}
	(1) The condition in the lemma is implied by increasing in inclusion sense. To see this, suppose $\Upsilon(s_{1}) \supseteq \Upsilon(s_{0})$ then take $s_1'=s_0'$ with any $s_0' \in \Gamma(s_{0},\delta)$. Then clearly (i) is met and (ii) follows as $\pi_{1} > 0$. By the inclusion, $s_1' \in \Upsilon(s_{1}) $.
	
	(2) The condition is met in the \ref{exa:NCG}. To see this, let $s_1' = - \bar{f}(s_{0}) + \bar{f}(s_{1})   + s_0'$ where $\bar{f}(s) = f(s) + (1-d) s$. (i) holds as $- \bar{f}(s_{0}) + \bar{f}(s_{1})  > 0$; (ii) holds as $U(  \bar{f}(s_{1})  - s_1' ) = U(  \bar{f}(s_{0})  - s_0' ) $. We need to verify  $s_1' \in \Upsilon(s_{1}) = [(1-d) s_{1} , (1-d) s_{1} + f(s_{1})]$. To check this, note that 
	\begin{align*}
		s_{0}'\in\Upsilon\left(s_{0}\right)\implies s_{0}'\leq\bar{f}\left(s_{0}\right)\iff\bar{f}\left(s_{1}\right)-\left[\bar{f}\left(s_{0}\right)-s_{0}'\right]\leq\bar{f}\left(s_{1}\right)\implies s_{1}'\leq\bar{f}\left(s_{1}\right)\end{align*}
	and
	\begin{align*}
		s_{0}'&\in\Gamma\left(s_{0},\delta\right)\implies s_{0}'\geq\left(1-d\right)s_{0}\implies f(s_{1})-f\left(s_{0}\right)+\left[s_{0}'-\left(1-d\right)s_{0}\right]\geq0\\&\iff\bar{f}\left(s_{1}\right)-\left[\bar{f}\left(s_{0}\right)-s_{0}'\right]\geq\left(1-d\right)s_{1}\iff s_{1}'\geq\left(1-d\right)s_{1}	\end{align*} $\triangle$

	(3) This condition holds in the (un)fitness model as well. To verify this, let $s_1' = - (1-d)s_0 + (1-d)s_1 + s_0'$ --- exercise for the same amount of time.  (i) holds strictly as $s_1 > s_0$. (ii) holds as $R(s_1 , s_1'-(1-d)s_1) - C(s_1' - (1-d) s_1) \geq R(s_0 , s_0'-(1-d)s_0) - C(s_0' - (1-d) s_0)$. We only need to verify $s_1' \in \Upsilon(s_1) = [(1-d)s_1 , (1-d)s_1 +1] $. To check this, recall that $0 \leq s_0' - (1-d)s_0\leq 1$, so
	\begin{align*}
		(1-d)s_1 \leq - (1-d)s_0 + (1-d)s_1 + s_0' \leq  (1-d)s_1 +1.
	\end{align*} $\triangle$
\end{remark}

Here are the results under the third approach. Given the nature of the proof, it requires conditions that ensure optimal choices are interior.

\begin{lemma} \label{lem:VF.MVT}
		%
	
	For any $s_{0} < s_{1}$ in $\mathbb{S}$ such that $\Gamma(s, \delta) \subseteq \Upsilon^{o}(s)$ for all $s \in (s_{0},s_{1})$, there exists a $c \in (s_{0},s_{1})$ such that  
	\begin{align*}
		& \max_{Y \in \Gamma(c, \delta)} \partial^{+}_{1} \pi(c,Y) =   \max\{	\partial^{+} V(c, \delta) ,\partial^{-} V(c, \delta) \} \\
		& \geq	\frac{V(s_1, \delta) - 	V(s_0, \delta)}{s_{1} - s_{0}} \geq \\
		& \min\{	\partial^{+} V(c, \delta) ,\partial^{-} V(c, \delta) \} = \min_{Y \in \Gamma(c, \delta)} \partial^{-}_{1} \pi(c,Y)
	\end{align*}
	
\end{lemma}

\begin{proof}
	Let $\frac{	V(s_1, \delta) - 	V(s_0, \delta) }{s_{1} - s_{0}} = : D$. By construction, $G(s): = V(s,\delta) - D s$ is such that $G(s_{1})=G(s_{0})$. Since $G$ is continuous, the function $G$ achieves a minimum or a maximum in $(s_{0},s_{1})$; denote the point as $c$. 
	
	%
	%

	By assumption,  $\Gamma(c, \delta) \subseteq \Upsilon^{o}(c)$, so by Proposition \ref{pro:VF.diff}, the left and right derivatives of $s \mapsto V(s,\delta)$ (and thus of $G$) exist. If $c$ is a minimizer then $\partial^{-} G(c) \leq 0 \leq \partial^{+} G(c)$, and if $c$ is a maximizer then $\partial^{-} G(c) \geq 0 \geq \partial^{+} G(c)$. These inequalities imply that $\partial^{-} V(c, \delta) \leq D \leq \partial^{+} V(c, \delta)$ and $\partial^{-} V(c, \delta) \geq D \geq \partial^{+} V(c, \delta)$ respectively. Hence,
	\begin{align*}
		\min\{	\partial^{+} V(c, \delta) ,\partial^{-} V(c, \delta) \}  \leq D \leq \max\{	\partial^{+} V(c, \delta) ,\partial^{-} V(c, \delta) \}.
	\end{align*}
	This implies 	
	\begin{align*}
		\min\{	\partial^{+} V(c, \delta) ,\partial^{-} V(c, \delta) \}  \leq  \frac{	V(s_1, \delta) - 	V(s_0, \delta) }{s_{1} - s_{0}}  \leq \max\{	\partial^{+} V(c, \delta) ,\partial^{-} V(c, \delta) \}.
	\end{align*}
	
	By Proposition \ref{pro:VF.diff} and the fact that $ \min_{Y \in \Gamma(c, \delta)} \partial^{-}_{1} \pi(c,Y) \leq  \max_{Y \in \Gamma(c, \delta)} \partial^{+}_{1} \pi(c,Y)$ (Assumption \ref{ass:pi.properties}(i)), the outer inequality in the lemma follows. 
\end{proof}

\begin{corollary} \label{cor:VF.incr.v2}
	For any $s_{0} < s_{1}$ in $\mathbb{S}$ such that $\Gamma(s, \delta) \subseteq \Upsilon^{o}(s)$ for all $s \in (s_{0},s_{1})$, $V(s_1, \delta) > V(s_0, \delta)$. 
\end{corollary}

\begin{proof}
	By Lemma \ref{lem:VF.MVT}, $V(s_1, \delta) - 	V(s_0, \delta)  \geq \min_{Y \in \Gamma(c, \delta)} \partial^{-}_{1} \pi(c,Y) (s_{1} - s_{0})$. By Assumption \ref{ass:pi.properties}(i) $\partial^{-}_{1} \pi(c,Y)  > 0$ so the desired result follows. 
\end{proof}

\subsection{Differentiability of the Value Function}

\begin{lemma}\label{lem:VF.diff.cont}
 For any $s \in  Range (\Gamma_{\delta} \cap \Upsilon^{o})$ the following are true:
	\begin{enumerate}
		\item 	$s \mapsto V(s,\delta)$ is smooth. That is, $\partial^{+} V(s,\delta) = \partial^{-} V(s,\delta) =  \pi_{1} (s,s')$ for any  $s' \in \Gamma(s, \delta)$. 
		\item  $\Gamma(s, \delta)$ is a singleton and $\partial^{+}_{1} \pi(s,\Gamma(s, \delta)) =  \partial^{-}_{1} \pi(s,\Gamma(s, \delta)) = : \pi_{1} (s,\Gamma(s, \delta))$.
	\end{enumerate}
\end{lemma}

\begin{proof}[Proof of Lemma \ref{lem:VF.diff.cont}]
	For any $s \in Range \Gamma(\cdot, \delta)$, it follows that there exists a $s_{-1}$ such that $s \in \Gamma(s_{-1}, \delta)$. Hence, $\pi(s_{-1},s) + \delta V(s,\delta) \geq \pi(s_{-1},s') + \delta V(s',\delta) $ for any $s' \in \Upsilon(s_{-1})$. Since $s \in  \Upsilon^{o}(s_{-1})$, the previous inequality holds for $s'=s + \Delta$  and  $s'=s - \Delta$ for sufficiently small $\Delta>0$. By Proposition \ref{pro:VF.diff}, $\partial_{+} V(s,\delta)$ exists and thus
	\begin{align*}
		0 \geq  \pi_{2}(s_{-1},s) + \delta \partial^{+} V(s,\delta)~and~0 \leq  \pi_{2}(s_{-1},s) + \delta \partial^{-} V(s,\delta).
	\end{align*}
	The two inequalities imply that $\partial^{-} V(s,\delta) \geq  \partial^{+} V(s,\delta)$. 
	
	On the other hand, by Proposition \ref{pro:VF.diff} and Assumption \ref{ass:pi.properties}(i), $\partial^{-} V(s,\delta) = \min_{Y \in \Gamma(s, \delta)} \partial^{-}_{1} \pi(s,Y) \leq  \partial^{-}_{1} \pi(s,Y) \leq  \partial^{+}_{1} \pi(s,Y) \leq \max_{y \in \Gamma(s, \delta)}  \partial^{+}_{1} \pi(s,Y)=\partial^{+} V(s,\delta) $.
	
	Therefore, $\partial^{-} V(s,\delta) =  \partial^{+} V(s,\delta) =  \partial^{+}_{1} \pi(s,Y) =  \partial^{-}_{1} \pi(s,Y)$ for any $Y \in \Gamma(s, \delta)$. 
	
	We conclude the proof by showing that  $\Gamma(s, \delta)$ is a singleton. This readily follows by the previous result which imply that $\min_{Y \in \Gamma(s, \delta)}  \pi_{1}(s,Y) = \max_{Y \in \Gamma(s, \delta)}  \pi_{1}(s,Y) $, which by Assumption \ref{ass:pi.properties}(iii) can only hold if $\Gamma(s, \delta)$ is a singleton. 
\end{proof}

\begin{lemma}\label{lem:VF.delta.diff} 
	$\delta \mapsto V(s,\delta)$ is continuously differentiable and for any $s \in \mathbb{R}_{+}$,
	\begin{align*}
		\frac{d V(s,\delta)}{d\delta} = \sum_{t=0}^{\infty} \delta^{t} V(s^{0}_{t+1}, \delta) = V(s_1, \delta) + \delta    \frac{d V(s_1, \delta)}{d\delta}   .
	\end{align*}
	for some $(s_{t})_{t}$ such that $s_{0} = s$ and $s_{t+1} \in  \Gamma(s_t, \delta)$.
\end{lemma}

\begin{proof} [Proof of Lemma \ref{lem:VF.delta.diff} ] 
	For any $s \geq 0$ and $\delta \in [0,1)$ and $\Delta \in \mathbb{R}$, 
	\begin{align*}
		V(s, \delta + \Delta) - 	V(s, \delta) \geq &  \delta (V(s_{1}, \delta+\Delta)  - V(s_1, \delta)) + \Delta   V(s_{1}, \delta+\Delta) 
	\end{align*}
	for any $s_{1} \in \Gamma(s, \delta)$. 	Applying this again to $s_{1}$, it follows that 
	\begin{align*}
		V(s, \delta + \Delta) - 	V(s, \delta) \geq   \delta^{2} (V(s_{2}, \delta+\Delta)  - V_{\delta}(s_{2})) + \Delta (  V(s_{1}, \delta+\Delta)  - \pi(s,s_{1}) ) + \Delta \delta V(s_{2}, \delta+\Delta)  
	\end{align*}
	for any $s_{2} \in \Gamma(s_{1}, \delta)$.
	
	By iterating in this fashion, one obtains for any $T>1$, 
	\begin{align*}
		V(s, \delta + \Delta) - 	V(s, \delta) \geq   \delta^{T+1} (V(s_{T+1}, \delta + \Delta) - V(s_{T+1}, \delta)) + \Delta \sum_{t=0}^{T} \delta^{t} V(s_{t+1}, \delta + \Delta)
	\end{align*}
	where $s_{0} : = s$ and any $(s_{t})_{t}$ such that $s_{t+1} \in  \Gamma(s_t, \delta)$. 
	
	We now show that $\lim_{T \rightarrow \infty}  \delta^{T+1} (V(s_{T+1}, \delta + \Delta) - V(s_{T+1}, \delta))  = 0$. If $\mathbb{S}$ is bounded this is trivial because $V$ is bounded and $\delta< 1$. In the more general case, under Assumption \ref{ass:growth}, it follows that  there exists a constant $K$ and a $0 \leq a<1$ such that for any $s_{t}$, $V(s_{t},\delta) = \frac{V(s_{t},\delta) }{\varphi(s_{t},\delta)}  \varphi(s_{t},\delta) \leq K \varphi(s_{t},\delta)$ and $\varphi(s_{t},\delta) \leq (a/\delta) \varphi(s_{t-1},\delta)$. Thus, iterating in this fashion it follows that $V(s_{t},\delta)  \leq K (a/\delta)^{t} \varphi(s_{0},\delta) = O((a/\delta)^{t})$. Therefore, 
	\begin{align*}
		\delta^{T+1} (V(s_{T+1}, \delta + \Delta) - V(s_{T+1}, \delta))  = O(a^{T+1}),
	\end{align*}
and since $0 \leq a<1$ the desired result follows.  
	
Therefore,
	\begin{align*}
		V(s, \delta + \Delta) - V(s, \delta) \geq   \Delta \sum_{t=0}^{\infty} \delta^{t} V(s_{t+1}, \delta + \Delta).
	\end{align*}
	Taking $\Delta>0$ and taking limits it follows that 
	\begin{align*}
		\partial_{+} V(s, \delta) : = \limsup_{\Delta \downarrow 0} \frac{V(s, \delta + \Delta) - V(s, \delta) }{\Delta}  \geq \limsup_{\Delta \rightarrow 0}   \sum_{t=0}^{\infty} \delta^{t} V(s_{t+1}, \delta + \Delta).
	\end{align*}
	Also, taking $\Delta < 0$ and taking limits it follows that,
	\begin{align*}
		\partial_{-} 	V(s, \delta) : = \liminf_{\Delta \uparrow 0} \frac{V(s, \delta + \Delta) - 	V(s, \delta) }{\Delta}  \leq \liminf_{\Delta \rightarrow 0}   \sum_{t=0}^{\infty} \delta^{t} V(s_{t+1}, \delta+\Delta).
	\end{align*}

	By Lemma \ref{lem:VF.contraction}, $\delta \mapsto V(s, \delta)$ is continuous. Hence since $\sum_{t=0}^{\infty} \delta^{t} V(s_{t+1},\delta + \Delta ) = O( \sum_{t=0}^{\infty} a^{t} ) < \infty$, by DCT, it follows that $ \lim_{\Delta \rightarrow 0}    \sum_{t=0}^{\infty} \delta^{t} V(s_{t+1}, \delta + \Delta)  =  \sum_{t=0}^{\infty} \delta^{t} V(s_{t+1}, \delta) $ and so
	\begin{align}\label{eqn:VF.bar.bounds-1}
		\partial_{+} 	V (s) \geq  \sum_{t=0}^{\infty} \delta^{t} V(s_{t+1}, \delta)   \geq \partial_{-} V(s, \delta) ,
	\end{align}
	for any $(s_{t})_{t}$ such that $s_{0} = s$ and $s_{t+1} \in  \Gamma(s_t, \delta)$. 
	
	Similarly,
	\begin{align*}
		V(s, \delta + \Delta) - 	V(s, \delta) \leq  \delta (V(s_{1}, \delta+\Delta)  - V(s_1, \delta)) + \Delta V(s_{1}, \delta+\Delta)
	\end{align*}
	for any $s_{1} \in\Gamma(s, \delta + \Delta)$. By iterating in the same way as before, it follows that 	
	\begin{align*}
		V(s, \delta + \Delta) - V(s, \delta) \leq   \Delta \sum_{t=0}^{\infty} \delta^{t} V(s_{t+1}, \delta+\Delta)
	\end{align*}
	for any $(s_{t})_{t}$ such that $s_{0} = s$ and $s_{t+1} \in  \Gamma(s_{t}, \delta + \Delta)$. 
	
	For each $\Delta$ take $s_{1} = : s^{\Delta}_{1}$ such that $s^{\Delta}_{1} \in \Gamma(s, \delta + \Delta)$ and $\lim_{\Delta \rightarrow 0} s^{\Delta}_{1} = s^{0}_{1}$; this last property holds because $s\in [0,1]$ and any sequence has a convergent subsequence. By Lemma \ref{lem:PF.cont}, $\delta \mapsto \Gamma(s, \delta)$ is UHC, so $s^{0}_{1} \in \Gamma(s, \delta)$. Now take $s_{2} : = s^{\Delta}_{2}$ such that $s^{\Delta}_{2} \in \Gamma(s^{\Delta}_{1}, \delta + \Delta)$ and $\lim_{\Delta \rightarrow 0} s^{\Delta}_{2} = s^{0}_{2}$. By Lemma \ref{lem:PF.cont}, $(s,\delta) \mapsto \Gamma(s, \delta) $ is UHC; this fact and the fact that $\lim_{\Delta \rightarrow 0} s^{\Delta}_{1} = s^{0}_{1}$, imply that $s^{0}_{2} \in \Gamma(s^{0}_{1}, \delta)$. By iterating in this fashion we obtain a sequence $(s^{\Delta}_{t})_{t}$ such that pointwise in $t$, $\lim_{\Delta \rightarrow 0} s^{\Delta}_{t} = s^{0}_{t}$ and $s^{0}_{t+1} \in \Gamma(s^{0}_{t}, \delta)$. 
	
	By continuity of $\pi$, $\lim_{\Delta \rightarrow 0} \pi(s^{\Delta}_{t},s^{\Delta}_{t+1}) = \pi(s^{0}_{t},s^{0}_{t+1})$ (pointwise on $t$). By Lemma \ref{lem:VF.contraction}, $(s,\delta) \mapsto V(s,\delta)$ is continuous, so $\lim_{\Delta \rightarrow 0} V(s^{\Delta}_{t+1}, \delta+\Delta) = V(s^{0}_{t+1}, \delta)$ (pointwise on $t$). Therefore,
	\begin{align*}
		\lim_{\Delta \rightarrow 0}  V(s^{\Delta}_{t+1}, \delta+\Delta)   =   V(s^{0}_{t+1}, \delta).
	\end{align*}
	
	Since $\sum_{t=0}^{\infty} \delta^{t} < \infty$ and $V_{\delta}$ is also uniformly bounded, it follows by the DCT that
	\begin{align*}
		\partial_{+} 	V(s, \delta)  : = \limsup_{\Delta \downarrow 0}	\frac{V(s, \delta + \Delta) - V(s, \delta)}{\Delta} \leq & \limsup_{\Delta \downarrow 0} \sum_{t=0}^{\infty} \delta^{t} V(s^{\Delta}_{t+1}, \delta+\Delta) 	\leq  \sum_{t=0}^{\infty} \delta^{t}   V(s^{0}_{t+1}, \delta) .
	\end{align*}
	
	Analogous reasoning to that above yields
	\begin{align*}
		\partial_{-} V(s, \delta)  \geq  \limsup_{\Delta \downarrow 0} \sum_{t=0}^{\infty} \delta^{t} V(s^{\Delta}_{t+1}, \delta+\Delta) 	\geq  \sum_{t=0}^{\infty} \delta^{t} V(s^{0}_{t+1}, \delta).
	\end{align*}
	
	Combining these displays with the expression in \ref{eqn:VF.bar.bounds-1}, it follows that 
	\begin{align}\label{eqn:VF.bar.bounds-2}
		\partial_{+} 	V(s, \delta) =  \sum_{t=0}^{\infty} \delta^{t}  V(s_{t+1}, \delta)  = \partial_{-} V(s, \delta) ,
	\end{align}
	for some $(s_{t})_{t}$ such that $s_{0} = s$ and $s_{t+1} \in  \Gamma(s_t, \delta)$.  
\end{proof}

\section{Properties of the optimal correspondence}
\label{app:PF.properties}

Recall 
\begin{align*}
	(s,\delta) \mapsto \Gamma(s, \delta) : = \arg\max_{s' \in \Upsilon(s)} \pi(s,s') + \delta V(s',\delta).
\end{align*}

\subsection{Topological Properties of the Optimal Correspondence}

\begin{lemma}\label{lem:PF.cont}
	$(s,\delta) \mapsto \Gamma(s, \delta)$ is non-empty-, compact-valued, and UHC.
\end{lemma}

\begin{proof}[Proof of Lemma \ref{lem:PF.cont}]
	By Lemma \ref{lem:VF.contraction} and Assumptions \ref{ass:pi.properties} and \ref{ass:C.properties}, $(s,\delta, s') \mapsto \pi(s,s') + \delta V(s',\delta)$ is continuous and $s \mapsto \Upsilon(s)$ is compact-valued and continuous. Thus, by the ToM, $(s,\delta) \mapsto \Gamma(s, \delta)$ is non-empty-, compact-valued, and UHC.
\end{proof}

The next lemma essentially shows that $\Gamma_{\delta}$ is a function at fixed points. 

\begin{lemma}\label{lem:PF.FP.single}
	If $s \in \Gamma(s, \delta) \cap \Upsilon^{o}(s)$, then $s=\Gamma(s, \delta)$.
\end{lemma}

\begin{proof}[Proof of Lemma \ref{lem:PF.FP.single}]
	Observe that $s$ trivially belongs to $Range \Gamma(\cdot, \delta) \cap \Upsilon^{o}$. Then, by Lemma \ref{lem:VF.diff.cont} $\Gamma(s, \delta)$ is a singleton thereby implying the desired result.  
\end{proof}

\subsection{Monotonicity of the Optimal Correspondence}

Recall that a correspondence is function-like if its graph has an empty interior. 

\begin{lemma}\label{lem:PF.incr}
	$s \mapsto \Gamma(s, \delta)$ is non-decreasing --- in the sense that $\max \Gamma(a, \delta) \leq \min \Gamma(b, \delta)$ for any $a<b$ --- and function-like. 
\end{lemma}

\begin{proof}[Proof of Lemma \ref{lem:PF.incr}]
	Observe that $(s,s') \mapsto F(s,s') : = \pi(s,s') + \delta V(s',\delta)$ satisfies increasing differences since for any $s_{0} > s_{1}$, $D(s') : = F(s_{0},s') - F(s_{1},s') = \pi(s_{0},s') - \pi(s_{1},s') $. Taken derivative with respect to $s'$ it follows that $D'(s') = \pi_{2}(s_{0},s') - \pi_{2}(s_{1},s') $ which is positive by Assumption \ref{ass:pi.properties}(iii). By Assumption \ref{ass:C.properties}, $\Upsilon$ is non-decreasing in the strong set order  sense.  Hence, by the Milgrom-Shannon Theorem (\cite{milgrom1994monotone}), for any $a < b$ and any $y \in \Gamma(a, \delta)$ and $y' \in \Gamma(b, \delta)$ it follows that $\min\{y,y'\} \in \Gamma(a, \delta)$ and $\max\{y,y'\} \in \Gamma(b, \delta)$. 
	
	We now show that $\max \Gamma(a, \delta) \leq \min \Gamma(b, \delta)$. Suppose not, suppose there exists a $y \in \Gamma(a, \delta)$ and $y' \in \Gamma(b, \delta)$ such that $y'<y$. By the previous result, $y' \in \Gamma(a, \delta)$ and $y \in \Gamma(b, \delta)$. So, $y,y' \in \Gamma(a, \delta) \cap \Gamma(b, \delta)$ which implies that 
	\begin{align*}
		\pi(b,y) + \delta V_{\delta}(y) =	\pi(b,y') + \delta V(y',\delta)~and~\pi(a,y) + \delta V(y,\delta) =	\pi(a,y') + \delta V(y',\delta).
	\end{align*}  
	This implies that $	\pi(b,y) - 	\pi(b,y') =\pi(a,y) -  	\pi(a,y')$. By the mean value theorem, this equality implies
	\begin{align*}
		\int_{0}^{1} \pi_{2}(b,\tau y' + (1-\tau)y) d\tau (y-y') = 	\int_{0}^{1} \pi_{2}(a,\tau y' + (1-\tau)y) d\tau (y-y'). 
	\end{align*}
	However, by Assumption \ref{ass:pi.properties}(iii), $s \mapsto\pi_{2}(s,s')$ is increasing, so $ \pi_{2}(a,\tau y' + (1-\tau)y) <  \pi_{2}(b,\tau y' + (1-\tau)y)$ for any $\tau$ and $a<b$. Therefore, the previous display cannot hold thereby yielding a contradiction. 
	
	We now show that $\Gamma(\cdot,\delta)$ is function-like. Otherwise, there exists a $(s,y) \in Graph(\Gamma(\cdot, \delta)$ that has an open neighborhood also in the graph. Thus one can find a pair $(s+a,y-a)$ with $a>0$ such that $\Gamma(s+a, \delta) \ni y-a$. But we found $y-a \in \Gamma(s+a, \delta)$ and $y \in \Gamma(s, \delta)$ such that $y-a = \min\{y, y-a\} \in 	\Gamma(s+a, \delta)$ which contradicts the previous results.
\end{proof}

Recall that $\delta \mapsto \Gamma(s, \delta)$  is no-decreasing (in the strong set sense) if for any $\delta'>\delta$, any $s \in \mathbb{R}_{+}$ and any $y \in \Gamma(s, \delta)$ and $y' \in \Gamma(s, \delta')$ it follows that $\min\{y,y'\} \in \Gamma(s, \delta)$ and $\max\{y,y'\} \in \Gamma(s, \delta')$.

\begin{lemma}\label{lem:PF.delta.incr}
	Take a $s \in \mathbb{S}$ such that $\Gamma(s, \delta) \subseteq \Upsilon^{o}(s)$ and \footnote{ The set $\Gamma^{1}_{\delta}(\Upsilon^{o}(s)) : = \{\Gamma(s', \delta) \colon s' \in \Upsilon^{o}(s)\}$, and $\Gamma^{l}_{\delta}(s) : = \Gamma( \Gamma^{l-1}_{\delta}(s), \delta)$ for any $l>1$. For $l=0$,  $\Gamma^{0}_{\delta}(\Upsilon^{o}(s)) : = \Upsilon^{o}(s)$.}
	\begin{align}\label{eqn:condition.V.delta.incr}
		V(\cdot,\delta)~is~increasing~over \cup_{l=0}^{\infty} \Gamma^{l}(\Upsilon^{o}(s),\delta).
	\end{align}
	Then, for any $(y,y') \in \Gamma(s, \delta) \times \Gamma(s, \delta')$ it follows that $\max\{ y, y'\} \in \Gamma(s, \delta')$ and $\min\{y,y'\} \in \Gamma(s, \delta)$ for any $\delta ' >\delta$.
	
	Moreover,\footnote{For a set $S$, $\min S : = \min\{ s \colon s \in S\}$ and analogously with $\max S$.}
	\begin{align}\label{eqn:PF.delta.incr}
		\min \Gamma(s, \delta') \geq \max \Gamma(s, \delta).
	\end{align}
\end{lemma}

\begin{proof}[Proof of Lemma \ref{lem:PF.delta.incr}]
	Let $(s',\delta) \mapsto W(s',\delta) : = \delta V(s',\delta)$.  Since $s$ is such that  $\Gamma(s, \delta) \subseteq \Upsilon^{o}(s)$ , it follows that $\Gamma(s, \delta) = \arg\max_{s'  \in \Upsilon(s)} \pi(s,s') + W(s',\delta) = \arg\max_{s'  \in \Upsilon^{o}(s)} \pi(s,s') + W(s',\delta) $. By Milgrom-Shannon theorem, if $W$ is (a) quasi-super modular and (b) has the single-crossing property then $\delta \mapsto \Gamma(s, \delta)$ is non-decreasing. Thus, we need to establish properties (a) and (b). For this it suffices to show that $s' \mapsto \frac{d W(s',\delta) }{d \delta}$ is increasing. Observe that, by Lemma \ref{lem:VF.delta.diff}, 
	\begin{align*}
		s' \mapsto 	\frac{d W(s',\delta) }{d \delta} = V(s',\delta) + \delta 	\frac{ d V(s',\delta)}{d \delta}  =  V(s',\delta) + \delta \left( V(s'',\delta) + \delta 	\frac{ d V(s'',\delta)}{d \delta}   \right),
	\end{align*}
	for some $s'' \in  \Gamma(s', \delta)$. 
	
	Now consider $s'_{1} > s'_{0}$ in $\Upsilon^{o}(s)$. By the condition in the lemma, $V(s'_{0},\delta) <  V(s'_{1},\delta)$. By Lemma \ref{lem:PF.incr}, $s \mapsto \Gamma(s, \delta)$ is non-decreasing so $s''_{0} \leq s''_{1}$ for any $s''_{l} \in \Gamma(s'_{l}, \delta)$  for $l=0,1$. Thus $V(s''_{0},\delta) \leq V(s''_{1},\delta) $ by our assumption as $s''_{0},s''_{1}$ belong to $\Gamma^{2}(s, \delta)$. By Lemma \ref{lem:VF.delta.diff}, iterating in this fashion establishes that $	\frac{ d V(s''_{0}, \delta)}{d \delta}  \leq 	\frac{ d V(s''_{1}, \delta)}{d \delta} $. Thus $\frac{d W(s'_{0},\delta) }{d \delta} < \frac{d W(s'_{1},\delta) }{d \delta} $ as desired.

	We now show that $\min \Gamma(s, \delta') \geq \max \Gamma(s, \delta)$ for any $\delta'>\delta$. We do this by contradiction, i.e., suppose there exists an $a \in \Gamma(s, \delta')$ and a $b \in \Gamma(s, \delta)$ such that $a < b$. By the first part of this lemma, $b \in  \Gamma(s, \delta')$ and $a \in \Gamma(s, \delta)$; so, $a,b \in \Gamma(s, \delta') \cap \Gamma(s, \delta)$. By this implies that 
	\begin{align*}
		\pi(s,a) + \delta V(a, \delta) = \pi(s,b) + \delta V(b, \delta)~and~\pi(s,a) + \delta' V_{\delta'}(a) = \pi(s,b) + \delta' V_{\delta'}(b)
	\end{align*}
	thus, by re-arranging terms it follows that $W(a,\delta') - W(a,\delta) = W(b,\delta') - W(b,\delta)$. By the mean value theorem this implies that
	\begin{align*}
		\int_{0}^{1} \frac{dW(a,\delta + t (\delta'-\delta)))}{d\delta}dt  = 	\int_{0}^{1} \frac{dW(b,\delta + t (\delta'-\delta)))}{d\delta}dt.
	\end{align*}
	However, we establishes that $y \mapsto \frac{dW(y,.)}{d\delta}$ was increasing and since $a<b$ this implies that $\frac{dW(a,\delta + t (\delta'-\delta)))}{d\delta} < \frac{dW(b,\delta + t (\delta'-\delta)))}{d \delta}$ for all $t$, a contradiction. 
\end{proof}

\begin{remark}
	The condition in the lemma is a high level condition that can be verified in many different ways. For instance, if $\Upsilon$ is non-decreasing in the inclusion sense then by Lemma \ref{lem:VF.incr} the condition immediately follows. 
	
	If $\Upsilon$ is not non-decreasing in the inclusion sense, the condition in the lemma can still be verified following the results in Appendix \ref{app:VF.mono}. $\triangle$
\end{remark}

\subsection{Characterization of Basins of Attraction of $\Gamma$}
\label{app:Gamma.BoA}

Recall that for any function $F : \mathbb{S}\rightarrow \mathbb{R}$ and for any $e \in \mathbb{S}$ let 
\begin{align*}
	\mathcal{B}^{+}[F](e) :  = \{  s \in \mathbb{S} \colon s<e~and~\forall s' \in (s,e)~F(s') > 0   \} \cup \{ e\}\\
	and~\mathcal{B}^{-}[F](e) : = \{  s \in \mathbb{S} \colon s>e~\forall s' \in (e,s)~F(s') < 0   \} \cup \{ e\}.
\end{align*}

That is, $	\mathcal{B}^{+}[F](e) $  is the set of all points $s \in \mathbb{S}$ such that for any point between $s$ and $e$, the function $s \mapsto F(s)$ is positive --- this set includes $e$ as a convention that simplifies the exposition. 

We refer to $\mathcal{B}[F](e)=\mathcal{B}^{-}[F](e) \cup 	\mathcal{B}^{+}[F](e)$ as the \emph{Basin of attraction (BoA) for $F$}; this terminology is justified by the following result.

\begin{proposition}\label{pro:Basins.chactarization}
	For any $e \in \mathcal{R}^{o}[\bar{\Gamma}(\cdot, \delta)]$,  $\mathcal{B}(e,\delta) = 	\mathcal{B}[\bar{\Gamma}(\cdot, \delta)](e)$. 
\end{proposition}

\begin{proof}[Proof of Proposition \ref{pro:Basins.chactarization}]
	Let $l(e) : = \inf \{s \colon s \in \mathcal{B}^{+}[\bar{\Gamma}(\cdot, \delta)](e) \}$. Clearly, $\mathcal{B}^{+}[\bar{\Gamma}(\cdot, \delta)](e) = (l(e),e]$. Similarly,  let $u(e) : = \sup \{s \colon s \in \mathcal{B}^{-}[\bar{\Gamma}(\cdot, \delta)](e) \}$. Clearly, $\mathcal{B}^{-}[\bar{\Gamma}(\cdot, \delta)](e) = [e,u(e))$.

	We first show that  $	\mathcal{B}(e,\delta) \supseteq \mathcal{B}^{+}[\bar{\Gamma}(\cdot, \delta)](e) \cup \mathcal{B}^{-}[\bar{\Gamma}(\cdot, \delta)](e)$. To do this, take any $s \in \mathcal{B}^{+}[\bar{\Gamma}(\cdot, \delta)](e)$. By definition $s>l(e)$ and $\Gamma(s, \delta) > s$. Therefore, since $s \mapsto \Gamma(s, \delta)$ is non-decreasing (Lemma \ref{lem:PF.incr} in Appendix \ref{app:PF.properties}) any flow starting in $s$ is such that $\phi_{\delta}(t,s) \geq  \phi_{\delta}(t-1,s)$ for any $t \in \mathbb{N}$. Therefore it must have a limit point which we denote as $a$. Since $s \leq e$, $\phi_{\delta}(t,s) \leq \phi_{\delta}(t,e) = e$, which implies that $a \leq e$. If $a<e$ it means that $a$ is a fixed point of $\Gamma(\cdot, \delta)$ that is in $(s,e)$, but this contradicts the definition of $\mathcal{B}^{+}[\bar{\Gamma}(\cdot, \delta)](e) $. Thus $a=e$ thereby showing that $	\mathcal{B}(e,\delta) \supseteq \mathcal{B}^{+}[\bar{\Gamma}(\cdot, \delta)](e) $. Analogous arguments can show that $\mathcal{B}(e,\delta) \supseteq \mathcal{B}^{-}[\bar{\Gamma}(\cdot, \delta)](e) $ and thus the desired inclusion holds.
	
	We now show that $	\mathcal{B}(e,\delta) \subseteq \mathcal{B}^{+}[\bar{\Gamma}(\cdot, \delta)](e) \cup \mathcal{B}^{-}[\bar{\Gamma}(\cdot, \delta)](e)$. We do this by contradiction, i.e., suppose there exists a $s_{0} \in \mathcal{B}(e,\delta)$ but $s_{0} \notin \mathcal{B}^{+}[\bar{\Gamma}(\cdot, \delta)](e) \cup \mathcal{B}^{-}[\bar{\Gamma}(\cdot, \delta)](e)$. This means that either (a) $s_{0} < e$ but there exists a $a \in \Gamma(s_{0}, \delta)$ such that $a \leq s_{0}$; or $s_{0} > e$ but there exists a $a \in \Gamma(s_{0}, \delta)$ such that $a \geq s_{0}$.
	
	Suppose (a) holds. Clearly, $s_{0}=a$ is a direct contradiction to the fact that $s_{0} \in \mathcal{B}(e,\delta)$, so lets take $a < s_{0}$. This means there exist at least one flow such that $\phi(1,s_{0}) \leq s_{0}$. By Lemma \ref{lem:PF.incr} in Appendix \ref{app:PF.properties} this implies that  $\phi(t,s_{0}) \leq s_{0}$ for all $t$. But since $e>s_{0}$, this implies that  $(\phi(t,s_{0}))_{t}$ does not converge to $e$,  contradicting the assumption that $s_{0} \in \mathcal{B}(e,\delta)$. If (b) holds analogous arguments also show a contradiction to the assumption that $s_{0} \in \mathcal{B}(e,\delta)$. 
	
	Hence, $	\mathcal{B}(e,\delta) \subseteq \mathcal{B}^{+}[\bar{\Gamma}(\cdot, \delta)](e) \cup \mathcal{B}^{-}[\bar{\Gamma}(\cdot, \delta)](e)$, thereby establishing the desired result.
\end{proof}

\section{Sufficient Conditions for Locator functions to have well-separated roots}
\label{app:separate.roots}


\begin{lemma}\label{lem:separate.roots}
	Suppose the locator function is three times continuously differentiable and 
	\begin{itemize}
		\item For any $s$ such that $\mathcal{L}_{1}(s,\delta) = 0$, then $|\mathcal{L}_{11}(s,\delta) |  > 0$. 
	\end{itemize}
	Then for any root $r$, there exists an open neighborhood such that no other roots are in it. If, in addition, there exists a $M$ finite such that $|\mathcal{L}(s,\delta)| > 0$ for all $|s|>M$, then the roots are finite.
\end{lemma}

\begin{proof}
We say $s$ is a critical point of the locator function if $\mathcal{L}_{1}(s,\delta) = 0$. We now show that critical points are isolated, i.e., for any critical point $s$, there exists a $\epsilon_{s}>0$ such that $\mathcal{L}_{1}(s,\delta) \ne 0$ for all $s \in (s-\epsilon_{s},s+\epsilon_{s})$. We prove this by contradiction, i.e., suppose that there exists a sequence $(c_{n})_{n}$ of critical points that converge to a critical point $c$. Then by the MVT (and the fact that the locator function is three times differentiable over a compact domain $[-M,M]$)
	\begin{align*}
		\mathcal{L}_{1}(c_{n},\delta) - \mathcal{L}_{1}(c,\delta) = \mathcal{L}_{11}(c,\delta)  (c_{n}-c) + o(c_{n}-c), 
	\end{align*}
	but the LHS is zero as both are critical points, thereby implying that $ \mathcal{L}_{11}(c,\delta)  = 0$ which is a contradiction to our assumption. 
	
	We now show that roots are well-separated in the sense that for any root $r$ there exists a $\epsilon > 0$ such that $|r-r'|>\epsilon$ for any other root $r'$. We prove this claim by contradiction. I.e., suppose there exists a root, $r$, such that for any $\epsilon>0$ there exists another root within $\epsilon$ distance of $r$. This implies the existence of a sequence of roots $(r_{n})_{n}$ that converges to $r$.  For any interval $[r_{n},r_{n+1}]$ by Rolle's Theorem there exists a $c_{n}$ such that $\mathcal{L}_{1}(c_{n},\delta) = 0$. So we constructed a sequence of critical points $(c_{n})_{n}$. Since the sequence of roots converges, this sequence has an accumulation point given by$c_{\infty}$, which is also a critical point because of continuity of the first derivative of the locator function. But this implies that $c_{\infty}$ is a critical point that is not isolated, a contradiction. 
	
	If in addition there exists a $M$ finite such that $|\mathcal{L}(s,\delta)| > 0$ for all $|s|>M$, then roots are confined to $[-M,M]$. Since roots are well separated they must be finite. 
\end{proof}

\section{Proofs of Auxiliary Lemmas}
\label{OA:lemmas.aux}

\begin{proof}[Proof of Lemma \ref{lem:locator.boundary}]
	\textbf{Parts 1-2.} We only prove the first part because the proof of the second part is completely analogous and thus can be omitted.

	Either $\Gamma(\underbar{s},\delta) \ni \underbar{s}$ or $\Gamma(\underbar{s},\delta) > \underbar{s}$. If the latter holds, there is nothing to prove, so we proceed under the assumption $\Gamma(\underbar{s},\delta) \ni \underbar{s}$. Hence, it remains to show that in this case,  if $\mathcal{L}(\underline{s},\delta) > 0$ then there exists a $\varepsilon>0$ such that  $\Gamma(s, \delta) > s$ for all $s \in (\underbar{s},\underbar{s}+\varepsilon)$. 
	
	We do this by contradiction, i.e., suppose there exists a sequence $(s_{n})_{n}$ converging to $\underbar{s}$ such that for each $s_{n}$, there exists a $s'_{n} \in \Gamma(s_{n}, \delta)$ such that $s'_{n} \leq s_{n}$ for all $n$. 
	
	We claim that for each $s_{n}$, there exists a $\bar{\Delta} > 0$ such that $\pi(s_{n},s'_{n}) + \delta V(s'_{n}, \delta) \geq \pi(s_{n},s'_{n} + \Delta) + \delta V(s'_{n} + \Delta, \delta)$ for any $\Delta \in [0,\bar{\Delta}]$. Since $s'_{n}$ is optimal for $s_{n}$ the previous inequality will hold as long as $s'_{n}$ is not at the ``upper boundary" of $\Upsilon(s_{n})$. Indeed, $s'_{n}$ is not at the upper bounded because if $s'_{n} = \max \Upsilon(s_{n})$ then, since $\Upsilon^{o}(s_{n}) \ni s_{n}$, it follows that $s'_{n} > s_{n}$ a contradiction. 
	
	Re-arranging terms and taking limits where $\Delta$ converges to zero, it follows that by Proposition \ref{pro:VF.diff},
	\begin{align*}
		0 \geq \pi_{2}(s_{n},s'_{n})  + \delta \max_{y \in \Gamma(s'_{n} , \delta)} \pi_{1}(s'_{n},y).
	\end{align*} 
	
	Since this holds for each $n$, we can take limits
	\begin{align*}
		0 \geq \lim_{n \rightarrow \infty} \pi_{2}(s_{n},s'_{n})  + \delta \lim_{n \rightarrow \infty} \max_{y \in \Gamma(s'_{n} , \delta)} \pi_{1}(s'_{n},y).
	\end{align*} 
	
	Since $s'_{n} \leq s_{n}$ and $s_{n} \to \underbar{s}$, $(s'_{n})_{n}$ converges to $\underbar{s}$.  This and the fact that $\pi_{1}$ is continuous imply by the ToM that  $\lim_{n \rightarrow \infty} \max_{y \in \Gamma(s'_{n} , \delta)} \pi_{1}(s'_{n},y) =\max_{y \in \Gamma(\underbar{s} , \delta)}  \pi_{1}(\underbar{s},y)$. So finally, since $\pi_{2}$ is conitnuous, the previous display implies 
	\begin{align*}
		0 \geq  \pi_{2}(\underbar{s},\underbar{s})  + \delta   \max_{y \in \Gamma(\underbar{s} , \delta)}  \pi_{1}(\underbar{s},y) \geq  \mathcal{L}(\underbar{s},\delta).
	\end{align*} 
	where the last line follows because  $\Gamma(\underbar{s},\delta) \ni \underbar{s}$. 	But this violates the assumption that $\mathcal{L}(\underbar{s},\delta)> 0$ and thus arrived to a contradiction.  
	
	
	\bigskip
	
	\textbf{Parts 3-4.} We only prove the third part because the proof of the fourth part is completely analogous and thus can be omitted.  We prove the result by contradiction, i.e., suppose there exists a sequence $(s_{n})_{n}$ converging to $\underbar{s}$ such that for each $s_{n}$, there exists a $s'_{n} \in \Gamma(s_{n}, \delta)$ such that $s'_{n} \geq s_{n}$ for all $n$.   We claim that for each $s_{n}$, there exists a $\bar{\Delta} > 0$ such that $\pi(s_{n},s'_{n}) + \delta V(s'_{n}, \delta) \geq \pi(s_{n},s'_{n} - \Delta) + \delta V(s'_{n} - \Delta, \delta)$ for any $\Delta \in [0,\bar{\Delta}]$. The inequality follows from the fact that $s'_{n}$ is an argmax for $s_{n}$ provided $s'_{n}$ is not at the ``lower boundary" of $\Upsilon(s_{n})$. This is true because if $s'_{n} = \min \Upsilon(s_{n})$ then, since $\Upsilon^{o}(s_{n}) \ni s_{n}$, it follows that $s'_{n} < s_{n}$ a contradiction. 
	
	Re-arranging terms, dividing by $-\Delta$, and taking limits where $\Delta$ converges to zero, it follows that by Proposition \ref{pro:VF.diff},
	\begin{align*}
		0 \leq \pi_{2}(s_{n},s'_{n})  + \delta \min_{y \in \Gamma(s'_{n} , \delta)} \pi_{1}(s'_{n},y).
	\end{align*} 
	
	Since this holds for each $n$, we can take limits
	\begin{align*}
		0 \leq \lim_{n \rightarrow \infty} \pi_{2}(s_{n},s'_{n})  + \delta \lim_{n \rightarrow \infty} \min_{y \in \Gamma(s'_{n} , \delta)} \pi_{1}(s'_{n},y).
	\end{align*} 
	
	By going to a subsequence if necessary, $(s'_{n})_{n}$ converges to $\underbar{s}'$. Since $\Gamma$ is compact-valued and UHC (see Lemma \ref{lem:PF.prop.text}), it follows that $\underbar{s}' \in \Gamma(\underbar{s},\delta)$, and moreover, by continuity of $\pi_{1}$ and the ToM, $\lim_{n \rightarrow \infty} \min_{y \in \Gamma(s'_{n} , \delta)} \pi_{1}(s'_{n},y) = \min_{y \in \Gamma(\underline{s}',\delta)} \pi_{1}(\underbar{s}',y)$. This, the fact that $\pi_{2}$ is continuous, imply that the the previous display equals
	\begin{align*}
		0 \leq  \pi_{2}(\underbar{s},\underbar{s}')  + \delta  \min_{y \in \Gamma(\underline{s}',\delta)} \pi_{1}(\underbar{s}',y).
	\end{align*} 
	
	Finally, since $\Gamma(\underbar{s},\delta) = \underbar{s}$ by assumption, $\underbar{s}=\underbar{s}'$, so 
	\begin{align*}
		0 \leq  \pi_{2}(\underbar{s},\underbar{s})  + \delta  \min_{y \in \Gamma(\underline{s},\delta)} \pi_{1}(\underbar{s},y) = \pi_{2}(\underbar{s},\underbar{s})  + \delta  \pi_{1}(\underbar{s},\underbar{s}) =  \mathcal{L}(\underbar{s},\delta).
	\end{align*} 
	But this violates the assumption that $\mathcal{L}(\underbar{s},\delta)< 0$ and  we thus arrived to a contradiction.

\end{proof}

\begin{proof}[Proof of Lemma \ref{lem:root.unique}]
	\textsc{Part 1.} By Lemma \ref{lem:locator.boundary}, for which all assumptions holds, we now that either  $\Gamma(\underline{s}, \delta) > \underline{s}$ or  $\Gamma(s, \delta) > s$ for all $s$ sufficiently close to  $\underline{s}$. Hence,  either $\bar{\Gamma}(\cdot, \delta) > 0$ over $\mathbb{S}^{o}$, and $e$ is not a "true" fixed point and $c = \underline{s}$; or, $\bar{\Gamma}(\cdot, \delta)$ changes signs in $\mathbb{S}^{o}$. By Lemma \ref{lem:PF.incr} in Appendix \ref{app:PF.properties}, $\bar{\Gamma}(\cdot, \delta)$ cannot ``jump down", so in this case: $\bar{\Gamma}(s, \delta) = 0$ for some $s$ in the interior of $\mathbb{S}$. By Theorem \ref{thm:FP.characterization} and the uniqueness assumption, such element is $e$. Moreover, over $[\underline{s},e)$, $\bar{\Gamma}(\cdot, \delta) > 0$ and $\bar{\Gamma}(e, \delta) = e$. Let $c > e$ be the largest element such that for all $s \in [e,c)$, $\bar{\Gamma}(s, \delta) < 0$. It is clearly that $ [\underline{s},c) \subseteq  \mathcal{B}(e,\delta)$. Moreover, if $c < \bar{s}$, then $c$ is a Skiba point as otherwise $c$ would be a fixed point but it contradicts the assumption of uniqueness. Finally, for any $s \in (c, \bar{s}]$, $\bar{\Gamma}(s, \delta)>0$, thus $s \in \mathcal{B}(\bar{s},\delta)$.
	
	\bigskip
	
	\textsc{Part 2.}  By analogous arguments to those presented in Part 1, either $\Gamma(\overline{s}, \delta) < \overline{s}$ or   $\Gamma(s, \delta) < s$ for all $s$ sufficiently close to  $\bar{s}$. So, either $e$ is not a "true" fixed point,  and so $c = \bar{s}$;  or $e$ is a fixed point of $\Gamma(\cdot, \delta)$. By following an analogous reasoning to that in part 1, we can conclude that there exists a $c$ such that  $[\underline{s},c) \subseteq \mathcal{B}(\underbar{s},\delta)$ and $(c,\bar{s})  \subseteq \mathcal{B}(e,\delta)$.
\end{proof}

\begin{proof}[Proof of Lemma \ref{lem:piSCVA.SSiffRoot}]
	By Theorem \ref{thm:FP.characterization} on side of the inclusion holds, so we only show the other --- i.e., if $s \in \mathcal{R}^{o}[\mathcal{L}(\cdot,\delta)]$, then $s$ is a (interior) steady state.

	Suppose not. That is, suppose there exists a $s \in \mathbb{S}^{o}$ such that $\gamma(s) : = \Gamma(s, \delta) \ne s$ but $\mathcal{L}(s,\delta) = 0$. We first consider the case $\gamma(s)>s$.

	By strict concavity of $\pi$ it is well-known that $s \mapsto V(s,\delta)$ is also strictly concave. Then by the FOC, which are sufficient,
	\begin{align*}
		\pi_{2}(s,\gamma(s)) + \delta \pi_{1}(\gamma(s),\Gamma(\gamma(s), \delta)) \geq 0 
	\end{align*}
	(with equality if $\gamma(s) \in \Upsilon^{o}(s)$). Since, by strict concavity, $s \mapsto V'(s,\delta) : = \pi_{1}(s, \Gamma(s, \delta) )$ is strictly decreasing. This fact, the fact that $\gamma(s)>s$, and the previous display thus imply
	\begin{align*}
		\pi_{2}(s,\gamma(s)) + \delta \pi_{1}(s,\gamma(s))>0. 
	\end{align*}

   This result and the condition in the lemma imply $0 < \max_{a \in \Upsilon(s) \colon a > s} sign \left\{  \pi_{2}(s,a) + \delta \pi_{1}( s ,a  )  \right\}  \leq	sign \left\{ \mathcal{L}(s,\delta) \right\} $. But this contradicts that $\mathcal{L}(s,\delta) > 0$.

	Consider now the case $\gamma(s)<s$. Analogous arguments imply 
	\begin{align*}
		\pi_{2}(s,\gamma(s)) + \delta \pi_{1}(s,\gamma(s)) < 0.
	\end{align*}
	By the condition in the lemma, 	$\min_{a \in \Upsilon(s) \colon a < s} sign \left\{  \pi_{2}(s,a) + \delta \pi_{1}( s ,a  )  \right\}  \geq	sign \left\{ \mathcal{L}(s,\delta) \right\} $. This implies that $\mathcal{L}(s,\delta) < 0$, a contradiction. 
\end{proof}

\begin{proof}[Proof of Lemma \ref{lem:Basin.characterization}]
	
	We first consider the case where $e$ is unstable. By Theorem \ref{thm:FP.characterization}, $\mathcal{L}_{1}(e,\delta) > 0$. Therefore, $ \mathcal{B}[\mathcal{L}(\cdot,\delta)](e) = \{e\}$, and thus there is nothing to prove. 
	
	So, henceforth, we take $e$ to be stable. By Proposition \ref{pro:Basins.chactarization}, it is sufficient to show 
	 \begin{align*}
	 	  \mathcal{B}[\bar{\Gamma}(\cdot, \delta)](e) \supseteq \mathcal{B}[\mathcal{L}(\cdot,\delta)](e). 
	 \end{align*}


	For any $s \in	\mathcal{B}[\mathcal{L}(\cdot,\delta)](e)$, either $s<e$ and $\mathcal{L}(s',\delta) > 0$ for all $s' \in (s,e)$, or  $s>e$ and $\mathcal{L}(s',\delta) < 0$ for all $s' \in (e,s)$ (if $s=e$ there is nothing to prove). We consider the first case --- the second case is omitted as the proof is completely analogous. 
	
	Given that $s<e$, to show that $s \in	  \mathcal{B}[\bar{\Gamma}(\cdot, \delta)](e) $ it suffices to show that $\bar{\Gamma}(s', \delta)> 0$ for any $s' \in (s,e]$. We prove this statement by contradiction; i.e., there exists a $s' \in (s,e]$ such that $\bar{\Gamma}(s', \delta) \leq 0$.

	If $\bar{\Gamma}(s',\delta)=0$, then $s' \in \mathcal{R}^{o}[\bar{\Gamma}(\cdot,\delta)]$. By theorem \ref{thm:FP.characterization} this implies that $s' \in \mathcal{R}^{o}[\mathcal{L}(\cdot,\delta)]$, but this contradicts the fact that $s \in	\mathcal{B}[\mathcal{L}(\cdot,\delta)](e)$. Thus $\bar{\Gamma}(s',\delta) < 0$. 
	
	Therefore,  $\Gamma(s', \delta) < s'$. Under our assumption in the text, $e$ is isolated, thereby implying there exists an open neighborhood of $e$, $(e-\gamma,e+\gamma)$ for some $\gamma>0$, such that $s \mapsto \bar{\Gamma}(s,\delta)$ is monotonic. Moreover, since $e$ is taken to be stable, $s \mapsto \bar{\Gamma}(s,\delta)$ is decreasing monotonic, thereby implying $\Gamma(s,\delta) > s$ for all $s \in (e-\gamma,e)$. 	So, by taken $\gamma$ sufficiently small, we obtained that   $\Gamma(s', \delta) < s'$ and $\Gamma(s'',\delta) > s''$ for $(s',s'')$ such that $s' <  e-\gamma < s'' < e$. Since $s \mapsto \bar{\Gamma}(s,\delta)$ is assumed to be a function, it is a continuous one (Lemma \ref{lem:PF.cont} in Appendix \ref{app:PF.properties}). Thus, by Bolzano, there exists a $c \in (s',s'')$ such that  $\Gamma(c, \delta) = c$. But by Theorem \ref{thm:FP.characterization} this implies that $c \in \mathcal{R}[\mathcal{L}(\cdot,\delta)]$. However, since $s<s'$ and $s'' < e$, we found a point $c$ that  $c \in \mathcal{R}[\mathcal{L}(\cdot,\delta)]$ and $c \in (s,e)$. This fact contradicts the assumption that  $s \in	\mathcal{B}[\mathcal{L}(\cdot,\delta)](e)$.\footnote{The assumption of $s \mapsto \Gamma(s, \delta)$ being a function is used to obtained this contradiction. Otherwise, the point $c$ could be a Skiba point (a point where $\Gamma(\cdot, \delta)$ "jumps") and will not be ``picked up" by $\mathcal{L}$. } 
\end{proof}

\section{Sufficient Conditions for Condition \ref{eqn:SCVA.cond-t}}\label{app:SCVA.cond}

\begin{lemma}\label{lem:suff.cond.SSiffRoot}
	Condition \ref{eqn:SCVA.cond-t}  in the Proposition \ref{pro:SCVA.FP.characterization} is implied by either one of the following conditions:
	\begin{enumerate}
		\item There exists functions $F,G_{1},G_{2}$ such that $F>0$ and $\pi_{1}(s,s') = F(s,s') G_{1}(s) $ and $\pi_{2}(s,s') = F(s,s') G_{2}(s) $. 
		\item 
		\begin{align}\label{eqn:SCVA.cond-1}
			& \sup_{a \in \Upsilon(s) \colon a > s} \pi_{2}(s,a) - 	\pi_{2}(s,s)  + \delta ( \pi_{1}(s,a) - \pi_{1}(s,s) ) \leq 0,\\\label{eqn:SCVA.cond-1b}
			and~& \inf_{a \in \Upsilon(s) \colon a < s} \pi_{2}(s,a) - 	\pi_{2}(s,s)  + \delta ( \pi_{1}(s,a) - \pi_{1}(s,s) ) \geq 0
		\end{align}
		\item $\pi_{22} + \delta \pi_{12} \leq 0$. 
	\end{enumerate}
\end{lemma}

\begin{proof}
	\textbf{(1).} By the functional form assumption, $\mathcal{L}(s,\delta) = F(s,s) (G_{2}(s) + \delta G_{1}(s))$. Since $F>0$, $sign \{ \mathcal{L}(s,\delta)   \} = sign\{  G_{2}(s) + \delta G_{1}(s) \}$. On the other hand, $sign \left\{  \pi_{2}(s,a) + \delta \pi_{1}( s ,a  )  \right\}  = sign \left\{ F(s,a)  (G_{2}(s) + \delta G_{1}(s) ) \right\} = sign\{ G_{2}(s) + \delta G_{1}(s)  \}$. Thus condition \ref{eqn:SCVA.cond-t} readily holds (with equality)
	
	\bigskip

	\textbf{(2).}  Take any $s \in \mathbb{S}^{o}$ such that $0 = \mathcal{L}(s,\delta)$. Then, for any $a \in \mathbb{S}$,
	\begin{align*}
		0 = &\mathcal{L}(s,\delta) 	  =  \pi_{2}(s,a)  + \delta \pi_{1}(s,a)    - \{ \pi_{2}(s,a) - 	\pi_{2}(s,s)  + \delta ( \pi_{1}(s,a) - \pi_{1}(s,s) ) \} .
	\end{align*}
If $a>s$, then by condition  \ref{eqn:SCVA.cond-1}, the term in the curly brackets is negative, so $ \pi_{2}(s,a)  + \delta \pi_{1}(s,a) < 0$. Since this holds for any $a>s$, it follows that $\sup_{a \in \Upsilon(s) \colon a > s} \pi_{2}(s,a)  + \delta \pi_{1}(s,a) \leq 0$.

If $a<s$, then by condition  \ref{eqn:SCVA.cond-1b}, the term in the curly brackets is positve, so   $ \pi_{2}(s,a)  + \delta \pi_{1}(s,a) > 0$. Since this holds for any $a<s$, it follows that $\inf_{a \in \Upsilon(s) \colon a < s} \pi_{2}(s,a)  + \delta \pi_{1}(s,a) \geq 0$. 

Hence, condition \ref{eqn:SCVA.cond-t} holds. 
	
%
	
	\bigskip  
	
	\textbf{(3).} We show that  $\pi_{22} + \delta \pi_{12} \leq 0$ implies the conditions in part 2. Take the case $a>s$. By the MVT, condition \ref{eqn:SCVA.cond-1} can be cast as 
	\begin{align*}
		\sup_{a \in \Upsilon(s) \colon a > s} \int_{0}^{1} (\pi_{22}(s,s+t(a-s)) + \delta \pi_{12}(s,s+t(a-s)) )dt \times (a-s) \leq 0
	\end{align*}
	which is implied by
	\begin{align*}
		\sup_{a \in \Upsilon(s) \colon a > s} \int_{0}^{1} (\pi_{22}(s,s+t(a-s)) + \delta \pi_{12}(s,s+t(a-s)) )dt  \leq 0.
	\end{align*}
	For the second condition in  \ref{eqn:SCVA.cond-1} a similar expression holds: 
	\begin{align*}
		\inf_{a \in \Upsilon(s) \colon a > s} \int_{0}^{1} (\pi_{22}(s,s+t(a-s)) + \delta \pi_{12}(s,s+t(a-s)) )dt  \geq 0.
	\end{align*}
	
	A sufficient condition for these inequalities is
	\begin{align}
		(\pi_{22} + \delta \pi_{12})(s,s') \leq 0
	\end{align}
	for any $s \in \mathcal{R}^{o}[\mathcal{L}(\cdot,\delta)]$ and $s' \in \Upsilon(s)$. 
	
\end{proof}

\section{Extensions of Application \ref{exa:intertemporal}} \label{app:ie.extension}

\subsection{Import foreign variety}

Suppose we allow for foreign variety of good $1$, and set the utility from good $c_0$, domestic $c_1$ and foreign variety $c_1^*$ is
\[
U(c_0, c_{1},c_{1}^{*})=u\left(c_{1}\right)+u^{*}\left(c_{1}^{*}\right)+c_0,
\]
and assume that the price of foreign good $1$ is fixed at $p_{1}$. Assume that this economy is endowed with 1 unit of inelastic label supply, so Assumptions \ref{ass:C.properties} and \ref{ass:growth} hold trivially. Next Lemma verifies the Assumption \ref{ass:pi.properties} in this extension.
\begin{lemma}\label{lem:ie.import}
	If $\eta(c_1^*):=-\frac{u^{*'}(c_1^*)}{u^{*''}(c_1^*) c_{1}^{*}}$ is always larger than $1$, Assumption \ref{ass:pi.properties} holds.
\end{lemma}

\begin{proof}
	By Envelope Theorem, we have the same $\pi_{1}(s,s')=u'(c_{1}(s,s'))H'(s)F(s')$,
	which is positive. Besides the good $1$ export decision, there is
	another consumption allocation problem. Given each $c_{1},s, s'$:
	\[
	\max_{c_{1}^{*}}u^{*}\left(c_{1}^{*}\right)+c_0,\quad s.t.\quad p_{1}c_{1}^{*}+c_0=p(e_{1})e_{1}+G(1-s').
	\]
	Basically, the planner should leverage the revenue from exporting
	good $1$ and (potentially) $0$ to finance its consumption of foreign
	variety of good $1$ and good $0$. Even though the economy is not
	necessarily able to achieve an interior solution, our framework can always
	handle it. 
	
	The interior solution of the consumption allocation problem requires
	$u^{*'}(c_{1}^{*})=p_{1}$, which means consumption of foreign variety
	of good $1$ is constant. If the export revenue is larger than $p_{1}c_{1}^{*}$,
	\[
	p(e_{1})e_{1}+G(1-s')>p_{1}c_{1}^{*},
	\]
	the first order condition for good $1$ export decision, $u'(c_{1})=\frac{\varepsilon -1}{\varepsilon }p\left(e_{1}\right)$,
	holds and analysis in the Lemma \ref{lem:ie.pi12} remains valid. 
	
	If the economy cannot afford to purchase enough foreign variety of
	good $1$ to achieve $u^{*'}(c_{1}^{*})=p_{1}$, we can approach the
	monotonic condition differently. We have
	\[
	\pi_{12}(s,s')= u'(c_{1}(s,s'))H'(s) \left[F'(s')-\frac{1}{\gamma(c_1) c_{1}(s,s')}\frac{\partial c_{1}(s,s')}{\partial s'}F(s') \right],
	\]
	while we have the first order condition for exporting decision $u'(c_{1})=u^{*'}(\frac{p(e_{1})e_{1}+G(1-s')}{p_{1}})\frac{p(e_{1})}{p_{1}}\left(1-\frac{1}{\varepsilon}\right)$.
	Fix $s$ and differentiate it, we have 
	\[
	\frac{\partial c_{1}(s,s')}{\partial s'}= \frac{\frac{1-\frac{1}{\varepsilon}}{p_{1}}\left\{ \frac{u^{*''}p^{2}(e_{1})}{p_{1}}\left[1-\frac{1}{\varepsilon}\right]+u^{*'}p'(e_{1})\right\} H(s)F'(s')-\frac{1-\frac{1}{\varepsilon}}{p_{1}}u^{*''}G'(1-s')\frac{p(e_{1})}{p_{1}}}{u''(c_{1})+\frac{1-\frac{1}{\varepsilon}}{p_{1}}\left\{ \frac{u^{*''}p^{2}(e_{1})}{p_{1}}\left[1-\frac{1}{\varepsilon}\right]+u^{*'}p'(e_{1})\right\} }.
	\]
	Substitute it into the $\pi_{12}$, we have
	\[
	\pi_{12}(s,s')= u'(c_{1})H'(s)F'(s') \left[1-\frac{F}{\gamma c_{1}}\frac{\frac{1-\frac{1}{\varepsilon}}{p_{1}}\left\{ \frac{u^{*''}p^{2}(e_{1})}{p_{1}}\left[1-\frac{1}{\varepsilon}\right]+u^{*'}p'(e_{1})\right\} H(s) - \frac{1-\frac{1}{\varepsilon}}{p_{1}}u^{*''}\frac{G'(1-s')}{F'}\frac{p(e_{1})}{p_{1}}}{u''(c_{1})+\frac{1-\frac{1}{\varepsilon}}{p_{1}}\left\{ \frac{u^{*''}p^{2}(e_{1})}{p_{1}}\left[1-\frac{1}{\varepsilon}\right]+u^{*'}p'(e_{1})\right\} }\right],
	\]
	Focus on the terms in the bracket, 
	\[
	1-\frac{1}{\gamma}\frac{\frac{1-\frac{1}{\varepsilon}}{p_{1}}\left\{ \frac{u^{*''}p^{2}(e_{1})}{p_{1}}\left[1-\frac{1}{\varepsilon}\right]+u^{*'}p'(e_{1})\right\} H(s) F - \frac{1-\frac{1}{\varepsilon}}{p_{1}}u^{*''}\frac{G'(1-s')F}{F'}\frac{p(e_{1})}{p_{1}}}{u''(c_{1})c_{1}+\frac{1-\frac{1}{\varepsilon}}{p_{1}}\left\{ \frac{u^{*''}p^{2}(e_{1})}{p_{1}}\left[1-\frac{1}{\varepsilon}\right]+u^{*'}p'(e_{1})\right\} c_{1}}.
	\]
	Divide $u'$ above and below the fraction line of the second term,
	and note that $u'=u^{*'}\frac{p(e_{1})}{p_{1}}\left(1-\frac{1}{\varepsilon}\right)$,
	\[
	1-\frac{1}{\gamma}\frac{\left\{ \frac{u^{*''}p(e_{1})}{u^{*'}p_{1}}\left[1-\frac{1}{\varepsilon}\right]+\frac{p'(e_{1})}{p(e_{1})}\right\} H(s) F - \frac{u^{*''}}{u^{*'}}\frac{G'(1-s')F}{F'}\frac{1}{p_{1}}}{\frac{u''(c_{1})c_{1}}{u'}+\left\{ \frac{u^{*''}p(e_{1})}{u^{*'}p_{1}}\left[1-\frac{1}{\varepsilon}\right]+\frac{p'(e_{1})}{p(e_{1})}\right\} c_{1}}.
	\]
	Use $e_{1}=H(s) F(s') - c_{1}$ and the definition of $\gamma,\eta$ to get 
	\[
	1-\frac{\left\{ -\frac{p(e_{1})}{\eta c_{1}^{*}p_{1}}\left[1-\frac{1}{\varepsilon}\right]-\frac{1}{\varepsilon e_{1}}\right\} \left(c_{1}+e_{1}\right)+\frac{1}{\eta p_{1}c_{1}^{*}}\frac{G'(1-s')F}{F'}}{-1+\left\{ -\frac{p(e_{1})}{\eta c_{1}^{*}p_{1}}\left[1-\frac{1}{\varepsilon}\right]-\frac{1}{\varepsilon e_{1}}\right\} \gamma c_{1}}
	\]
	Since we are at a corner in which there is no consumption of good
	$0$ then we have $p_{1}c_{1}^{*}=pe_{1}+G$. Substitute $p_{1}c_1^*$ to this term, we have
	\[
	\frac{1+\left\{ \frac{p}{\eta(pe_{1}+G)}\left[1-\frac{1}{\varepsilon}\right]+\frac{1}{\varepsilon e_{1}}\right\} \gamma c_{1}+\left\{ -\frac{p}{\eta(pe_{1}+G)}\left[1-\frac{1}{\varepsilon}\right]-\frac{1}{\varepsilon e_{1}}\right\} (c_{1}+e_{1})+\frac{1}{\eta\left(pe_{1}+G\right)}\frac{G'(1-x)F}{F'}}{1+\left\{ \frac{p}{\eta(pe_{1}+G)}\left[1-\frac{1}{\varepsilon}\right]+\frac{1}{\varepsilon e_{1}}\right\} \gamma c_{1}}
	\]
	Then, we can rearrange terms, and get 
	\[
	\frac{(\gamma-1)\left\{ \frac{p}{\eta(pe_{1}+G)}\left[1-\frac{1}{\varepsilon}\right]+\frac{1}{\varepsilon e_{1}}\right\} c_{1}+\left(1-\frac{pe_{1}}{\eta(pe_{1}+G)}\right)\left[1-\frac{1}{\varepsilon}\right]+\frac{1}{\eta\left(pe_{1}+G\right)}\frac{G'(1-x)F}{F'}}{1+\left\{ \frac{p}{\eta(pe_{1}+G)}\left[1-\frac{1}{\varepsilon}\right]+\frac{1}{\varepsilon e_{1}}\right\} \gamma c_{1}}.
	\]
	We note that if $\gamma(c_1)>1,\varepsilon>1,\eta(c_1^*)>1$, this expression
	is always positive. Henceforth, $\pi_{12}$ is still positive. The
	intuition for condition $\eta>1$ is that we need the utility of foreign
	variety of good $1$ of less curvature, so that more production of
	good $1$ will induce more purchasing of foreign variety rather than
	too much domestic variety. And then the marginal utility of good $1$
	will not decrease too fast.
\end{proof}

\subsection{Closed economy with CES preference}

We allow constant elasticity of substitution between good $0$ and $1$ in a closed economy. Suppose the economy is endowed with 1 unit of labor, so $\mathbb{S}=[0,1]$. The current utility for the closed economy is 
\[
U(c_0,c_1)=\frac{\gamma}{\gamma-1}\left(c_0^{\frac{\sigma-1}{\sigma}} + c_{1}^{\frac{\sigma-1}{\sigma}} \right)^{\frac{\gamma-1}{\gamma} \frac{\sigma}{\sigma-1}},
\]
so we can rewrite the per-period payoff function as follows
\[
\pi(s,s')=\frac{\gamma}{\gamma-1} \left( \left(G(1-s')\right)^{\frac{\sigma-1}{\sigma}} +  \left(H(s) F(s')\right)^{\frac{\sigma-1}{\sigma}}  \right)^{\frac{\gamma-1}{\gamma}\frac{\sigma}{\sigma-1}}.
\]
Obviously, this setting satisfies Assumptions \ref{ass:C.properties} and \ref{ass:growth}, and the following Lemma provides a sufficient condition to ensure Assumption \ref{ass:pi.properties} holds.
\begin{lemma}\label{lem:ie.closed}
	If $1<\gamma<\sigma$, the closed economy with CES preference satisfies Assumption \ref{ass:pi.properties}.
\end{lemma}

\begin{proof}
	It follows that $\pi_{1}(s,s')=\left( \left(G(1-s')\right)^{\frac{\sigma-1}{\sigma}} +  \left(H(s) F(s')\right)^{\frac{\sigma-1}{\sigma}}   \right)^{\frac{\gamma-1}{\gamma}\frac{\sigma}{\sigma-1}-1}H(s)^{-\frac{1}{\sigma}} F(s') ^{\frac{\sigma-1}{\sigma}} H'(s)>0,$
	and
	\begin{align*}
		\pi_{12}(s,s')=\left(c_{0}^{\frac{\sigma-1}{\sigma}}+c_{1}^{\frac{\sigma-1}{\sigma}}\right)^{\frac{\gamma-1}{\gamma}\frac{\sigma}{\sigma-1}-2} (HF)^{-\frac{1}{\sigma}} \left\{ (\frac{1}{\sigma} - \frac{1}{\gamma}) H'F\cdot (H^{\frac{\sigma-1}{\sigma} } F^{\frac{1}{\sigma}} F' - G^{-\frac{1}{\sigma}} G' ) +\frac{\sigma-1}{\sigma} H' F' \left(c_{0}^{\frac{\sigma-1}{\sigma}}+c_{1}^{\frac{\sigma-1}{\sigma}}\right) \right\}
	\end{align*}
	Next, let's focus on terms within the brace, which can be simplified
	as
	\[
	(1-\frac{1}{\gamma}) (HF )^{\frac{\sigma-1}{\sigma}}H'F'  + \left(\frac{1}{\gamma}-\frac{1}{\sigma}\right) FG^{\frac{-1}{\sigma}}G' + H'F'G^{\frac{\sigma-1}{\sigma}}\frac{\sigma-1}{\sigma}.
	\]
	It is positive when $1<\gamma<\sigma$.
\end{proof}

\end{document}